\documentclass[journal]{IEEEtran}
\usepackage{color} 
\usepackage[acronym]{glossaries}
\usepackage{hyperref}
\usepackage[noadjust]{cite}

\usepackage{subfigure}

\ifCLASSINFOpdf
   \usepackage[pdftex]{graphicx}
   \graphicspath{{../pdf/}{../jpeg/}}
   \DeclareGraphicsExtensions{.pdf,.jpeg,.png}
\else
   \usepackage[dvips]{graphicx}
   \graphicspath{{../eps/}}
   \DeclareGraphicsExtensions{.eps}
\fi
\begin{document}
%
\title{Satellite Communications in the New Space Era: \\ A Survey and Future Challenges}
%
%
%

\author{Oltjon~Kodheli, Eva~Lagunas, Nicola~Maturo, Shree~Krishna~Sharma, Bhavani~Shankar, Jesus~Fabian~Mendoza~Montoya, Juan~Carlos~Merlano~Duncan, Danilo~Spano, Symeon~Chatzinotas, Steven~Kisseleff, Jorge~Querol, Lei~Lei, Thang~X.~Vu, George Goussetis 

\thanks{This work was supported in part by the Luxembourg National Research Fund (FNR) under the following projects: ASWELL, FLEXSAT, DISBUS, ROSETTA, PROCAST, SIERRA, COHESAT and SATIOT.}
        
\thanks{O. Kodheli, E. Lagunas, N. Maturo, S. K. Sharma, B. Shankar, J. F. Mendoza Montoya, J. C. Merlano Duncan, D. Spano, S. Chatzinotas, S. Kisseleff, J. Querol, L. Lei, T. X. Vu  are with Snt - Interdisciplinary Centre for Security, Reliability and Trust, University of Luxembourg, 29 Avenue J.F. Kennedy,Luxembourg City L-1855, Luxembourg (e-mail: oltjon.kodheli@uni.lu)}

\thanks{G. Goussetis is with School of Engineering and Physical Sciences, Institute of Sensors, Signals and Systems, Edinburgh EH14 4AS, U.K.}

}


%
%

\markboth{IEEE COMMUNICATIONS SURVEYS \& TUTORIALS (DRAFT)}%
{Kodheli \MakeLowercase{\textit{et al.}}: Satellite Communications in the New Space Era: A Survey and Future Challenges}
%



 \newacronym{ats}{ATS}{Analog Transparent Scheme}
 \newacronym{dts}{DTS}{Digital Transparent Scheme}
 \newacronym{gw}{GW}{Gateway}
 \newacronym{rf}{RF}{Radio Frequency}
 \newacronym{fso}{FSO}{Free Space Optics}
 \newacronym{fsl}{FSL}{Free Space Loss}
 \newacronym{cflos}{CFLOS}{Cloud-Free Line Of Sight}
 \newacronym{hpa}{HPA}{High Power Amplifier}
 \newacronym{los}{LOS}{Line of Sight}
 \newacronym{imux}{IMUX}{Input Multiplexing}
 \newacronym{omux}{OMUX}{Output Multiplexing}
 \newacronym{ogs}{OGS}{Optical Gateway Station}
 \newacronym{ogsn}{OGSN}{Optical Ground Station Network}
 \newacronym{lnoa}{LNOA}{Low Noise Optical Amplifier}
 \newacronym{edfa}{EDFA}{Erbium-Doped Fiber Amplifier}
 \newacronym{agc}{AGC}{Automatic Gain Control}
 \newacronym{sim}{SIM}{Subcarrier Intensity Modulation}
 \newacronym{mimo}{MIMO}{Multiple Input Multiple Output}
 \newacronym{ook}{OOK}{On-Off Keying}
 \newacronym{ad}{A/D}{analog-to-digital}
 \newacronym{da}{D/A}{digital-to-analog}
 \newacronym{snr}{SNR}{Signal-to-Noise Ratio}
 \newacronym{fer}{FER}{Frame Error Rate}
 \newacronym{fec}{FEC}{Forward Error Correction}
 \newacronym{si}{SI}{Scintillation Index}
 \newacronym{ase}{ASE}{Amplified Stimulated Emission}
 \newacronym{adc}{ADC}{analog-to-digital converter}
 \newacronym{dac}{DAC}{digital-to-analog converter}
 \newacronym{lms}{LMS}{Least Mean Squares}
 \newacronym{osnr}{OSNR}{Optical Signal-to-Noise Ratio}
 \newacronym{dc}{DC}{Direct Current}
 \newacronym{srrc}{SRRC}{Square-Root Raised Cosine}
 \newacronym{ldpc}{LDPC}{Low Density Parity Check}
 \newacronym{osr}{OSR}{Oversampling Rate}
 \newacronym{hts}{HTS}{High Throughput Satellites }
 \newacronym{vhts}{VHTS}{Very High Throughput Satellites }
 \newacronym{rin}{RIN}{Relative Intensity Noise}
 \newacronym{ut}{UT}{User Terminal}
 \newacronym{rc}{RC}{Repetition Coding}
\newacronym{ilwc}{ILWC}{Integrated Liquid Water Content}

\maketitle

\begin{abstract}
Satellite communications (SatComs) have recently entered a period of renewed interest motivated by technological advances and nurtured through private investment and ventures. The present survey aims at capturing the state of the art in SatComs, while highlighting the most promising open research topics. Firstly, the main innovation drivers are motivated, such as new constellation types, on-board processing capabilities, non-terrestrial networks and space-based data collection/processing. Secondly, the most promising applications are described i.e. 5G integration, space communications, Earth observation, aeronautical and maritime tracking and communication. Subsequently, an in-depth literature review is provided across five axes: i) system aspects, ii) air interface, iii) medium access, iv) networking, v) testbeds \& prototyping. Finally, a number of future challenges and the respective open research topics are described.   
\end{abstract}


%
\IEEEpeerreviewmaketitle


\section{Introduction} \label{sec1}

Since their inception, Satellite Communications (SatComs) have found a plethora of applications, including media broadcasting, backhauling, news gathering etc. Nowadays, following the evolution of Internet-based applications, SatComs are going through a transformation phase refocusing the system design on data services, namely broadband SatComs. The main motivation is a) the rapid adoption of media streaming instead of linear media broadcasting and b) the urgent need to extend broadband coverage to underserved areas (e.g. developing countries, aero/maritime, rural). Furthermore, a major milestone of the 5th generation of communication systems (5G) is the integration and convergence of diverse wired and wireless technologies. In this context, SatComs pave the way for seamless integration targeting specific use cases which can take advantage of their unique capabilities. In parallel, private ventures have led the development of a multitude of manufacturing and launching options, previously only reserved for governments and a handful of large international corporations. This initiative named New Space has spawned a large number of innovative broadband and earth observation missions all of which require advances in SatCom systems. 

\begin{figure*}[!t]
\centering
\includegraphics[width=180mm,keepaspectratio]{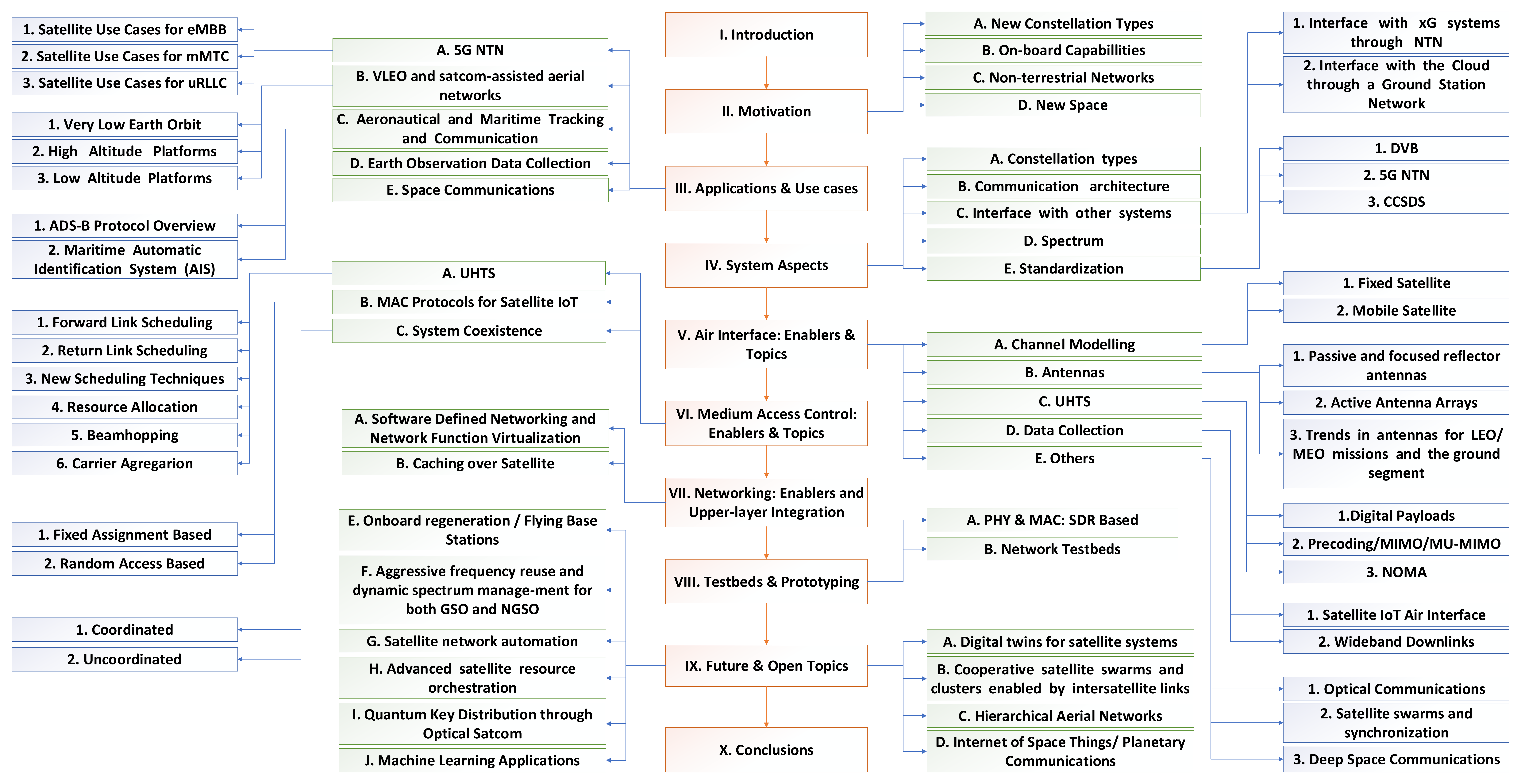}
\caption{Structure of the paper and topic classification}
\label{structure}
\end{figure*}

The purpose of this survey is to describe in a structured way these technological advances and to highlight the main research challenges and open issues. In this direction, Section \ref{sec2} provides details on the aforementioned developments and associated requirements that have spurred SatCom innovation. Subsequently, Section \ref{sec3} presents the main applications and use cases which are currently the focus of SatCom research.  The next four sections describe and classify the latest SatCom contributions in terms of 1) system aspects, 2) air interface, 3) medium access techniques 4) networking and upper layers. When needed, certain preliminaries are provided in a tutorial manner to make sure that the reader can follow the material flow without reverting to external sources. Section \ref{sec8} surveys communication testbeds which have been developed in order to practically demonstrate some of the advanced SatCom concepts. The last section is reserved for highlighting open research topics that are both timely and challenging. To improve the material flow we provide the structure of the paper in Fig. \ref{structure} and the list of acronyms in Table \ref{list_of_acr}.


\section{Motivation} \label{sec2} 

\subsection{New Constellation Types} \label{sec2a}
Traditionally, Geostationary (GEO) satellites have been mainly used for SatComs since they avoid fast movement between the terminals and the satellite transceiver and they allow for wide coverage using a single satellite. Multibeam satellite systems have been specifically developed to allow efficient frequency reuse and high-throughput broadband rates across the coverage area, not unlike their terrestrial cellular counterparts.  However, new more ambitious constellation types are currently being developed, motivated by advanced communication technologies and cheaper launch costs. 

In this direction, there has recently been a tremendous interest in developing large Low Earth Orbit (LEO) constellations that can deliver high-throughput broadband services with low latency. This constellation type has been the holy grail of SatComs since Teledesic first proposed it 25 years ago \cite{teledesic2}. However, it appears that now the relevant manufacturing and launching processes have matured and a viable implementation and deployment may be within grasp. Multiple companies, such as SpaceX, Amazon, OneWeb, TeleSAT, have already announced large LEO plans including thousands of satellites and some have already launched demo satellites. As of January 2020, SpaceX has deployed 242 satellites to build its Starlink constellation, with the goal to reach nearly 12000 satellites by mid-2020 \cite{starlink}.

\begin{table*}[!t] \caption{List of Acronyms}  \label{list_of_acr}
\centering
 \begin{tabular}{l l | l l}
 \hline
 \textbf{Acronyms} & \textbf{Definitions} & \textbf{Acronyms} & \textbf{Definitions} \\ 
 \hline
 3GPP & The 3rd Generation Partnership Project & LLO & Low Lunar Orbit \\
 5G & The 5th Generation of Mobile Communication Systems & LoRa & Long Range  \\
 ACM & Adaptive Coding and Modulation & LoS & Line of Sigh  \\
 ADC & Analog to Digital Converter & LPWAN & Low-power Wide Area Networks \\
 ADS-B & Automatic Dependent Surveillance-Broadcast &  LSA & Licensed Shared Access \\
 AFR & Array Feed Reflector &  LTE & Long Term Evolution \\
 AIS & Automatic Identification Systems & LUT  & Look up Table  \\
 ATC & Air Traffic Controller & MAC & Medium Access Control Layer \\
 ATCRBS & Air Traffic Control Radar Beacon System &  MBAs & Multibeam Antennas \\
 ATM & Air Traffic Management &  MEO & Medium Earth Orbit \\
  ATSC & Advanced Television Systems Committee  &    MFPB & Multi Feed per Beam \\
 AWGN & Additive White Gaussian Noise & MF-TDMA & Multi-Frequency Time Division Multiple Access  \\
 BS & Base Station  & MIMO & Multiple Input Multiple Output \\
 BATF & Backhauling and Tower Feed & mMTC & Massive MAchine Type Communications \\
 CA & Carrier Aggregation & ML & Machine Learning \\
 CCSDS & Consultative Committee for Space Data System &  MSS &  Mobile Satellite Services\\
 COOM & Communication on The Move  & MU-MIMO & Multi-User Multiple-Input Multiple-Output \\
 CN & Core Network & MUI & Multi-User Interface \\
 CR & Cognitive Radio & NB-IoT & Narrowband Internet of Things \\
 CSI & Channel State Information & NCC & Network Control Center \\
 CSS & Chirp Spread Spectrum & NFV & Netowrk Function Virtualization \\
 D2D & Device-to-Device & NGSO & Non Geostationary Orbit \\
  DRA & Direct Radiating Array & NOMA & Non-orthogonal Multiple Access \\
   DSP & Digital Signal Processing & NTN & Non Terrestrial Network  \\
    DSS & Dynamic Spectrum Sharing & OBP & On-board Processing \\
 DVB & Digital Video Broadcasting &   OFDM & Orthogonal Frequency Division Multiplexing \\
 DTPs & Digital Transparent Processors  & OMA & Orthogonal Multiple Access\\
 EBU & European Broadcasting Union & PAPR & Peak to Average Power Ratio\\
  EDRS & European Data Relay System & PHY & Physical Layer\\
  EPC  & Evolved Packet Core & QoS & Quality of Service \\
 EHF & Extremely High Frequency & RC & Repetition Coding \\
 eMTC  & enhanced Machine Type Communication & RF & Radio Frequency \\
 eMBB & Enhanced Mobile Broadband & RN & Relay Node \\
 EO & Earth Observation & RTT & Round Trip Time \\
 ESA & European Space Agency &  SatComs & Satellite Communications  \\
 ETSI & European Telecommunications Standards Institute & SC-FDM & Single Carrier Frequency Division Multiplexing  \\
 FAA & Federal  Aviation  Administration &  SDMA & Space Division Multiple Access\\
  FEC & Forward Error Correction & SDMB & Satellite Digital Multimedia Broadcasting\\
  FPGA & field programmable gate arrays & SDN  & Software Defined Networking\\
  GEO & Geostationary Orbit & SDR  & Software Radio\\
 GNSS & Global Navigation Satellite System &  SFPB & Single Feed per Beam \\
 GPS & Global Positioning System & SIC  & Successive Interference Cancellation \\
  GW & Gateway &  SLP & Symbol Level Precoding \\
  HAPs &  High Altitude Platforms & SINR & Signal to Interference plus Noise Ratio \\
  HEO & Highly Elliptical Orbit & SNR & Signal to Noise Ratio \\
  HTS & High Throughput Satellite & SOTM & Satellite on The Move \\
  HYMP & Hybrid Multiplay & TDRSS & Tracking and Data Relay Satellite System \\
  IETF & Internet Engineering Task Force  & THEF & Trunking and Headend Feed \\
   IFF & Identification Friend or Foe & TT\&C & Tracking and Command \\
  IMO & International Maritime Organization  & TWTAs & Traveling Wave Tube Amplifiers \\
  IMT & International Mobile Telecommunications  &  UAT & Universal Access Transceive \\
    IOT & In Orbit Testing  & UAVs  & Unmanned Aerial Vehicles \\
    IoT & Internet of Things & UCSS & Unipolar Coded Chirp Spread Spectrum  \\
  ISL & Inter-satellite Link  &  UE & User Equipment  \\
  ISS & International Space Station &  UHTS & Ultra High Throughput Satellite \\
  ISTB & Integrated Satellite-Terrestrial Backhaul   & UNB & Ultra Narrowband Signal \\
   KPI & Key Performance Indicators & uRLLC & Ultra Reliable and Low Latency Communications \\
 LAPs &  Low Altitude Platforms & USRP & Universal Software Radio Peripheral\\
 LDM &  Layered Division Multiplexing & UT & User Terminal \\
 LDPC & Low Density Parity Check  & VLEO  & Very Low Earth Orbit  \\
  LEO &  Low Earth Orbit & VNE  & Virtual Network Embedding   \\

 \hline
 \end{tabular}
\end{table*}

\begin{figure*}[!t]
\centering
\includegraphics[width=150mm,keepaspectratio]{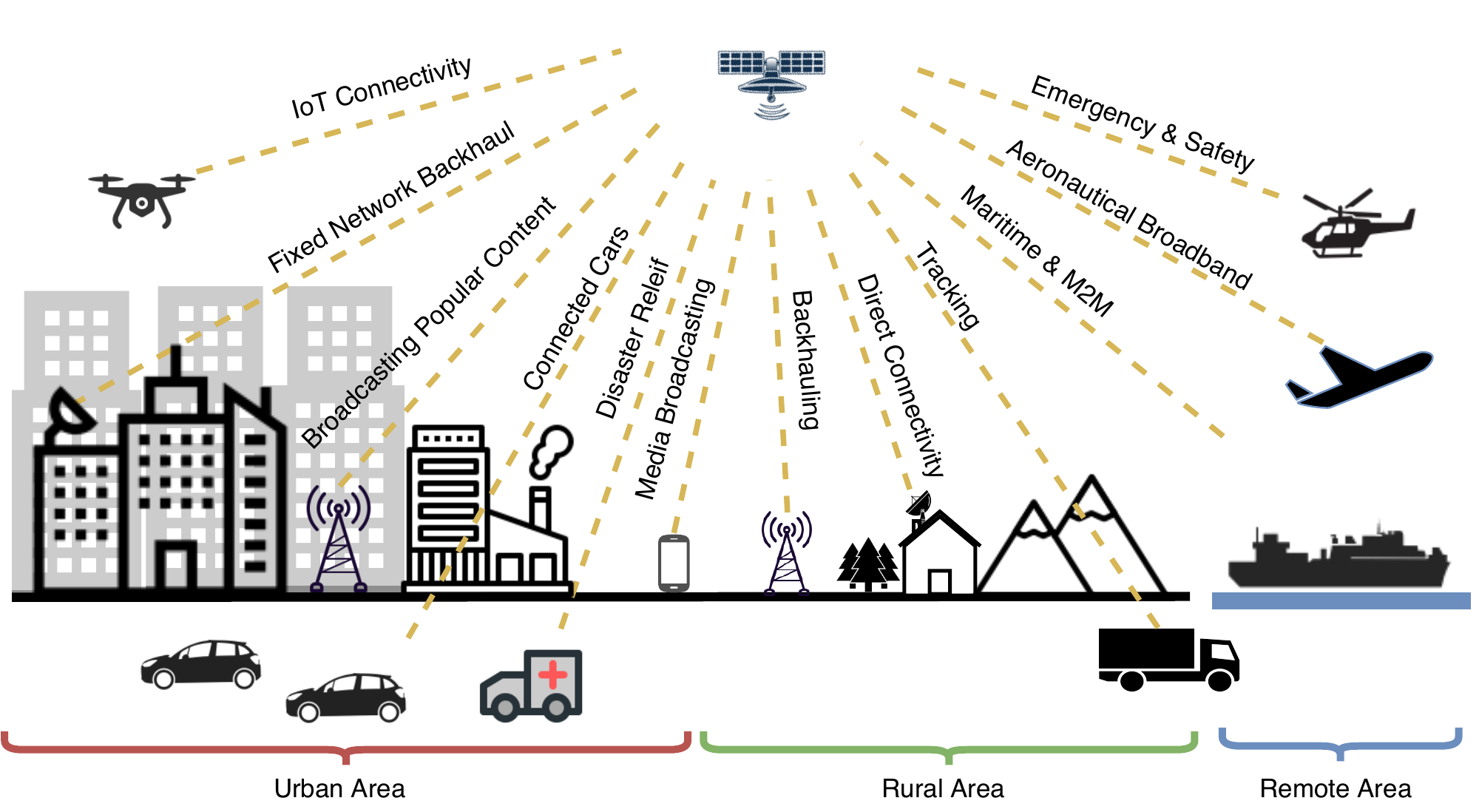}
\caption{The role of satellites in the 5G ecosystem}
\label{sat5g}
\end{figure*}

Moreover, we turn our focus to Medium Earth Orbit (MEO) where a constellation of 20 satellites (O3B) has been placed in a circular orbit along the equator at an altitude of 8063 km. Each satellite is equipped with twelve mechanically steerable antennas to allow tracking and handover of terminals. The next generation of O3B satellites is planned to use an active antenna (see Section \ref{sec5b2}) which can generate thousands of beams along with an on-board digital transparent processor (see Section \ref{digital_Payloads_Section}). This constellation type is unique since it manages to hit a trade-off between constellation size and latency. 

Finally, the proliferation of new constellation types has given rise to hybrid constellations which combine assets in different orbits. One such example is the combination of MEO and GEO connectivity, where the terminals can seamlessly handover between the two orbits \cite{sesref}. Another example is the backhauling of LEO satellite data through higher orbit satellites \cite{edrs, audacy}.

\subsection{On-board Capabilities}
Traditionally, the on-board processing capabilities have been the limiting factor for advanced SatCom strategies. Firstly, the majority of satellites operate as a relay which frequency-converts, amplifies and forwards and thus the on-board processing has to be waveform agnostic. Secondly, there is usually a large path loss to combat and a limited power supply which is tightly correlated with the satellite mass and launch cost. Thirdly, employed on-board components and technologies have to be ultra-reliable and robust since there is very little chance of repairing/replacing after the asset is put in orbit. 
Nevertheless, recent advances in the efficiency of power generation as well as the energy efficiency of radio frequency and digital processing components have allowed for enhanced on-board processing which can enable innovative communication technologies, such as flexible routing/channelization, beamforming, free-space optics and even signal regeneration (see Section \ref{sec5c}). Furthermore, space-hardened software-defined radios can enable on-board waveform-specific processing which can be upgraded during the satellite lifetime. Finally, cheap launching cost and conveyor-belt manufacturing allow for deploying more risky/innovative approaches while keeping up with the latest evolutions in communication technology. 

\subsection{Non Terrestrial Networks}
Non Terrestrial Networks (NTN) is a term coined under 5G standardization to designate communication systems that include satellites, Unmanned Aerial Systems (UAVs) or High Altitude Platforms (HAPs). The main objective of this initiative is to seamlessly integrate these assets into the 5G systems by studying their peculiarities in terms of architecture and air interface. More importantly, the relevant stakeholders would like to valorize unique characteristics of NTNs, such as wide coverage, multicast capabilities and the complementarity with local terrestrial infrastructure. Furthermore from a deployment point-of-view, the cost can be largely decreased by using 5G chipsets/systems and tapping into economies of a larger scale. In this direction, a number of promising use cases have been put forward (see Section \ref{sec3a}) and specific adaptation points of the current 5G standards have been suggested through the relevant working groups, focusing on air interface compatibility and architectural integration (see Section \ref{sec4c1}).

\subsection{New Space}
New Space does not refer to a specific technology, but it rather implies a new mentality towards space. It originated from three main aspects: 1) space privatization, 2) satellite miniaturization, 3) novel services based on space data. Privatization refers to the manufacturing and especially the launching of satellites by private companies, such as SpaceX and Rocket Lab, in contrast to the traditional institutional approach. In parallel, satellite and component miniaturization allowed easy access to space by multiplexing multiple cube/micro/nano-satellites into a single launcher. The combination of the two first aspects has lead to the latter, by allowing quick and relatively inexpensive access to space. In this direction, a wealth of data collection constellations have made it into orbit, spanning a wide range of services e.g. earth observation, radio frequency (RF) monitoring, asset tracking, sensor data collection etc. Bringing our focus back to communication aspects, New Space has inspired new opportunities in terms of collecting data from ground sensors directly via satellites, i.e. Satellite Internet of Things. Currently, tens of private companies are building demonstrators and competing to launch a viable commercial service. Almost all such ventures rely on low earth orbits and this raises additional communication challenges in efficiently downlinking the collected data back to the ground for processing. Conventionally, each such venture would require an extensive network of earth stations for high availability. However, cloud-based services (e.g. Amazon Web Services) have rolled out ground station networks that can be shared among the various constellations, while providing easy access to high performance computing for the data processing (see Section \ref{sec4c2}).   


\section{Applications \& Use Cases} \label{sec3}

The aim of this section is to outline and briefly describe some of the most relevant applications and use cases where SatComs can play a significant role. 

\subsection{5G Non Terrestrial Network} \label{sec3a}

5G will be more than just an evolution of the previous standards, embracing a wide new range of applications so as to satisfy future important market segments, such as automotive
and transportation sectors, media and entertainment, e-Health, Industry 4.0, etc., \cite{5gppref1}, \cite{5gppref2}. Three major groups of 5G use cases are defined by ITU-R for International Mobile Telecommunications (IMT) for 2020 and beyond (IMT‐2020) \cite{ITUref}: enhanced mobile broadband (eMBB), massive machine-type communication (mMTC) and ultra-reliable and low latency communications (uRLLC). The role that the satellites can play in the 5G ecosystem is crucial and has been widely recognized. The 3rd Generation Partnership Project (3GPP) initiated new activities in March 2017 to study the role of the satellites in the 5G, and two study items (SI) have already been concluded \cite{3gppref1}, \cite{3gppref2}. After two years of a study phase, it is now approved that NTN will be a new key feature of 5G and a work item (WI) will start from January 2020 \cite{3gppnews}.

Three major groups of use cases for NTN 5G systems have been defined by the 3GPP \cite{3gppref3}. Firstly, NTN can significantly enhance the \textbf{5G network reliability} by ensuring service continuity, in cases where it cannot be offered by a single or a combination of terrestrial networks. This is especially true in case of moving platforms (e.g. car, train, airplane etc.) and mission-critical communications. Secondly, NTN can guarantee the \textbf{5G service ubiquity} in un-served (e.g. desert, oceans, forest etc.) or underserved areas (e.g. urban areas), where a terrestrial network does not exist or it is too impractical/cost-ineffective to reach. Last but not least, NTN can enable the \textbf{5G service scalability} due to the efficiency of the satellites in multicasting or broadcasting over a very wide area. This can be extremely useful to offload the terrestrial network, by broadcasting popular content to the edge of the network or directly to the users.  
A more detailed list of the satellite use cases for each 5G service group can be found below and an illustration is shown in Fig. \ref{sat5g}.

\subsubsection{Satellite use cases for eMBB}
The authors in \cite{usecases1} come up with a consolidated list of satellite-based 5G uses cases for the eMBB service, as listed hereafter.

\begin{itemize}
    \item Backhauling and tower feed (BATF): In this use case the satellite provides a complementary role by backhauling the traffic load from the edge of the network or broadcasting the popular content to the edge, hence optimizing the operation of the 5G network infrastructure;
    \item Trunking and head‐end feed (THEF): The satellite ensures a direct 5G connectivity in remote areas where a terrestrial infrastructure is difficult or impossible to implement;
    \item Hybrid multiplay (HYMP): The satellite enables 5G service into home/office premises in underserved areas via hybrid terrestrial-satellite broadband connections;
    \item Communications on the move (COOM): The satellite provides a direct or complementary connectivity to support 5G service on board moving platforms, such as aircraft, vessels, and trains.
\end{itemize}

\subsubsection{Satellite use cases for mMTC}
The massive machine-type communication, also known as internet of things (IoT), has to do with low complexity and extremely cheap devices (sensors/actuators) able to generate and exchange information. Even though small in nature, the traffic generated by these IoT devices, will have a significant impact on the network load. 
Therefore, the satellites can help to offload the terrestrial IoT network through backhauling, or provide service continuity in cases where a terrestrial network cannot reach.
This group of uses cases can be categorized into two smaller subgroups depending on the type of application that the satellite can support and on how the IoT sensors are distributed on Earth. 

\begin{itemize}
    \item Wide area IoT services: This use case has to do with applications based on a group of IoT devices distributed over a wide area and reporting information to or controlled by a central server. Typical applications where the satellite can play a role include:
    \begin{itemize}
         \item Energy: Critical surveillance of oil/gas infrastructures (e.g. pipeline status)
         \item Transport: Fleet management, asset tracking, digital signage, remote road alerts
         \item Agriculture: Livestock management, farming
    \end{itemize}
    \item Local area IoT services: The IoT devices in this kind of applications are used to collect local data and report to the central server. Some typical applications can be a smart grid sub-system (advanced metering) or services to on-board moving platforms (e.g. container on board a vessel, a truck or a train).
\end{itemize}

\subsubsection{Satellite use cases for uRLLC}

This 5G use case is expected to support services where the delay in the communication link (lower than 1 ms) and the reliability (1 packet loss in $10^5$ packets) is of utmost importance. Some typical application examples include autonomous driving, remote surgery, factory automation etc. It is clear that the satellite, regardless of the selected orbit altitude, is not able to directly support these services due to the increased latency in the communication link. However, it can be crucial in certain cases by playing a supporting role. A typical example is content broadcast over a wide area and intelligently cached locally (either at the network edge or directly at the terminals), which can enable the low "perceived" latency (lower than 1 ms) at the user side. Also, if we consider the autonomous driving use case, the satellites can be extremely useful for car software updates, traffic updates, etc., because of its broadcasting over a wide area capabilities.  

\subsection{VLEO and SatCom-assisted Aerial Networks}

\begin{figure}[!t]
\centering
\includegraphics[width= 0.9\linewidth,keepaspectratio]{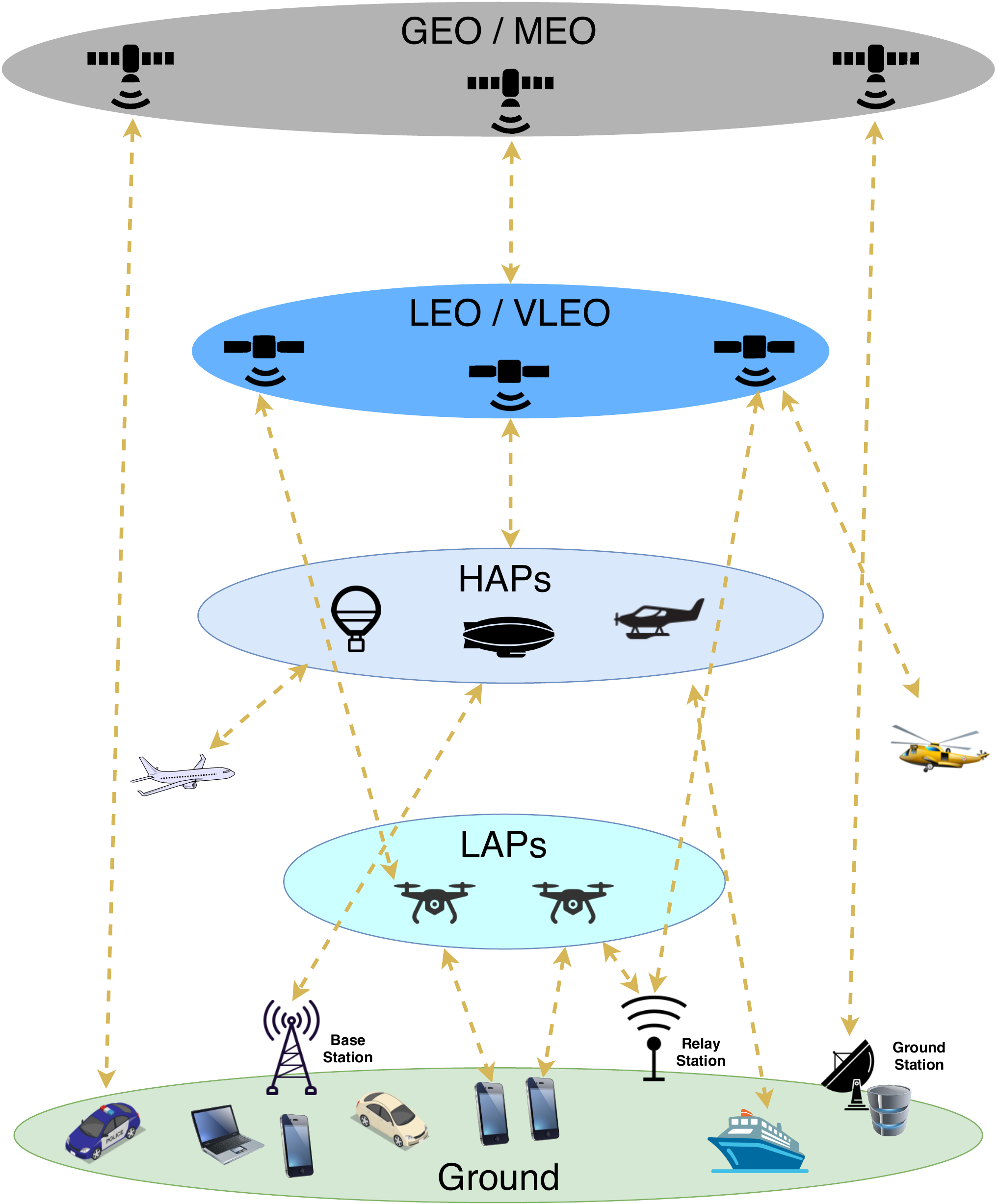}
\caption{Multi-layer communications architecture.}
\label{multi-layer-comms}
\end{figure}

During the last years, intermediate layers of communications systems between terrestrial and traditional satellite segments have emerged thanks to the technological advance of the aerial and miniaturized satellite platforms. Regardless of the application, these new platforms can be classified according to their operation altitude. Three major groups can be distinguished: Very Low Earth Orbit (VLEO) satellites, High Altitude Platforms (HAPs), and Low Altitude Platforms (LAPs). Their respective altitude ranges are  \cite{VLEO-comms,aerial-comms} 100 to 450 km for VLEO, 15 to 25 km for HAPs, and 0 to 4 km for LAPs. The advent of these new platforms enables a new multi-layer communications architecture \cite{LEO-MSS-HAPS} with multiple inter-layer links capable to overcome the most challenging scenarios. Fig. ~\ref{multi-layer-comms} shows a schematic approach of this new multi-layer communications paradigm. The following subsections summarize the benefits ad challenges of LAPs, HAPs and VLEO satellites.

\subsubsection{Very Low Earth Orbit}
VLEO platforms operate closer to the Earth than LEO satellites. This allows them to be simpler, smaller, and thus, cheaper \cite{VLEO-comms}. However, such low altitudes contain a denser part of the atmosphere, and therefore, larger aerodynamic forces. This can be seen as a challenge, but they can also represent an opportunity for orbit and attitude control \cite{VLEO_attitude}. Moreover, the increased drag represents a shortening of the orbital lifetime, but this also means a more frequent fleet replacement of smaller and cheaper spacecrafts, thus, becoming more responsive to technology and market changes \cite{VLEO-comms2}. Several private companies such as SpaceX, OneWeb or Telesat are planning to launch their Mobile Satellite Services (MSS) at VLEO.

\subsubsection{High Altitude Platforms}
HAPs have the potential to complement conventional satellite networks. Indeed, they are also known as High Altitude Pseudo-Satellites \cite{happiest,haps-teleo}. Due to their working altitude, HAPs have the potential to provide communications services at a regional scale. There are two main ways of cooperation between satellites and HAPs according to \cite{happiest,haps-teleo}:

\begin{itemize}
\item \textit{Backhauling}: HAPs can be an intermediate element between the satellite and the ground receiver. This two-step downlink communication will have a first hop between satellite and HAPs, and a second hop between HAPS and ground. The former is prone to the use of high bandwidth optical links, as it suffers little atmospheric effects, whereas the latter has a much shorter path than the satellite height, which improves the link budget enabling smaller antennae (cost-saving) or wider bandwidth (revenue increase).

\item \textit{Trunking}: HAPs have a good balance between regional coverage and reduced signal degradation. This triggers their use as low-cost deployment solution for broadcast or multicast services, allowing the users to directly connect within its coverage area and going to the satellite for inter-coverage communications.
\end{itemize}

Despite their promising applications, HAPs are still facing some major challenges for their deployment at a global scale, although they have been successfully deployed in emergency scenarios \cite{Loon1, Loon2}. One of the main challenges is the limited autonomy, especially in higher latitudes due to the reduced amount of daylight hours. Another is the weather conditions since high wind speeds may drag HAPs away from their operating area and low temperatures reduce the lifetime of the batteries. However, their benefits have been studied in depth in \cite{aerial-comms,happiest,haps-teleo}. The following list highlights the main advantages of the use of HAPs in communication networks:

\begin{itemize}
    \item \textit{Geographical coverage}: HAPs provide an intermediate coverage range between terrestrial and satellite systems.
    \item \textit{Fast deployment}: aerial base-stations can be deployed for operation within hours. They can be a supplement or complement to the existing terrestrial and satellite communications networks when they are overloaded or in case of failure.
    \item \textit{Reconfiguration}: HAPs can be operated for long periods, but they can also return to the ground for reconfiguration.
    \item \textit{Propagation delay}: the propagation delay ($\sim$50-85~$\mu$s) is significantly lower compared to the GEO ($\sim$120~ms), MEO ($\sim$15-85~ms) and even LEO satellites ($\sim$1.5-3~ms), offering important advantages for delay-sensitive applications.
    \item \textit{Less infrastructure}: a simple aerial platform can serve a large number of terrestrial cells, limited by its antenna technology.
\end{itemize}

\subsubsection{Low Altitude Platforms}
Unmanned Aerial Vehicles (UAVs) are the most prominent example of LAPs, but other systems, such as tethered balloons \cite{LAP-example}, have been also used for communication purposes. UAVs are expected to be an important component of the near-future wireless networks. They can potentially facilitate wireless broadcast and support high rate transmissions\cite{UAV-comms,UAV-comms2}. The main benefits of UAVs (and LAPs) are similar to the HAPs ones, but at a cellular level: fast and flexible deployment, strong line-of-sight (LoS) connection links, and additional design degrees of freedom with the autonomous and controlled mobility. Moreover, UAV-enabled aerial base stations may establish, enhance, and recover cellular coverage in real-time for ground users in remote, densely populated, and disastrous areas.

Despite the technological maturity of UAVs, UAV-based communication networks have not been widespread because of several limiting factors such as cost constraints, regulatory frameworks, and public acceptance \cite{aerial-comms}. The use of autonomous UAVs as 5G aerial base stations or as relays in a multi-layer vertical architecture is also a major research topic \cite{5G-Sky}. The technical challenges to be overcome are:

\begin{itemize}
    \item Improve the operation range and safety of the drones.
    \item Integrate trustfully beyond-visual-line-of-sight communication.
    \item Assessing the applicability of all 5G capabilities in UAV base stations.
\end{itemize}

Further work related to wireless communications using HAPs and LAPs may be found in \cite{HAPs_Further1} and \cite{HAPs_Further2}.

\subsection{Aeronautical and Maritime Tracking and Communication} \label{sec3c}

In addition to the above-mentioned uses cases, satellites can also play an important role in the aeronautical and maritime tracking systems. These systems share many similarities with other kinds of Device-to-Device (D2D) communications and the IoT. Such similarities are the very low data rates, the sporadic nature of the communications, and the simplicity of the protocols.

\subsubsection{Automatic Dependent Surveillance-Broadcast}
\begin{figure}[!t]
\centering
\includegraphics[width=85mm,keepaspectratio]{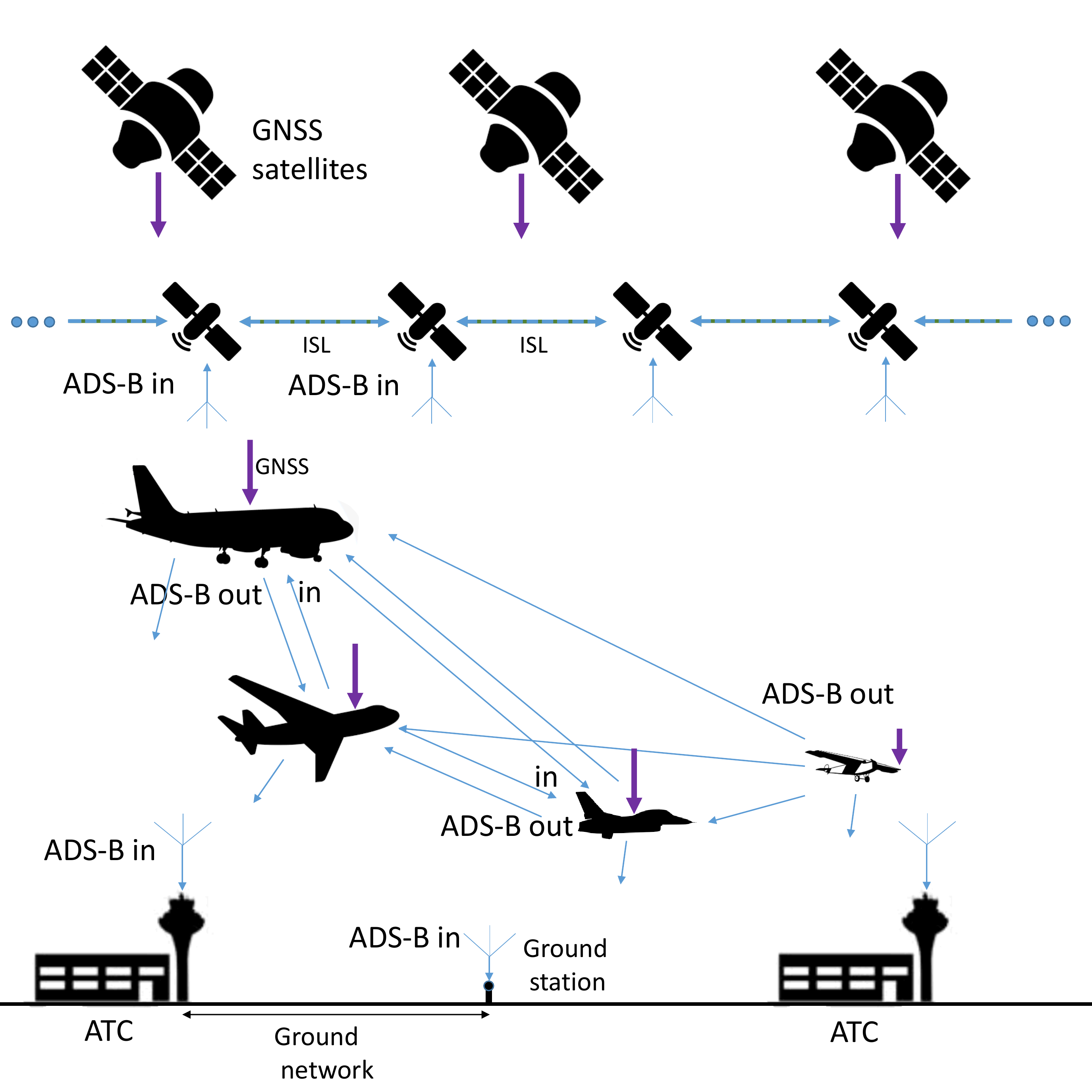}
\caption{ADS-B hierarchy, with integration of the satellites in low orbit into the scenario performing ADS-B reception. }
\label{adsb_hierarchy}
\end{figure}

Air transportation has been continuously increasing in the last years up to a number of two billion passengers per year in 2018, and is expected to continue growing to more than eight
billion by 2037 \cite{InternationalAirTransportAssociation2018}. This exponential increase may cause a shortage of the radio resources and collapse the Air Traffic Management (ATM) system, on which will rely the safety of billions of future passengers \cite{Strohmeier2014}.
Next generation air traffic management systems are increasingly supported on Automatic Dependent Surveillance-Broadcast (ADS-B). Although ADS-B is not mandatory yet in all the regions of the world, it will be operational in most of the flying aircrafts by 2020. 
The ADS-B system is based on the capability of the aircraft to navigate to a destination (typically using Global Navigation Satellite System (GNSS) data and barometric altitude), communicate to an air traffic controller, and to participate in cooperative surveillance to air traffic control for separation and situational awareness services. ADS-B is automatic as it requires no human intervention, and it is dependent on the data coming from the aircraft navigation system.
The ADS-B signals are received by the available sensors, which are connected in the ATM network. These sensors have been usually deployed on ground in the proximity of the Air Traffic Controller (ATC). However, as the under-the-horizon transmission is not feasible, ground-based ADS-B receivers cannot accurately receive signals from flights passing over areas without ground stations, such as in the middle of the oceans or in the Arctic regions. As a result, a large part of the airspace still remains unsupervised \cite{Francis2011,Strohmeier2014} and the ground stations become congested by the workload they require to process. For these reasons, during the last years, it is proposed to implement space-based ADS-B receivers using a LEO constellation of small satellites which can become part of the complete ATM relay network. In this way is possible to achieve low latency and secure global ADS-B coverage \cite{Francis2011,Knudsen2014}. An illustration of a satellite-based ADS-B system is shown in Fig. \ref{adsb_hierarchy}.

Frequent and reliable ADS-B communication from space allows to improve the efficient use of the aerospace and increase the aircraft security. This is explained by the fact that the aircraft climbing trajectory is optimized for a given security constraints, saving millions of dollars per year in  fuel consumption \cite{juan_RTCA-DO-144A}.
However, this comes up with the cost of having an increased amount of data generated which has to be routed towards the control centers. Some specialized companies offer the services of satellite based ADS-B reception and networking. Some examples are SPIRE \cite{SpireWeb} and Aireon \cite{juan_adsb_aireon}. Both companies provide global air traffic surveillance system using a space-based ADS-B network, and with the help of cloud computing.

\subsubsection{Maritime  Automatic Identification System} \label{sec3c2}

The Automatic Identification System (AIS) is currently used on ships as a short-range tracking system and it is regulated by the International Maritime Organization (IMO) \cite{juan_IMO2018}. It provides the vessels and shore stations with information on identification and positioning on real-time in order to avoid ship collision accidents. Despite having been specified in the late part of the twentieth century it has only gained popularity over the last decade due to the use of satellite-based receivers which provides global coverage, improved response times and more reliability \cite{juan_Yang2019}. Since 2008, satellites equipped with AIS receivers have been able to detect AIS signals transmitted by AIS transceivers on a global scale. Space-based AIS receptions open the possibility of unmanned transoceanic journeys, convenient for the transport of hazardous materials, which consequently enables the elongation of the duration of non-time-critical journeys, optimize the fuel consumption or even allows the direct use of electrical or solar power \cite{juan_Winther2014}, \cite{juan_FernandezArguedas2018}.
Additionally, these satellites serve as supplementary data sources for vessels and coastal authorities in busy port areas where conventional AIS receivers may not be able to cope with the large volume of ocean traffic \cite{juan_Kaluza2010}. Satellite-based AIS provides an easy way for  collecting AIS data on a global scale in almost real-time \cite{juan_7947788, juan_ais_Sahay2017}. Commercial exploitation of space AIS has been carried out during the last decade by companies such as SpaceQuest, Elane, ExactEarth, Marine Traffic, ORBCOMM, and SPIRE \cite{SpireWeb}). 

\subsection{Earth Observation Data Collection}

Traditionally, Earth Observation (EO) has been used by Governmental or International agencies to report the weather, monitor the oceans, detect changes in vegetation and analyse the damage done by natural disasters like earthquakes or hurricanes. It provides objective data on what really happens, showing trends and changes over time in a way that could never be observed from the ground.

However, in the last few years the space industry is experiencing a trend towards investment in so-called “agile” space activity as opposed to traditional “big space” government's  program. Agile space has the potential to open up space program to a wider, more flexible range of players, such as universities, companies and developing countries. The agile space sector is broadly split into two segments: upstream and downstream. Upstream space is focused on hardware, launchers, rockets and satellites, whereas downstream space data activities take information from the upstream and turn it into useful applications for business. 

Private space data collection and space data analytic companies, like SPIRE \cite{SpireWeb}, are proposing new type of services by combining together satellite technology to collect information and modern data analysis techniques (e.g. machine learning). A field where satellite information collection and machine learning data analytic can be very effective is the field of logistics. Consider, for instance, the task of monitoring the number of containers that are moved in an harbor during the day. An effective way to accomplish this objective is to take pictures of the harbor container storage zone trough a fleet of small LEO satellite. Then, these pictures are sent back to Earth, where they are processed using some machine learning technique in order to efficiently count the number of containers that have been moved between the different satellite passages. This allows to get a count of the total number of containers moved in that harbor during the day \cite{PlanetWeb}.

While the LEO orbits guarantee some advantages for EO purposes, it also poses some challenges from the telecommunication point of view. First of all, satellites in LEO orbit move relatively fast and because of this they can guarantee coverage of a certain area only for a few minutes each several hours. Hence, to guarantee continuous coverage a large fleet of satellites is needed. For the same reason, a Gateway (GW) can stay in contact with the satellite for a very limited amount of time. To guarantee full-time connectivity between the ground and the satellites' fleet either a large number of GWs must be built all around the globe, or inter-satellite link (ISL) capabilities must be implemented in the satellites. More details on this matter can be found in Section \ref{sec4b}.


\subsection{Space Communications}

Telecommunications play a fundamental role in space exploration. Seeing Apollo 11 land on the Moon, downloading Pluto’s pictures from New Horizon, receiving scientific data on 67-p/Churyumov-Gerasimenko comet from Rosetta, commanding Voyager 1 to turn its camera and take a photograph of Earth from a record distance of about 6 billion kilometers- all these and many other incredible achievements would have been impossible without very efficient communication systems between us and our space explorers.

The Space Exploration age began in 1957 with the launch of the Sputnik, and until now has been carried out mainly by either robotics missions, or very short human missions outside the Earth orbit, as in the case of the Apollo Program. 
The paradigm shift that we see today in space activities is best encapsulated by the term ‘Space 4.0’, where the different space agencies are planning to have stable human presence in other celestial bodies of our solar system. One of the most promising in this sense is the ‘Moon Village’ concept developed by ESA \cite{ESAmoonVILLAGE}, which seeks to transform this paradigm shift into a set of concrete actions and create an environment, where both international cooperation and the commercialisation of space can thrive. Such an ambitious goal will not be achievable if we are not able to guarantee high capacity and very reliable communication between Earth and these human outposts in the solar system.

Until now, all the space exploration activities have been carried out by National or Transnational Space Agencies (e.g NASA, ESA, ROSCOMOS, JAXA, etc.), but this scenario is going to change soon with the private sector entering the space exploration sector. In particular, there is a huge interest from young start-ups to exploit the resource available on asteroids and on the Moon. Asteroids contain, in fact, a huge amount of minerals, including gold, platinum, cobalt, zinc, tin, lead, indium, silver, copper, iron, and various rare-Earth metals. For millennia, these metals have been mined from the Earth's crust, and they have been essential to economic and technological progress.
In addition, there are thought to be many asteroids and comets that are largely composed of water ice and other volatiles (ammonia, methane, etc.). Water ice could be harvested to satisfy a growing demand for freshwater on Earth, for everything from drinking to irrigation and sanitation. Volatile materials could also be used as a source of chemical propellant like hydrazine, thus facilitating further exploration and mining ventures. In fact, in \cite{PlanRes} it is indicated that there are roughly 2 trillion metric tons (2.2 trillion US tons) of water ice in the Solar System.

The challenges related to Deep Space Communication and the available solutions are detailed in \ref{DeepSpaceComm}.

\nocite{monitor}



\section{System Aspects} \label{sec4}

This section covers the system aspects of a satellite communication system. Some preliminaries regarding SatComs are included in order to introduce terminology and facilitate the reader to follow the material flow.

\subsection{Constellation types} \label{sec4a}
A fundamental aspect of satellite constellations is the orbit's altitude, which severely affects the latency of the communication, the signal attenuation and the coverage. As anticipated, three basic orbit configurations are LEO, MEO, and GEO. The respective altitude ranges are 500 to 900 km for LEO, 5,000 to 25,000 km for MEO, and 36,000 km for GEO \cite{Elbert_book}. One might notice how extensive altitude ranges are not considered for the aforementioned satellite orbits. The reason for this is related to the Van Allen belts, which are regions containing energetic charged particles, most of which originate from the solar wind, that are captured by and held around the Earth by its magnetic field. The radiation levels of these zones are deemed to be unsuitable for the typical commercial satellites. Therefore, LEO, MEO, and GEO altitudes are such that the radiation field is within specific design constraints, consistent with achieving operating lifetimes in the range of 10 to 20 years.

A GEO satellite can cover about one third of the Earth’s surface, with the exception of the polar regions. This coverage includes more than 99\% of the world’s population and economic activity. The LEO and MEO orbits require more satellites to achieve such global coverage, since non-GEO satellites move in relation to the surface of the Earth, hence a higher number of satellites must be operating to provide continuous service. The fundamental trade-off is that the GEO satellites are farther and therefore are characterized by a longer path length to Earth stations, while the LEO systems promise short paths analogously to terrestrial systems. The path length introduces a propagation delay since radio signals travel at the speed of light. Depending on the nature of the service, the increased latency of LEO, MEO and GEO orbits may impose some degradation on the quality of the received signals or the delivered data rate. The extent to which this influences the acceptability of the service depends on several factors, such as the degree of interactivity, the delay of other components of the end-to-end system, and the protocols used
to coordinate information transfer and error recovery.

Another relevant characteristic of satellite orbits is the eccentricity. While for most SatCom services the orbits are circular, there are cases of elliptical orbits with high eccentricity, typically referred to as highly elliptical orbits (HEO).  Examples of inclined HEO include Molniya orbits and Tundra orbits. Such extremely elongated orbits have the advantage of long dwell times at a point in the sky during the approach to, and descent from, apogee. Bodies moving through the long apogee dwell appear to move slowly, and remain at high altitude over high-latitude ground sites for long periods of time. This makes these elliptical orbits useful for communications towards high latitude regions.

Besides the constellations of satellites orbiting around the Earth, it is also worth mentioning the existence of lunar orbiting satellites, which orbit around the moon. Among the different types of lunar orbits, low lunar orbit (LLO) are of particular interest for the exploration of the moon. Such orbits have an altitude below 100 km and a period of about two hours. Further, highly eccentric orbits are used for the moon exploration, too.

In general, in the design of a satellite constellation for SatCom services, it is important to assess a number of parameters and to evaluate their respective trade-offs. The principal performance parameter is the coverage, as the first requirement to guarantee the communication link is to reliably cover the regions of interest. Typically, the coverage of the satellite is assessed taking into account various practical restrictions, such as the minimum elevation angle for the user terminal and required service availability. Another fundamental performance parameter to be considered is the link latency, which is directly related to the constellation altitude, as previously mentioned. While high altitude constellations, such as GEO ones, allow wide coverage, they suffer a much higher latency compared to the lower altitude ones. Furthermore, satellites at lower altitudes move faster, which leads to higher Doppler frequency offset/drift and can be crucial for the design of the user equipment, especially for wideband links, as described in Section \ref{wideband_sec5d2}. This trade-off in the altitude choice clearly needs to be addressed taking into account the type of service to be provided. Concerning the cost of constellations, the principal parameter is clearly the number of satellites, thus it is important to achieve the desired performance keeping this number as low as possible. Also, the number of orbital planes affects the overall cost, as changes require large amounts of propellant. Ultimately, once the constellation altitude is selected based on the specific service to be provided, the constellation design aims at guaranteeing coverage in the regions of interest, using the lowest possible number of satellites and orbital planes. After that, the satellite payload and architecture are designed by taking into account the system requirements. More details on communication architecture can be found in Section \ref{sec4b}. In this context, the number of supporting ground stations and the frequency plan are decided with respect to the optimized resource allocation, such as signal bandwidth and transmit power. Based on the maximum resources needed in order to satisfy the service demand at all times the system architecture is finalized.

\subsection{Communication architecture} \label{sec4b}
\begin{figure}
\centering
\includegraphics[width=0.9\linewidth,keepaspectratio]{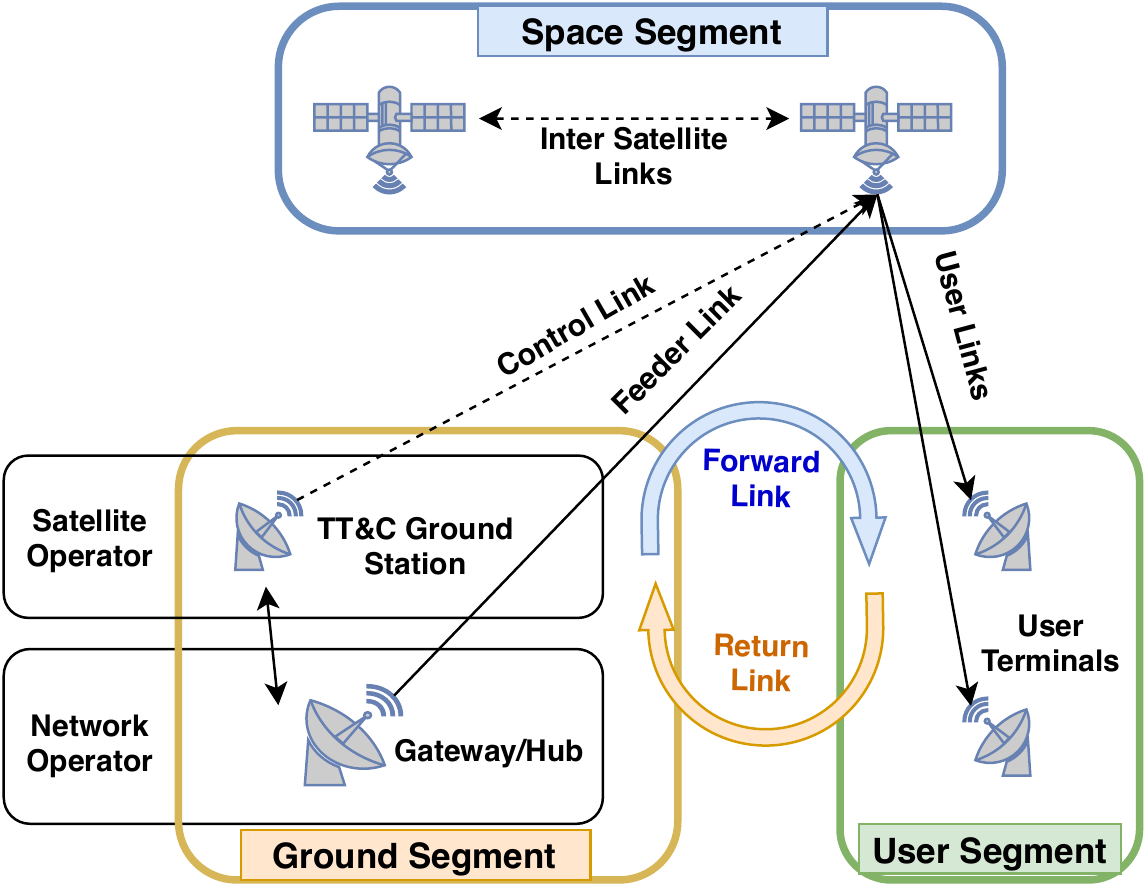}
\caption{SatCom System Architecture}
\label{syst_overview}
\end{figure}

The basic structure of a satellite communication system consists of a space segment that includes the satellite constellation, a ground segment including GW stations and large ground facilities for control, network operations and backhauling, and a user segment with the user terminals deployed on fixed and mobile platforms (e.g. airplanes and ships), see Fig. \ref{syst_overview}. The link between the GW station and the user terminal via intermediate satellite is named the forward link, whereas the link coming from the user through the satellite to the GW is referred to as return link. The link connecting the GW with the satellite (in both directions) is named the feeder link. The link connecting the satellite with the user terminal is referred to as user link.

\begin{figure}
\begin{centering}
\subfigure a){\includegraphics[width=0.35\textwidth]{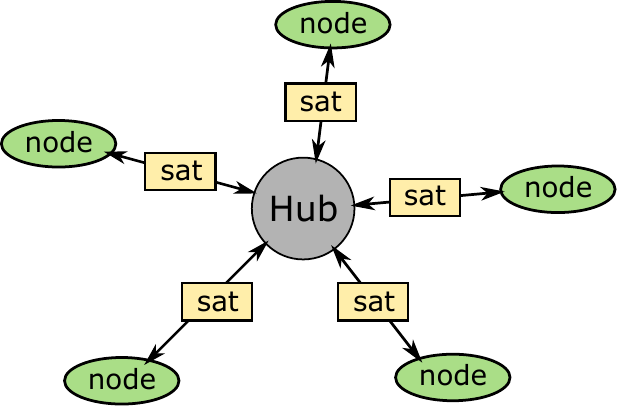}}
\subfigure b){\includegraphics[width=0.35\textwidth]{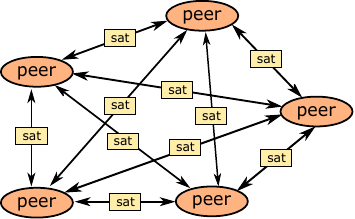}}
\caption{Communication topology: a) star; b) mesh.}
\label{topology}
\par\end{centering}
\end{figure}

\begin{figure*}[!t]
\begin{centering}
\subfigure a){\includegraphics[width=85mm,keepaspectratio]{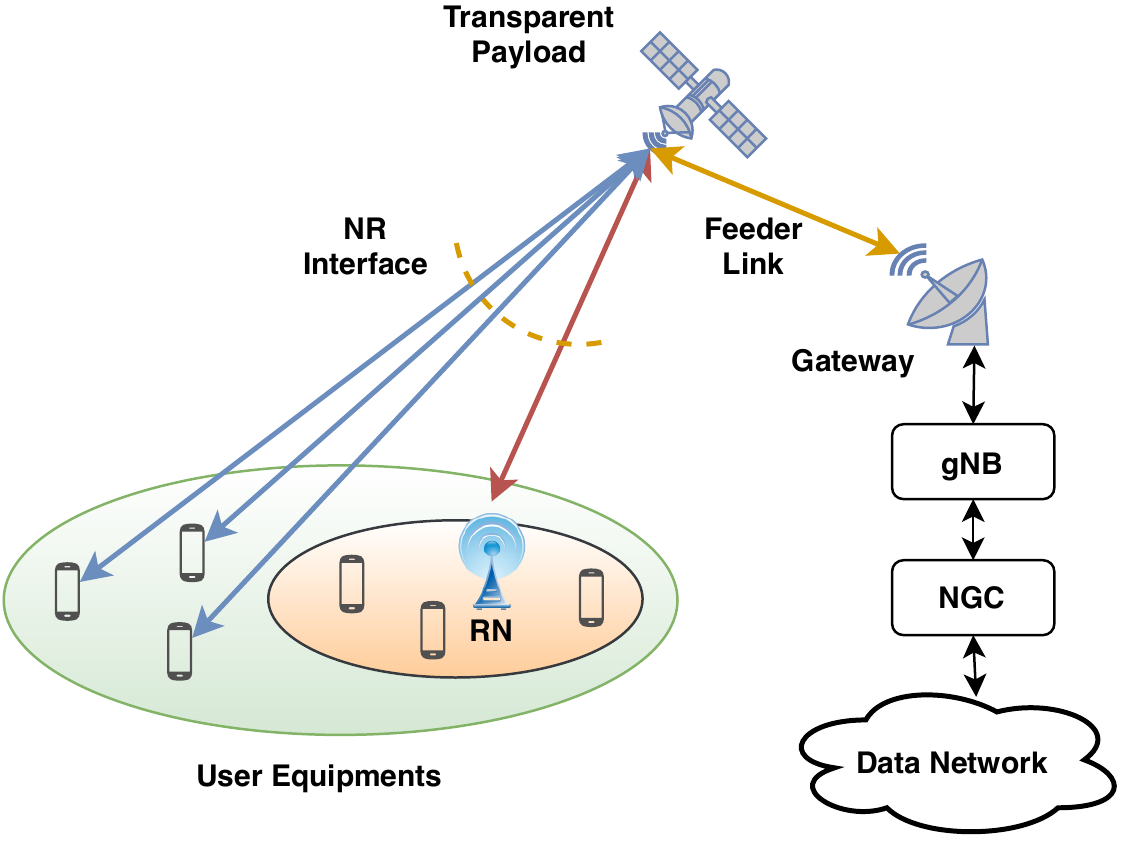}}
\subfigure b){\includegraphics[width=85mm,keepaspectratio]{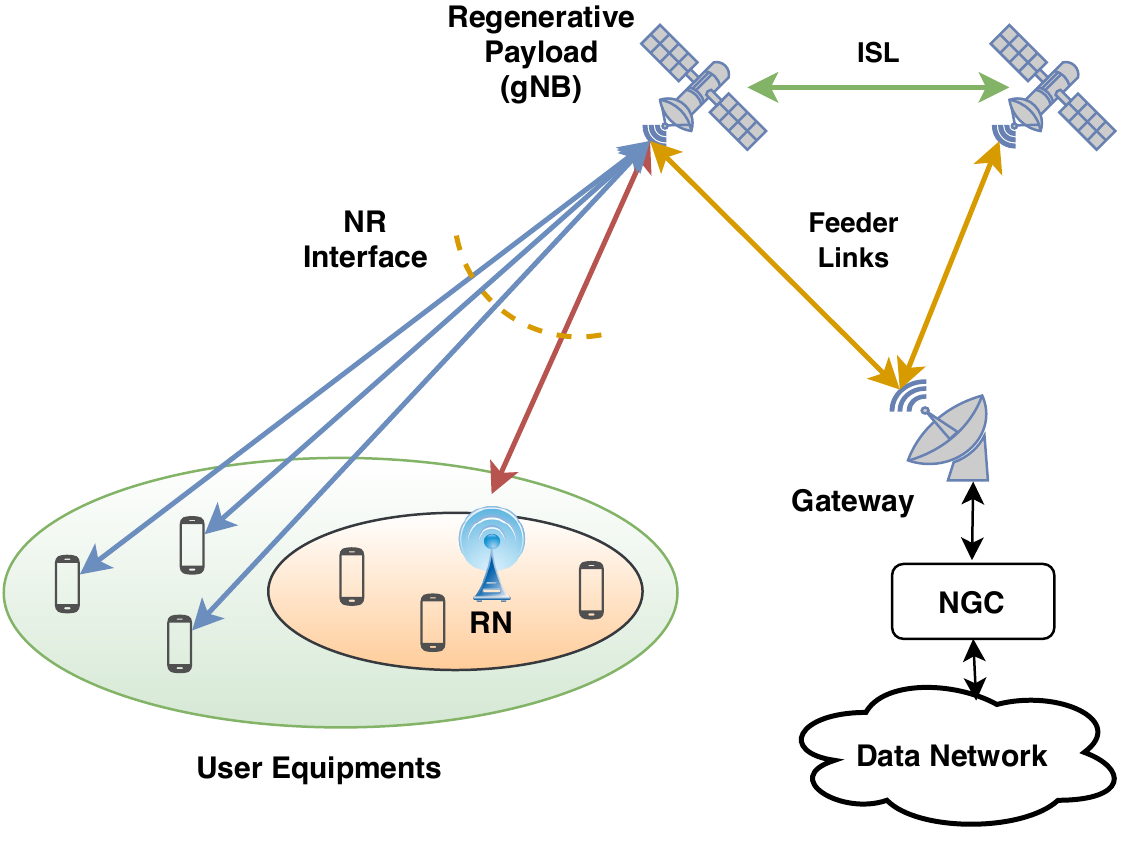}}
\caption{5G-NTN architecture options with a) transparent payload; b) regenerative payload.}
\label{5g_ntn_system}
\par\end{centering}
\end{figure*}

The control of the satellites is performed by the so-called Telemetry, Tracking and Command (TT\&C) stations. The main task of TT\&C stations is to monitor the status of the satellite sub-systems, run tests and update the configuration. Such control mechanisms are needed for the maintenance purposes and in order to keep the satellites on the respective orbits. Correspondingly, the operation of TT\&C stations falls into responsibility of the satellite operator. In contrast, the GW stations are run and maintained by the network operator, since they manage the network access and backhauling. As the coverage area of MEO satellites is typically larger than the coverage area of LEO satellites, LEO constellations require a substantially larger number of supporting GWs compared to MEO constellations. In contrast, GEO satellites require only one GW for backhauling due to their fixed position. For more information on ground segment, we refer to \cite{ground_segment_book}.

The communication topology depends primarily on the target application of the communication system. The two typical topologies are star and mesh. In both cases, the satellite acts as a relay between each node and the hub (backhaul) or between multiple peer nodes, respectively. Here, we differentiate between point-to-point and point-to-multipoint transmissions. The point-to-multipoint connectivity as in traditional broadcast services, internet connections via satellite and data collection from the sensors deployed on the earth surface, the star topology is used, where each terminal is connected to the hub via satellite on a single-hop basis, see Fig. \ref{topology}a). The data collection from the sensors deployed on board of the satellite (e.g. in earth observation applications), can be viewed as a special case of star topology, since the satellite acts both as a relay and as a signal source. For point-to-point connectivity as in video conferencing, the star topology would imply two-hop transmissions, which might be crucial with respect to the end-to-end latency of packet transmission. Hence, mesh topology is usually preferred, where each peer node can communicate with another peer node via satellite relay, see Fig. \ref{topology}b). However, this topology may require intelligent routing of data packets by the satellite. As an example, the mesh topology is employed by AIS (see Section \ref{sec3c2}). In addition, mesh topology has been recently proposed for various LEO and GEO satellite constellations based on optical ISLs in order to ensure a sufficient connectivity and cooperation between satellites. 

Upon being employed as a relay, the satellite can be either transparent or regenerative. Transparent satellites do not perform any signal processing besides amplification, spatial filtering and frequency conversion. Hence, the functionality of the satellite resembles an amplify-and-forward relay structure from traditional wireless communications. In contrast, regenerative satellites perform additional signal processing, e.g. decoding, interference cancellation, signal regeneration, etc., similar to decode-and-forward relaying. The payload of the satellite is designed accordingly. More details on the state-of-the-art types of digital payloads can be found in Section \ref{digital_Payloads_Section}.

In order to enhance the performance of satellite constellations, ISLs can be created, such that multiple satellites can cooperatively accomplish complicated missions. Correspondingly, the complexity of each satellite is reduced. Furthermore, ISLs can be employed for data offloading. The implementation of the ISLs can be done using traditional RF antennas or optical wireless technology. The latter is beneficial due to narrower beams generated by the employed lasers. A distinct advantage of this technique is the substantially reduced antenna size. The link can be established between multiple satellites of the same orbit (e.g. LEO-LEO) as well as between satellites of different orbits (e.g. GEO-LEO). A typical example for the latter is the use of GEO satellites as relays for the links between LEO satellites and GWs. As mentioned earlier, this technique is employed by specifically designed GEO satellite constellations, such as European Data Relay System (EDRS) or Tracking and Data Relay Satellite System (TDRSS) in order to improve the connectivity and coverage. Both systems are intended to provide the requested on-demand services in nearly realtime, especially for emergency applications. However, the coexistence of multiple satellites belonging to different orbital planes with coordinated and uncoordinated access is very challenging and therefore plays an important role in the system design/operation. We will discuss the inter-plane access technology in more detail in Section \ref{syscoex_section}.

\subsection{Interface with other systems} \label{sec4c}

\subsubsection{Interface with xG systems through NTN} \label{sec4c1}
From the system level point of view, in order to create an interface between the satellite and the 5G network, different architecture options have been identified within 3GPP studies for NTN \cite{3gppref2}. The different architecture options are categorized based on the payload type (e.g. transparent or regenerative) and the user access link type (direct or non-direct). They are illustrated in Fig. \ref{5g_ntn_system}. In case of a transparent payload, the satellite provides connectivity between the users and the base station, which is on ground. On the other hand, in case of a regenerative payload, the base station functionalities can be performed by the satellite. This option, even though it is more complex, would improve significantly the round trip time (RTT) of the communication. In addition, due to the regenerative payload, an ISL can be also established, which would be beneficial for hand-over procedures in case of a constellation of satellites (typically in LEO and MEO orbits). Both architecture options can ensure direct or non-direct access to the user equipments (UE) on ground. In the latter case, the access link to the users is provided by the relay nodes (RN), which are then connected to the base stations through the satellite link. The functionality of the RN and the air interface for the link between base stations and the RN is still under definition in the 3GPP. However, assuming that they would have a similar role as the RN in the Long Term Evolution (LTE) network, they can simplify the integration of the satellites in the 5G network, by aggregating the traffic coming from many users on ground. For interested readers \cite{5g_ntn_arch} provides a more detailed explanation of each component in these architecture options, whereas the challenges of a 5G satellite-terrestrial network integration and possible solutions can be found in \cite{5g_ntn_ch}\cite{5g_ntn_ch2}.

\subsubsection{Interface with the Cloud through a Ground Station Network} \label{sec4c2}

As already mentioned in Section \ref{sec2a}, it is expected that in the near future thousands of satellites will be in the LEO orbit. Therefore, the amount of data to be collected by these satellites will be tremendously high. In order to have access to this data, the interested customers must either build their own ground stations and antennas, or lease them from ground station providers. In addition, servers, storage and routing  capabilities are needed in order to store,  process and transport the data coming from the satellite. This requires a significant investment since the cost of each of the above mention components is high.  

\begin{figure}[t]
\centering
\centerline{\includegraphics[width=0.35\textwidth]{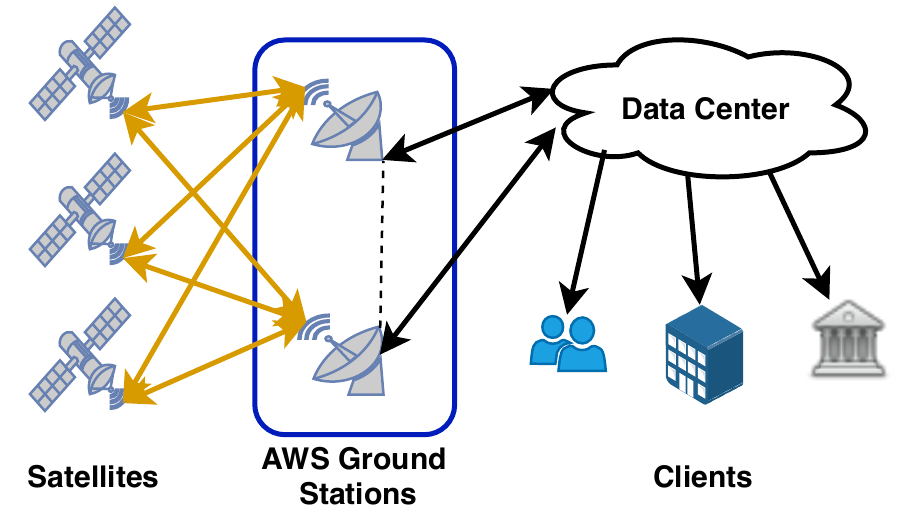}}
\caption{AWS Ground Station System Architecture}
\label{aws1}
\end{figure}

Through a Ground Station Network that can be shared among the various constellations, the data can be collected from the different satellites orbiting the Earth and stored in a central cloud. In such a case, the interested customers will only need to access the cloud, without the need for a long-term investment towards a personal ground station infrastructure. A typical example of such a system is the AWS Ground Station, which is an initiative launched by Amazon, and an illustration of the system architecture is shown in Fig. \ref{aws1}. Such a cloud based service solution, not only lowers the cost of sending data from space to Earth, but also it significantly reduces the data access delay \cite{aws}. 

\subsection{Spectrum}
\begin{figure}[t]
\centering
\centerline{\includegraphics[scale=0.3]{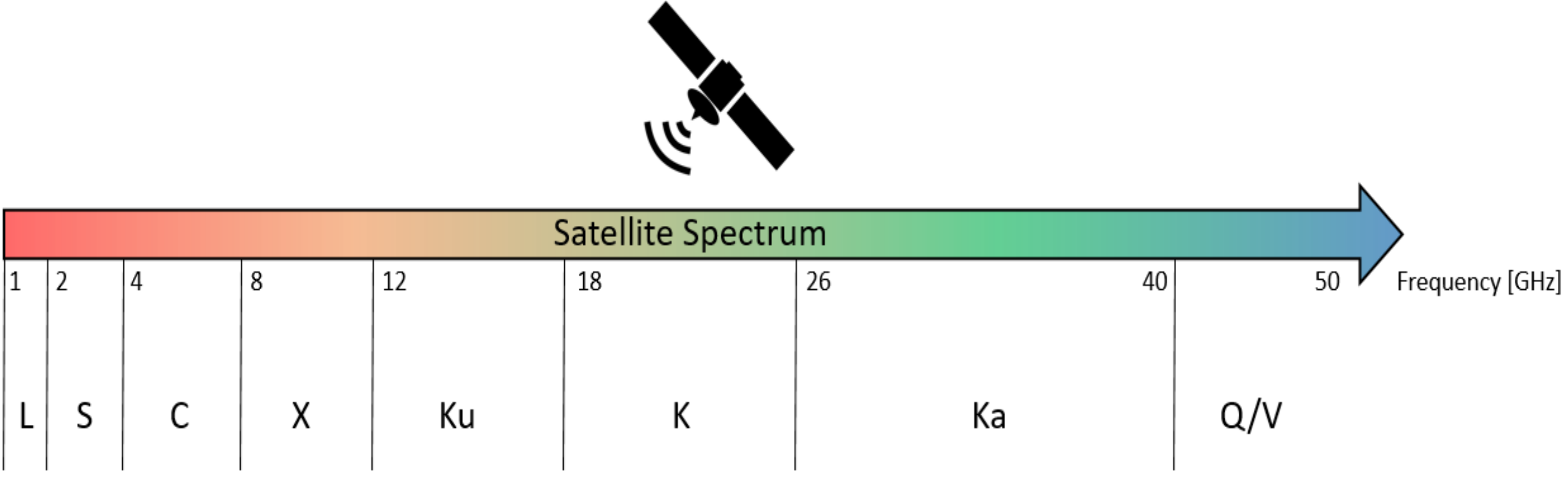}}
\caption{Satellite spectrum}
\label{SatSpect_Fig}
\end{figure}

Satellite communications operate in the Extremely High Frequency (EHF) band, in particular between 1-50 GHz. Different frequency bands are suitable for different climate conditions, types of service and types of users. For simplicity, the frequency bands used for satellites are identified by simple letters: (i) Lower frequencies (L, S, X and C-bands), and (ii) Higher frequencies (Ku, K, Ka, Q/V bands). A schematic illustration of the satellite spectrum is provided in Fig. \ref{SatSpect_Fig}.

Radio navigation systems, like GPS or Galileo, operate in L-band. S-band is used for weather radar, surface ship radar, and some communications satellites, especially those of NASA for communication with International Space Station (ISS) and Space Shuttle \cite{esoa}. L and S bands are also used for TT\&C. In particular, the frequency bands between 2-2.3 GHz  are  shared  co-equally  by  the space research, space  operation, and EO satellite services \cite{6689909}. Clearly, there is not much bandwidth available in lower bands, so it has become a costly commodity.

Satellite communications, especially TV broadcasting, predominately operate in C and Ku bands. Because of recent developments in satellite communications \cite{8766193,Perez2019} together with the conventional fixed spectrum allocation policy, congestion of C and Ku bands has become a serious issue. To enhance the spectral efficiency and leave room for new broadband applications, satellite systems have moved from single-beam to multi-beam satellites with smaller beam spots. In essence, the multi-beam satellite payloads are designed to allocate a fixed bandwidth segment to each beam according to a regular frequency reuse scheme and constant equal power. Therefore, the maximum system capacity of current multi-beam satellites is limited by the fractional frequency reuse factor. Since the same frequency band is shared by different beams, the problem of multiuser interference arises. Aggressive frequency reuse schemes have been shown to be a promising approach towards enhancing the spectral efficiency of satellite communications (see Section \ref{precoding_section}).


Due to the spectrum scarcity, satellite operators are moving from the conventional C-band and Ku-band to Ka-band, which offers much greater signal bandwidth than C and Ku bands altogether. However, Ka-band systems are much more susceptible to adverse weather conditions than Ku-band and especially C-Band. On the other hand, moving to higher frequencies allows for smaller antenna size thus promoting the use of multi-antenna systems.

The 5G system deployment have posed numerous challenges, mainly in terms of supporting very high data rates with low end-to-end delays \cite{gupta}. The success of 5G heavily depends on national governments and regulators, as they are responsible to provide the new spectrum bands and operational guidelines for 5G deployment. The main representatives of the digital technology industry have released a list of recommendations on the commercial spectrum for 5G in Europe \cite{5GPPP}, where the frequency band 3400-3800 MHz (C-band) is identified as a potential candidate for the initial deployment of 5G mobile service. 

C-band spectrum has been traditionally reserved exclusively for satellite use and the reallocation of C-band spectrum to other telecommunications would inevitably have an impact on the satellite systems. In this regard, C-band should be carefully assigned to new 5G systems so as to ensure the continuity of vital satellite communication services. In this context, it is worth citing the recent developments in EEUU, where a satellite alliance is proposing ways to clear the C-band spectrum and accommodate the 5G wireless services \cite{cbandA}.

Recently, moving  the feeder link from Ka-band to the Q/V-band (40/50 GHz) has been investigated as a solution to the Ka-band congestion \cite{kyrgiazos}. This migration, not only frees-up the whole Ka-band spectrum for the user link, but also provides higher bandwidth for feeder link that can accommodate a broadband HTS system. Unfortunately, weather impairments heavy affect Q/V band, claiming for the use of GW diversity techniques to ensure the required availability \cite{6856179,6666239}.

Last but not least, new Non Geostationary Orbit (NGSO) satellite systems are gaining momentum due to the low free space attenuation and small propagation delay of lower orbits \cite{braun}. Still, the available usable radio spectrum is limited and is costly for the NGSO satellite operators. This has led to the concept of spectrum coexistence of LEO/MEO satellites with the already existing GSO satellites \cite{ITURS1325} and/or the spectral coexistence among different NGSO satellites \cite{ITURS1431}.

Cognitive Radio (CR) is a well-known spectrum management framework to solve the spectrum scarcity, as it enables unlicensed systems to opportunistically utilize the underutilized licensed bands. Within the satellite communications context, CR has been considered in \cite{lagunas,chatzinotas}, where the non-exclusive Ka-band (17.7-19.7 GHz for Space-to-Earth and 27.5-29.5 GHz for Earth-to-Space) is considered for spectrum coexistence between incumbent terrestrial backhaul links and the non-exclusive satellite links.

In order to further improve the capacity and reliability of mobile wireless backhaul networks, the concept of seamlessly Integrated Satellite-Terrestrial Backhaul Network (ISTB) has been proposed in \cite{sansa,viasat,Sat5G}. In ISTB, satellite and the terrestrial system intelligently collaborate not only to enhance the backhaul network capacity but also to overcome current spectrum scarcity while reducing the spectrum licensing costs.

In both Satellite CR and ISTB scenarios, spectrum coexistence results in undesired interference which needs to be carefully addressed to truly leverage the full potential of such schemes (e.g. \cite{lagunas2,artiga,shaat}). 

For more information on system coexistence scenarios, the reader is referred to the subsection \ref{syscoex_section} of the present manuscript.

\subsection{Standardization} \label{sec4e}

\begin{figure*}[!t]
\centering
\includegraphics[width=160mm,keepaspectratio]{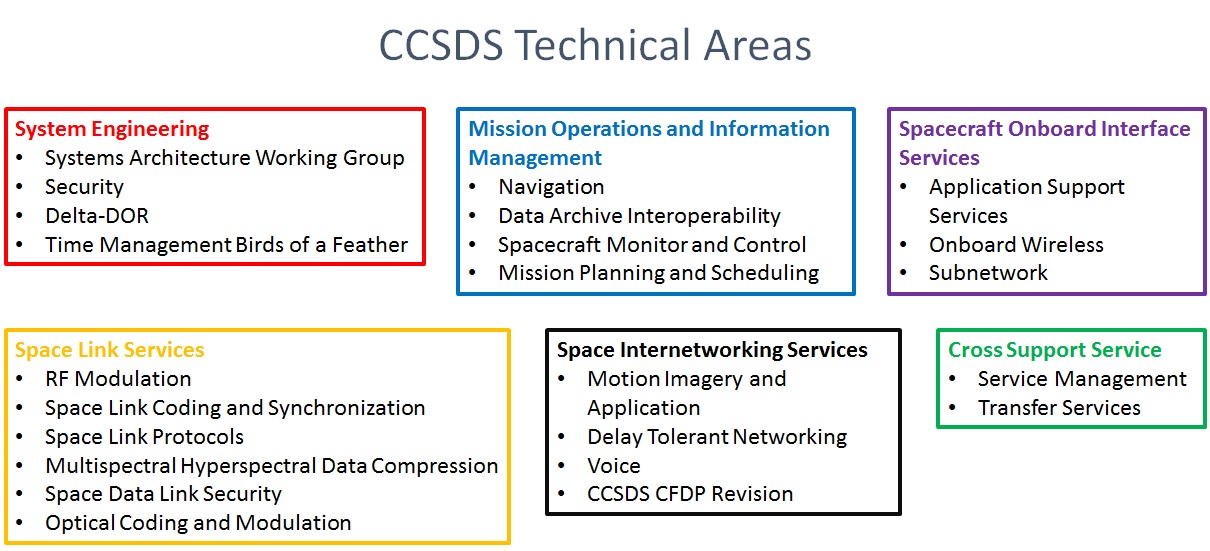}
\caption{Areas and Working Groups (topic) that are currently developing new standards in CCSDS.}
\label{CCSDS_Areas}
\end{figure*}

Standardization is also an important aspect of all the telecommunication systems. The usage of common open standards is fundamental to guarantee interoperability between devices from different manufacturers on both the transmitter and the receiver side. This reveals an open market with different manufacturers competing to offer the best possible devices in order to acquire more market shares. Apparently, this is hugely beneficial for the development of the technology and also for consumers. The main set of standards for SatComs can be found hereafter. 

\subsubsection{DVB}

Digital Video Broadcasting (DVB) \cite{DVB} is a set of international open standards for digital television. DVB standards are maintained by the DVB Project, an international industry consortium, and are published by a Joint Technical Committee (JTC) of the European Telecommunications Standards Institute (ETSI), European Committee for Electrotechnical Standardization (CENELEC) and European Broadcasting Union (EBU).

DVB standards are recognized as the most important set of standard for Television Broadcast and are widely used around the World, well beyond the European Border where the standards have been originally developed. The DVB standards cover different TV broadcast technology from satellite to cable to terrestrial television. They cover both the physical and data link layer of the ISO-OSI stack. The most important standards developed by DVB for the Physical Layer are:
i) DVB-S, DVB-S2 and DVB-S2X for Satellite TV;
ii) DVB-C and DVB-C2 for cable TV;
iii) DVB-T and DVB-T2 for terrestrial TV. 
For the second layer of the ISO-OSI stack, the most important DVB standards are:
i) DVB-MPEG; 
ii) DVB-GSE.

\subsubsection{5G NTN standardization} \label{sec4e2}

The standardization of 5G, like the previous mobile communications generations, is led by the 3GPP. Traditionally, satellite and terrestrial standardization have been separate processes from each other. However, in recent years, there has been an increasing interest from the satellite communication industry in participating in the 3GPP standardization effort for 5G, due to the market potential of an integrated satellite-terrestrial network. As a matter of fact, 3GPP initiated in March 2017, as part of Release 14, a study item in order to analyse the feasibility of satellite integration into 5G network \cite{3gppref3}. The initial goal was to bring together satellite operators and other companies to create aligned contributions in the support of NTNs in the 5G standardization. Two study items have already been concluded \cite{3gppref1} \cite{3gppref2}, where the role that the satellites can play in the 5G ecosystem has been studied. In addition, the challenges of a satellite-terrestrial network co-existence have been analysed taking into account different architecture options and all the layers of communication. After two years of a study phase, it is now approved from the 3GPP that NTN will be a new key feature of 5G and a work item (WI) will start from January 2020 \cite{3gppnews}. It is agreed that as a starting phase only the LEO and GEO satellite orbits will be considered having a transparent payload. Last but not least, initial studies on the support of IoT technologies, such as Narrowband Internet of Things (NB-IoT) and enhanced Machine Type Communications (eMTC), will be performed.

\subsubsection{CCSDS}

The Consultative Committee for Space Data System (CCSDS) was established in 1982 by the major space agencies of the world to provide a forum for solving common problems in the development and operations of space data systems. It develops recommended standards and practices for data and communications systems with two main aims: to promote interoperability and cross support among cooperating space agencies, so reducing operations costs by sharing facilities, and also reduce costs performing common data functions, by eliminating unjustified project-unique design and development within the various agencies. 

On the official CCSDS website, \cite{CCSDS01}, all standards and practices developed by CCSDS are published and available for free. CCSDS publications are organized into seven categories: Blue book, Magenta book, Green book, Orange book, Yellow book, Red book and Silver book. 

The Architectural Overview in Fig. \ref{CCSDS_Areas} shows the Areas and Working Groups (topics) that are currently developing new standards in CCSDS. As we can see there are six technical areas with twenty-three working groups, responsible to produce new standards. The working body responsible to define communications standards is the Space Link Service (SLS), composed of six working groups. It defines two main links between Earth and space probes: telemetry and telecommand.


\section{ Air Interface: Enablers \& Topics} \label{sec5}

This section covers the main technical aspects related to the air interface of a satellite communication system. 

\subsection{Channel Modelling}
Channel and propagation characteristics play a key role in determining the system design and determining the right techniques. These aspects are typically determined by the frequency of operation in addition to the system configuration. The satellite communication system provides a plethora of configurations using different frequencies. Hence, satellite systems encounter a variety of channels \cite{PDA1}. However, common to these models are the following:
\begin{itemize}
    \item Absence of scatterers near the satellite transmit antennas.
    \item Long term components
    \item Dynamic channel components
\end{itemize}
In the following, we present a canvas of the channel models encountered in satellite communications.
\subsubsection{Fixed Satellite}
The next generation satellite systems would be typically operating at frequencies higher than 10 GHz. According to \cite{PDA1}, such channels are characterized by line of sight (LoS); the satellite channel essentially corresponds to an Additive White Gaussian Noise (AWGN) channel. However, on top of this, propagation at the Ku and, especially, Ka-band is subjected to various atmospheric fading mechanisms (see references in \cite{PDA1} for details). These effects can be modelled as,
\begin{itemize}
    \item {\bf Long Term Channel Effects:}  The key constituents to this category include, attenuation due to precipitation, gaseous absorption and clouds, tropospheric scintillation, signal depolarization among others. The models for such effects typically involve {\em first-order statistics} \cite{PDA1}.
    \item{\bf Dynamic Channel Effects:} These effects determine the temporal properties of the AWGN channel when impacted by rain. Such models allow for the calculation of several second-order statistics, such as fade slope and
    fade duration.
\end{itemize}
Clearly, since the distance between the user terminal and the satellite is quite large compared to the distance between antennas (either on-board or on the ground). This, and the absence of scatterers near the satellite antennas, tend to make the fading among all the channels between the satellite and the user terminal correlated. This spatial correlation negatively impacts the use of \gls{mimo} for Fixed Satellites \cite{PDA1}, \cite{Bundeswehr}. Further, with regard to the rain-fading, the ground terminals need to be several miles away to ensure a significant decorrelation of fading. 
\subsubsection{Mobile Satellites}
As mentioned in \cite{PDA1}, the channel characteristics in mobile satellite systems differ from their FS counterparts since  mobility implies the presence of diffuse multipath (or Non-LoS) paths in addition to LoS. Narrowband and Broadband channel models have been proposed in the literature. Typically, these models involve a multi-state Markov model, with each state determining the parameters and nature of distribution of the corresponding channel. As a case in point, ITU Recommendation P.681 \cite{RD15} presents a narrowband 3-state Markov model with related state statistics. The states represent (i) Deep shadow state, (ii) Intermediate shadow state and (iii) a good state with very slow variations. The severity of mean attenuation increases with the shadow strength. Another approach for a narrowband 2-state land-mobile satellite channel model at 2.2 GHz is presented in \cite{RD17}. This approach assumes a Loo distributed RX signal defined by a parameter triplet
and the channel state statistics is determined by the User Terminal (UT) (vehicle) speed. Further, a wideband satellite channel model comprises the 2-state semi-Markov model for shadowing and the ITU multi-tap model \cite{RD19} for the multipath propagation. Typically, the channel properties are assumed quasi-stationary over short-time periods, and during these periods are represented by stationary stochastic processes.

\begin{figure}
\begin{centering}
\subfigure a){\includegraphics[width=0.22\textwidth]{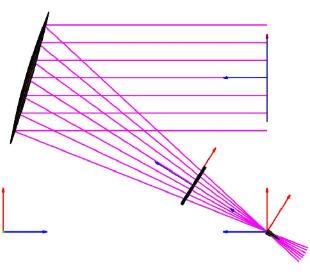}}
\subfigure b){\includegraphics[width=0.22\textwidth]{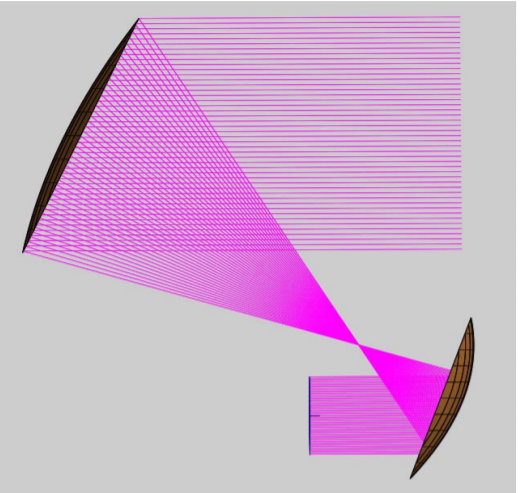}}
\caption{Imaging antenna systems: a) defocused and b) confocal}
\label{antenna_systems}
\par\end{centering}
\end{figure}

\subsection{Antennas} \label{sec5b}

\subsubsection{Passive and focused reflector antennas}
The shift from broadcast to broadband missions marks a transition from contour beam coverage, which is designed to serve a given geographical region, to multibeam antennas (MBAs) that by virtue of narrower beams enable both higher gains and frequency reuse thereby maximising the spectral efficiencies. The corresponding evolution of traditional passive antenna architectures for GEO missions has driven shaped reflector antennas towards single feed per beam (SFPB) and multiple feed per beam (MFPB) MBAs \cite{gouss_1, gouss_2}. For telecommunications missions in the Ku- and Ka-band, both SFPB and MFPB architectures involve an array of feed horns at the focal plane (i.e. focused architecture) of the reflector, often in an offset-fed parabolic architecture. 

In SFPB antennas, each beam is produced by the illumination from a single feed. SFPB MBAs provide high gain and low side-lobe level thereby leading to an advantageous carrier to interference ratio. On the other hand, the SFPB architecture typically requires 3 or 4 reflectors to achieve contiguous coverage \cite{gouss_1, gouss_2}, leaving little or no space to accommodate additional missions on the satellite. This is in order to accommodate for horns of relatively large electrical aperture (typically in the range 2-3 wavelengths) that lead to favourable antenna efficiency and reduced sidelobes. Consequently, the minimum displacement of adjacent feed horns leads to a large separation of the associated beams, which are then interleaved by beams produced by another reflector. 

In the alternative MFPB architecture, each beam is produced by a cluster of feeds. This allows adjacent feed clusters to overlap as individual feeds may contribute to multiple adjacent beams. An advantage of MFPB is therefore that contiguous coverage can be achieved with one or two main reflectors \cite{gouss_1, gouss_2}. On the other hand, MFPB antennas require more complex beam-forming network and can give rise to challenges in terms of e.g. the operating frequency bandwidth or lower aperture efficiency. Consequently, MFPB does not always represent the choice of preference within passive antenna architectures \cite{gouss_3}.

The SFPB and MFPB architectures outlined above, have so far been the most widely used architectures for GEO High Throughput Systems \cite{gouss_4}. The use of multiport amplifiers \cite{gouss_4a} is increasingly being deployed in these architectures to add flexibility in power allocation (e.g. Eutelsat 172B \cite{gouss_4b}). Larger reflectors enabled by deployable technologies are also being developed for delivering more directive beams for telecommunications and other missions \cite{gouss_4c}. Meanwhile, the pressing needs for flexibility in coverage and an ever-increasing number of beams is driving major efforts for the developments of active array solutions \cite{gouss_4, gouss_5, gouss_6}. Active arrays decouple the number of beams from the number of feeds and, in conjunction with beamforming technologies, enable both flexibility in the coverage as well as the power sharing between the beams \cite{gouss_6}. 

\begin{figure*}
\begin{centering}
\subfigure a){\includegraphics[width=0.3\textwidth]{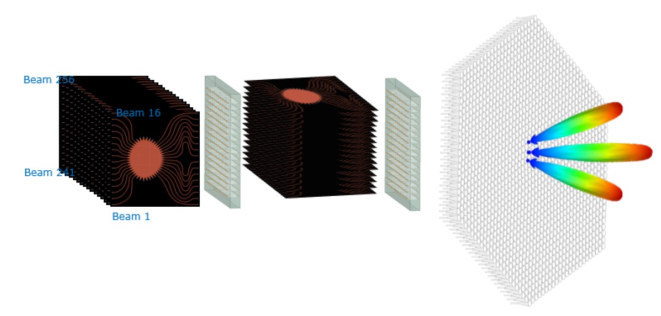}} 
\subfigure b){\includegraphics[width=0.3\textwidth]{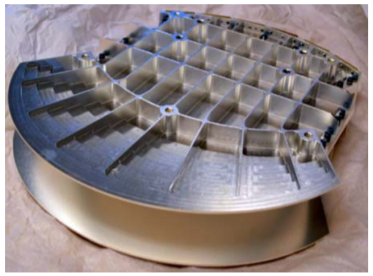}}
\subfigure c){\includegraphics[width=0.3\textwidth]{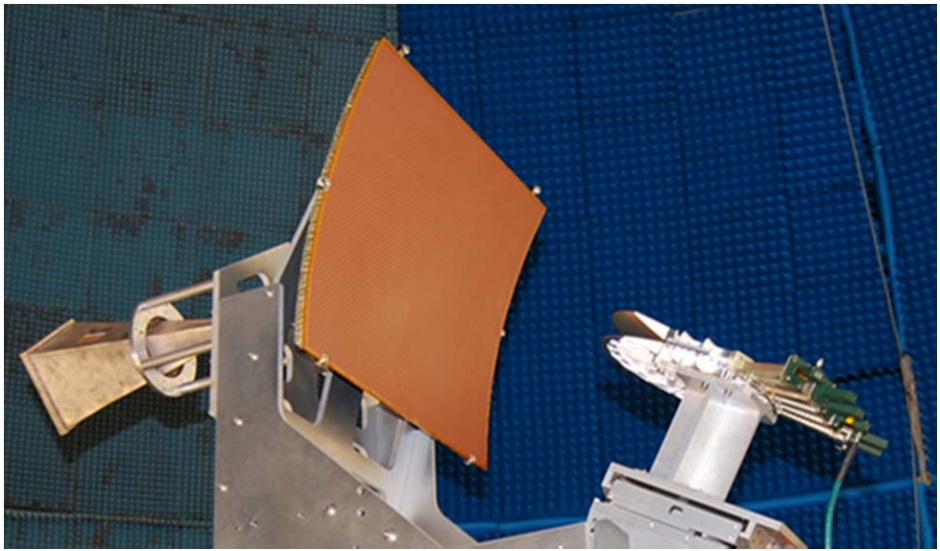}}
\caption{a) : Double layers of Rotman lenses producing pencil beams in two directions; b) Continuous parallel plate waveguide lens-like antenna; c)Prototype of a QOBF antenna \cite{gouss_12}}
\label{beamformer}
\par\end{centering}
\end{figure*}

\subsubsection{Active antenna arrays} \label{sec5b2}
By definition, in active antennas the amplifiers are integrated with the radiating elements. A marking difference from passive antennas is thus the distributed amplification of the radiating signal. The spatial RF power distribution and reduced peak RF power levels in active arrays improve reliability (including reducing multipaction thresholds \cite{gouss_6a}) and graceful degradation. Recent developments in wide bandgap semiconductor technologies (e.g. GaN) are promising to improve the relatively low power and thermal efficiencies of MMIC amplifiers at Ku-band and beyond \cite{gouss_6ai}, which is otherwise a natural choice for active arrays in place of the more traditional vacuum (travelling wave tube) amplifiers due to the advantageous integration. 

Active antennas can be deployed in either direct radiating array (DRA) or array fed reflector (AFR) architectures. The choice and associated trade-offs strongly depend on the system requirements. For LEO and MEO systems, the large field of view coupled and the reduced demands on gain favor DRA solutions. For GEO High Throughput Satellite (HTS) missions, the large electrical sizes required to achieve the gain targets are preferentially achieved with reflector-based geometries that provide magnification of the radiating aperture \cite{gouss_5}; the best active array architecture for GEO thus remains an open question \cite{gouss_4}. Two AFR architectures are primarily considered, namely the defocused parabolic reflector (typically in an offset configuration) or a combination of two parabolic reflectors, often referred to as a confocal reflector configuration, Fig. \ref{antenna_systems} \cite{gouss_6}. Both configurations typically require oversized reflectors to minimise spillover losses during beam scanning. Defocused reflectors provide pathways for trading off between RF performance and the number of feeds per beam. They typically suffer from higher aberrations compared to confocal systems but are compatible with a reduced number of feeds. Comparatively, they also provide advantages in terms of complexity and accommodation. 

The development of on-board digital processing capability (see Section \ref{digital_Payloads_Section}) is, together with distributed amplification, another major underpinning technology towards active arrays for telecommunication payloads. It enables control of analogue beamforming networks, direct digital beamforming or their hybrids. Reducing the number of active elements in an array provides advantages in terms of cost, complexity, thermal management as well as digital demands for the control of the phased array. Significant efforts are therefore placed in this context. 

Since the gain of an antenna is strongly linked with the size of the illuminating aperture, one technique to reduce the number of elements in an array without reducing the overall surface area is to increase their size. Antenna array theory suggests that in this case grating lobes are likely to appear \cite{gouss_6b}. In multibeam satellite systems, grating lobes can be tolerated as long as they do not compromise the system performance, primarily due to causing unwanted interference (e.g. typically by being kept outside the field of view of the Earth \cite{gouss_4}). Depending on the orbit, this can provide some margin to increase the size of radiating elements and thereby reduce their number for a fixed size of the illuminating aperture.

An alternative approach for reducing the number of elements in an active antenna is based on array thinning techniques. The latter relies on sparse and aperiodic arrays, which provide opportunities for a trade-off between side-lobe levels and number of elements \cite{gouss_7, gouss_8}. Sparse array design typically targets a suitable combination of the radiating element positions along an aperiodic lattice with the tapering of the excitation amplitude across the aperture. The size of the radiating elements can also be modulated along the aperture. Aforementioned techniques enable the control of the beamwidth while maintaining side lobe levels kept under an assigned value \cite{gouss_8}. A drawback of this approach is that the antenna development is in general bespoke to a mission and thereby complexity and costs do not necessarily scale down.

The realisation of lightweight and efficient reconfigurable antennas can also benefit from quasi-optical beamforming (QOBF) networks \cite{gouss_9, gouss_10, gouss_11}. An example of this is the Rotman lens, where beamforming is achieved by virtue of the phasing propagated waves in a parallel plate waveguide \cite{gouss_9}. A single Rotman lens offers beam steering along one axis, whereas two layers of stacked Rotman lenses offer the possibility to generate pencil beams that can be steered along 2 principal axes (see Fig. \ref{beamformer}. a). An example of Rotman lens development for GEO VHTS mission is presented in \cite{gouss_10}. In order to remediate the complexities associated with the discrete antenna ports of the Rotman lens as well as the losses from the dielectric substrate, a continuous parallel plate waveguide lens-like multiple-beam antenna is presented in \cite{gouss_11} (see Fig. \ref{beamformer}. b). A prototype involving this beamformer targeting LEO/MEO missions (see Fig. \ref{beamformer}. c) is presented in \cite{gouss_12} and an analysis of the performance of this solution in the presence and absence of beam-hopping under varying traffic scenarios is presented in \cite{gouss_13}. 

\subsubsection{Trends in antennas for LEO/MEO missions and the ground segment}
MEO and LEO satellites experience a spatial evolution of the traffic along the satellite orbit. Therefore, a reconfigurable antenna is needed to match
the satellite coverage with the spatial distribution of the traffic. The LEO constellations by Telesat, Starlink and Akash are aligned with this approach \cite{gouss_14}. The MEO constellation by O3B is also adopting a steerable beam approach \cite{gouss_15}. For smaller platforms, such as cubesats, the priorities in terms of the antenna selection are defined by the mission specifics, such as the limited capacity for on-board accommodation. A review of antenna solutions for smaller satellite platforms (e.g. cubesats) can be found in \cite{gouss_16}.

Antennas for ground terminals is also a rapidly evolving area of technology development. Of particular commercial interest remains the development of flat panel antennas with beam steering capability that enables satellite on the move (SOTM) applications as well as connectivity to non-GEO platforms \cite{gouss_17}. A number of mechanical, electronic and hybrid approaches are reported in the literature and are actively being pursued by a vibrant academic and industrial research community. They typically use two narrow-width arrays on a 2-axis positioner for tracking in azimuth and elevation. A significant disadvantage with this type of antennas is the broad beamwidth along the narrow plane of the aperture, which can lead to interference problems \cite{gouss_17c}. A mechanically steerable flat panel antenna product that overcomes the aforementioned high skew angle problem \cite{gouss_17c} and has successfully been deployed in the avionic industry is provided by \cite{gouss_18}. Alternative approaches that target to maintain performance while reducing costs include the use of nematic liquid crystals \cite{gouss_20}.

\subsection{Ultra High Throughput Satellites (UHTS)} \label{sec5c}

\subsubsection{Digital Payloads}
\label{digital_Payloads_Section}

Due to the diversification of markets, satellite communications need to meet enhanced demand for reliable and flexible connectivity at higher throughput. Novel architectures like multibeam systems, migration to higher frequencies and novel techniques such as precoding, predistortion, interference and resource management have already been considered \cite{SPM}. To fully exploit these developments in the emerging contexts, and impart flexibility in design, additional resources need to be considered. In this context, space-based assets are considered with on-board processing (OBP) being the widely accepted methodology.

\begin{figure*}[!t]
	\centering
	\includegraphics[width=140mm,keepaspectratio]{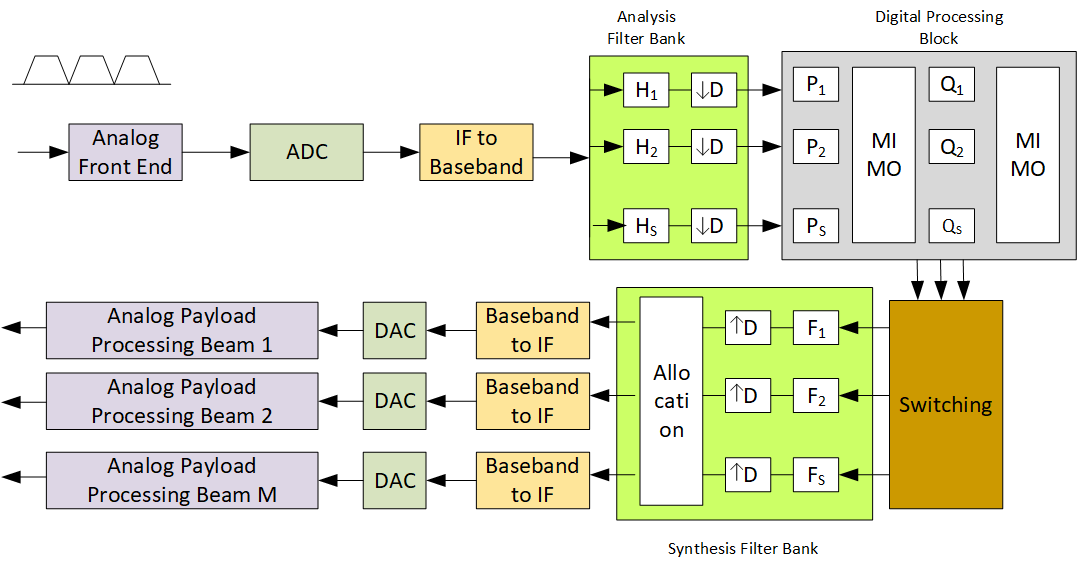}
	\caption{On-board processing architecture \cite{SPM}}
	\label{fig:OBP}
\end{figure*}

Providing digital processing on-board the satellite is not a new concept and has been discussed in the last decades \cite{ref_9}, \cite{ref_10}, \cite{ref_11}. A perusal of the literature indicates two key OBP paradigms.
\begin{enumerate}
    \item {\bf Digital Transparent Processors (DTPs)}: These processors sample the waveform and operate on the resulting digital samples; neither demodulation nor decoding is implemented \cite{ref_10}. DTP based processing results in payload designs agnostic to air-interface evolutions. They have been used in a number of missions including INMARSAT-4, SES 12 \cite{ref_13} and typical applications include digital beamforming, broadcasting/multicasting based on single channel copies among others.
    \item {\bf Regenerative Processing}: This methodology operates on the digital baseband data obtained after waveform digitization, demodulation and decoding. Missions like Iridium, Spaceway3, HISPASAT-AG1, incorporate regenerative processing mainly for multiplexing different streams, switching and routing. While regeneration generalizes DTPs and decouples the user and feeder links, the additional processing comes at a higher cost. Further, regenerative processing limits the flexibility to use newer transmission modes and can suffer from obsolescence of technology unless reprogrammable payloads are considered \cite{ref_10}.
\end{enumerate}
An interesting hybrid processing paradigm involves digitizing the entire waveform, but regenerating only a part for exploitation. In this context, the header packet is regenerated to allow for on-board routing in \cite{ref_10}.  This capability would radically change satellite networks and the services they can deliver. 

For the sake of exposition, in the following, we briefly present the structure of a DTP. This develops on the detailed work in \cite{TAES_DTP} and extends to cover novel processing. 
Fig. \ref{fig:OBP} presents a payload transponder employing DTP. Standard analog front-end receiver processing including antenna systems, analog beamforming network, Low noise amplifiers, down conversion (mixer, filter) and automatic gain control that appear before the digital processing is not detailed. The key components in OBP are listed below, the reader can refer to \cite{TAES_DTP}, \cite{SPM} for further information.
\begin{itemize}
    \item High Speed Analog to Digital Converters (ADC) and Baseband/IF Conversion.
    \item Channelizers comprising an analysis filter bank for de-multiplexing uplink signals and a synthesis filter bank to regenerate appropriate bands.
    \item Processing Block includes processing of individual streams like (de)modulation, decoding/ encoding as well as joint processing using Multiple-Input Multiple-Output (MIMO) technique. It also includes a Look-up Table (LUT) for predistortion, beamforming, precoding and spectrum calculation.
    \item Switching Block affects routing in spatial (e. g, from one beam to another), temporal (e.g.,store and forward) and spectral (e.g., frequency hopping) domains.
\end{itemize}
As mentioned earlier, on-board processing with digital payloads are being considered by satellite manufacturers and operators. With the emergence of novel signal processing and digital communications, digital payloads offer an ideal platform to overcome many of the short-comings of traditional on-ground processing. This includes reducing the latency and the inefficient use of resources, as well as enabling additional flexibility among others \cite{SPM}. Some examples of application involve on-board predistortion \cite{Nicolo_DTP} and energy-detection \cite{Christos_DTP}.

The payload is often seen as part of the end-to-end channel and its behaviour should be regularly measured. The so-called In-Orbit-Test (IOT) operation of the satellite payload consists in transmitting and receiving to and from the satellite a specifically designed test signal, mainly a spread spectrum signal, for the measurement and extraction of some key payload parameters such as, on-board filters responses, high power amplifier response, G/T, etc. The IOT operation is fundamental in several situations during the life-time of the satellite to verify and monitor the performance and functional requirements of the satellite payload.

Allowing customers to continuously monitor the status of the satellite transponder payload while, at the same time, avoiding interference with the traffic (when the satellite is in the operation phase) and with adjacent/nearby satellites (when in re-location phase) represents an important open challenge. In addition, the possibility of performing non-interfering tests in close or open loop modes, enables the utilization of the equipment in a distributed scenario.

Techniques to effectively monitor the on-board amplifiers and to identify possible degradation effects are open research topics with no close solution at the moment, in particular when dealing with wideband applications. In the latter, reducing the testing time and improving the accuracy of existing narrow-band-based techniques are of key importance.

While conventional IOT methods require the interruption of the main customer service \cite{5674689}, novel cognitive techniques based on spread spectrum signals are gaining momentum \cite{8761654,mazalli,Mishra_2015}.

\subsubsection{Precoding/MU-MIMO}
\label{precoding_section}

\begin{figure*}[!t]
	\centering
	\includegraphics[width=140mm,keepaspectratio]{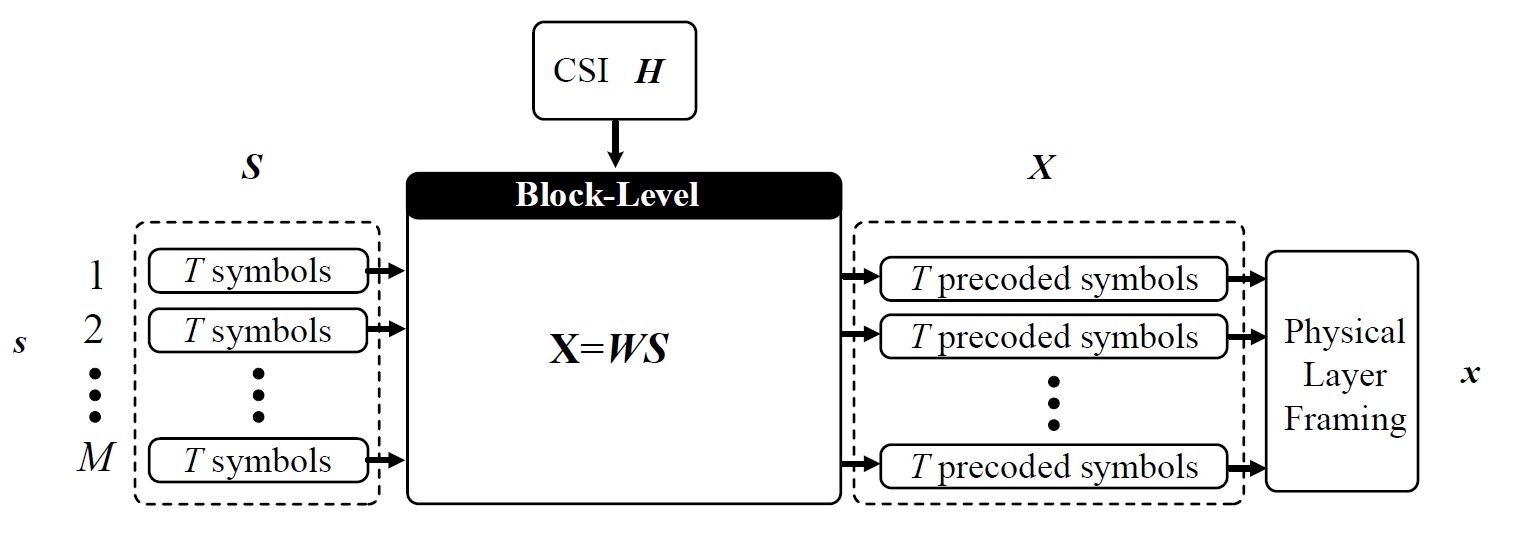}
	\caption{Schematic diagram for conventional linear precoding \cite{Alodeh2018_tutorial}. The CSI is used to compute the precoding matrix, denoted as $W$. The precoding matrix is then used to filter the input data streams.}
	\label{precoding_block_scheme}
\end{figure*}

The state-of-the-art in high throughput SatCom relies on multi-beam architectures, which exploit the spatial degrees of freedom offered by antenna arrays to aggressively reuse the available spectrum, thus realizing a space-division multiple access (SDMA) scheme \cite{Roy1997}. As a matter of fact, aggressive frequency reuse schemes are possible only if advanced signal processing techniques are developed, with the objective of handling the multi-user interference (MUI) arising in multi-beam systems and deteriorating their performance. Such signal processing techniques are commonly referred to as multi-user multiple-input multiple-output (MU-MIMO) and, in the satellite context, also as multi-beam joint processing \cite{Perez2019}. In this context, linear precoding (or beamforming) techniques have been a prolific recent area in the recent years, showing to be an effective way to manage the MUI while guaranteeing some specific service requirements \cite{Liu2011,Bjornson2014,Gershman2010,Bengtsson2001,Schubert2004}. The benefits of using precoding techniques for managing the interference at the gateway in SatCom are also considered in the most recent extensions of broadband multi-beam SatCom standards \cite{DVB_S2X}. 
The conventional precoding approach exploits the knowledge of the channel state information (CSI) in order to design a precoder to be applied to the multiple data streams, thus mitigating the MUI.  With the aid of precoding, a satellite user terminal can obtain a sufficiently high signal-to-interference-plus-noise ratio (SINR), even though the same bandwidth is reused by adjacent beams. This is possible because the precoder uses the channel knowledge to mitigate the interference toward the user terminals, and therefore a certain SINR value can be guaranteed for the users. 
Typically, the precoding matrix is computed at the satellite gateway. After that, the beam signals are precoded (by multiplying the data streams by the precoder matrix) and transmitted through the feeder link using a frequency-division multiplexing scheme. A schematic representation of the precoding operation, taking place at the gateway, is shown in Fig. \ref{precoding_block_scheme}.  Then, the satellite payload performs a frequency shift and routes the resulting radio signal over an antenna array that transmits the precoded data over a larger geographical area that is served by the multiple beams in the user link. It shall be stressed that the signals transmitted for different beams in the downlink (i.e., from the satellite to the user terminals) use the same bandwidth in a full frequency reuse fashion. This is made possible by the described precoding operation which counteracts the interference across multiple beams.

The computational complexity that is required to implement multibeam satellite precoding techniques can be considerable when the dimensions of multibeam satellite systems are high. This is often the case, as many current systems are characterized by several hundreds of beams. This complicates the precoding implementation because of the extremely large size of the precoding matrix that must be calculated. In this regard, low-complexity linear precoding techniques are of great interest, and this is a problem that deserves further attention and research. 

A different precoding strategy, known as symbol-level precoding, has been  considered more recently in the literature \cite{Spano2018IJSC}. In this approach, the transmitted signals are designed based on the knowledge of both the CSI and the data information (DI), constituted by the symbols to be delivered to the users. Since the design exploits also the DI, the objective of symbol-level precoding is not to eliminate the interference, but rather to control it so to have a constructive interference effect at each user. This approach has been shown to outperform the conventional precoding schemes, in terms of reduced power consumption at the gateway side for a given quality of service at the user terminals in terms of SINR.

A fundamental assumption of conventional precoding schemes is that independent data is addressed to each user, thus dealing with multiuser unicast systems. However, the physical layer design of DVB-S2X SatCom standard \cite{DVB_S2X} has been optimized to cope with the noise limited satellite channel, characterized by excessive propagation delays and intense fading phenomena. Therefore, long forward error correction (FEC) codes and fade mitigation techniques that rely on an adaptive link layer design (adaptive coding and modulation—ACM) have been employed. This implies that each frame accommodates several users, and therefore the communication system becomes a multicast one. Accordingly, the multicast framing structure hinders the calculation of a precoding matrix on a user-by-user basis, and ad-hoc precoding schemes need to be employed to address multicast systems. Precoding schemes for physical layer multicasting have been proposed in \cite{Sidiropoulos2006,Silva2009,christopoulos,Christopoulos2014}.

Another relevant challenge for the application of precoding in practical satellite systems is related to non-linearities. In fact, the on-board per-antenna traveling-wave-tube amplifiers (TWTAs) usually introduce non-linear effects, which result in a distortion on the transmitted waveforms. A typical solution to this problem
in single-user links relies on predistortion techniques, but their extension to multi-beam systems relying on precoding is not straightforward, because of the mutual correlation between the data streams induced by the precoding schemes. In this context, different precoding schemes have been proposed in the literature \cite{Mohammed2013,Studer2013,Spano2018IJSC,Spano2018TSP}, having the aim of enhancing the dynamic properties of the transmitted signals, such as the peak-to-average power ratio (PAPR), and therefore improving the signals robustness to non-linear effects. In particular \cite{Spano2018IJSC,Spano2018TSP} are based on symbol-level precoding.

Overall, the research on precoding has been developing quite fast in the recent years, and a number of practical challenges \cite{Alodeh2018_tutorial} have been addressed, with the aim of exploiting full frequency reuse in current communication systems. In this direction, besides the pure research, it is particular important the development of ad-hoc testbeds that allow in lab implementation and validation of precoding schemes. Considerable advances have been made in this regard, as further discussed in \ref{sec8}.

\subsubsection{Non Orthogonal Multiple Access}

As one of the promising 5G new radio techniques, non-orthogonal multiple access (NOMA) has attracted considerable research attention from both industry and academia over the past few years \cite{survey_noma}.
NOMA breaks the orthogonality in conventional orthogonal multiple access (OMA) such that multiple terminals can access the same time-frequency resource simultaneously, which improves the efficiency of spectrum utilization.
The resulted co-channel interference can be alleviated by performing multi-user detection and successive interference cancellation (SIC) at the receiver side.
In various 5G terrestrial scenarios, NOMA has demonstrated performance improvement over conventional OMA schemes \cite{survey_noma,leitwc_noma,leitvt_noma, 7562275}.
By observing its advantages in aggressive frequency reuse and suppressing interference, it is natural to further extend the NOMA applications beyond the cellular systems.
For instance, the advanced television systems committee (ATSC) has proposed a new type of multiplexing scheme,  i.e., layered division multiplexing
(LDM) which adopts NOMA principle, in the physical layer protocol standard ATSC 3.0 for  terrestrial digital TV broadcasting systems \cite{atsc_noma}.

In NOMA-based multi-beam satellite systems, 
\cite{perezneira2018noma} analyzed the applicability of integrating NOMA to satellite systems from a system-level point of view, and provided general approaches for cooperating NOMA with precoding.
In  \cite{caus_noma}, two suboptimal user-scheduling algorithms were proposed to maximize the capacity for over-loaded satellite systems. 
The numerical results showed that an appropriate user-grouping strategy is to pair the users with high-correlation channels.
In \cite{anyueoctr_noma}, a max-min fairness optimization problem was studied to apply NOMA to achieve a good match between the offered and requested capacity among satellite beams. 
The authors in \cite{beigi2018interference} proposed an overlay coding strategy to utilize the cooperative NOMA to mitigate interference in multi-beam satellite systems. 

In NOMA-enabled 5G terrestrial-satellite networks, 
\cite{jsacnoma2017} investigated a joint resource optimization problem for user pairing, beamforming design, and power allocation. 
In \cite{jsacsp2019_noma}, joint beamforming and power allocation for
NOMA based satellite-terrestrial networks were studied. Optimal solutions of beamforming weight vectors and power coefficients were developed.
In both works,
the satellite component is viewed as a supplement part to terrestrial networks. NOMA is applied within the terrestrial component.
In \cite{yan2018hybrid}, a cooperative NOMA scheme was proposed for satellite-terrestrial relay networks, where the user with better channel gain is viewed as a relay to help transmit data to the user with poorer channel condition. 
Outage probability and ergodic capacity were analyzed mathematically and shown the performance improvement of NOMA over OMA. 

In general, the solutions developed for terrestrial NOMA systems might not be simply applied to satellite systems, mainly due to the following reasons. 
Firstly, different channel propagation models may lead to different user grouping strategies. Unlike the cellular system, the users located in a beam typically undergo similar path loss towards the satellite \cite{SPM}.
The user paring in terrestrial NOMA is suggested grouping the users with large gaps of their channel gains \cite{survey_noma}. However, this paring strategy could be challenging to implement in satellite scenarios \cite{caus_noma}.
Secondly, simply applying the solutions developed from terrestrial NOMA may result in  high complexity for the satellite systems due to the presence of large amount of beams. 
Thirdly, compared to cellular systems, some distinctive characteristics in satellite systems can introduce new constraints and challenges, e.g.,  on-board power constraints, limited power supply, longer propagation delay, signal distortion, and mobility issues. 
Fourthly, the capability of flexibly allocating on-board resources is typically limited which introduces new dimensions in NOMA-satellite resource management \cite{anyueoctr_noma}.

\subsection{Data Collection}
\label{sec5d}
\subsubsection{Satellite IoT Air Interface} 

As mentioned in Section \ref{sec3}, the satellite can play an important role in the IoT services, more specifically in the so-called long-range IoT or low power wide are networks (LPWANs), by ensuring a global connectivity and service continuity. The three main technologies in the LPWAN family are the NB-IoT, Long-Range (LoRa) and Sigfox \cite{sat_iot_phy2}. Their PHY layer is quite different from each other and mostly driven by the need to satisfy important requirements, such as extended coverage, low power consumption, high network capacity etc., taking into account the technical peculiarities of the terrestrial infrastructure. More specifically, the NB-IoT uses a multicarrier modulation (OFDM in downlink and SC-FDM in uplink) for data transmission \cite{sat_iot_phy1}, SigFox utilizes an ultra-narrowband signal (UNB) with a differential binary phase shift keying (DBPSK) modulation \cite{sat_iot_phy3} and LoRa employs a chirp spread spectrum signal (CSS) \cite{sat_iot_phy4}. Since the satellite channel impairments are quite different from the terrestrial one, using the same air interface of the terrestrial IoT over a satellite link in order to collect the tremendous amount of data generated by the IoT devices, is not a trivial task. The increased delay in the satellite channel and the high amount of Doppler effects experienced, especially in the LEO orbit, imposes new challenges to the PHY interfaces of these technologies. 

In this context, the authors in \cite{sat_iot_mac13} stress out the impact of the Doppler effects on a LEO satellite-based NB-IoT system, while in \cite{sat_iot_phy7} the LoRa CSS signal over a LEO satellite is analyzed. To overcome the Doppler effects in such systems, the authors in \cite{sat_iot_phy5} come up with a new air interface for NB-IoT based on Turbo-FSK modulation, which was firstly introduced in \cite{sat_iot_phy6}.
Regarding LoRA, in \cite{sat_iot_phy8} a new acquisition method under increased Doppler effects is analyzed, while in \cite{sat_iot_phy9} a folded chirp-rate shift keying (FCrSK) modulation with strong immunity to Doppler effect is proposed. Other works in the literature focus on IoT over GEO orbits, where the main issue would be the increased round RTT of communication. In \cite{sat_iot_phy10} a new air interface for such a system is proposed while in \cite{sat_iot_phy11} and \cite{sat_iot_phy12} the authors present a novel waveform called Unipolar Coded Chirp Spread Spectrum (UCSS) that enables ultra-narrowband (uNB) communications of IoT nodes using a GEO satellite.

\subsubsection{Wideband Downlinks}
\label{wideband_sec5d2}
Commercial applications of satellite communication, such as observation satellites and LEO sensors, rely on extremely high data rates available during a short passage of the satellite. Hence, the recent trend of developing novel terminal modems capable of efficiently operating at very high symbol rates can be clearly observed. One of the main challenges for the modem design results from the assumed very large signal spectrum, which can be e.g. up to 1.5 GHz, if the whole Ka-band is utilized. Currently, the proposed terminal modems support up to 500 MHz for commercial high data rates, cf. \cite{mdm5000, hdrm, satix}. In order to enhance the symbol rate even further, the following design challenges need to be circumvented:
\begin{itemize}
\item parallel processing with a very high factor of parallelism, which can lead to access conflicts and performance degradation;
\item frequent trade-offs between performance, latency and complexity for the selection of signal processing and synchronization algorithms. In this context, high complexity may also lead to processing delays, which negatively affects the performance of the algorithms;
\item high frequency selectivity of the wideband communication, which may result from the limitations of the hardware, in particular cables and transponders. The magnitude of this effect typically increases with signal bandwidth;
\item additional impairments due to a large difference between the minimum and the maximum employed frequencies. In particular, clock frequency offset and drift due to the Doppler effect become substantial in wideband scenarios. 
\end{itemize}
These challenges have been recently tackled in \cite{Wideband_modem}, where a novel modem architecture for terminal modems with a substantially wider target signal bandwidth of up to 1.5 GHz has been proposed. The potential peak symbol rate can reach up to 1.4 Gsps, such that peak data rates of 5 Gbps and higher seem to be possible in future.

\subsection{Others}

\subsubsection{Optical Communications} 

A potential solution for solving the high bandwidth requirement on the feeder link is to move them to the Q/V-band (40/50 GHz)  \cite{gharanjik_large_2013,kyrgiazos_gateway_2014}, or even to the W-band (70/80 GHz) where bandwidths up to 5 GHz are available. However, given the demand trends, it would be a matter of short time before which these bandwidths also fall short of the requirement. A revolutionary solution is to move the feeder link from \gls{rf} frequencies to optical frequencies \cite{cowley_optical_2014,gharanjik_spatial_2014,dimitrov_digital_2014}. The high frequency \gls{rf} and optical approaches are challenging due to the attenuation by atmospheric phenomena (e.g. rain, clouds) whose severity increases with the frequency. In either case, a network of multiple gateways with appropriate switching capabilities is thus envisaged \cite{barrios_rivoli-tn1:_2015,gharanjik_spatial_2014}. Although optical communications is highly impaired compared to the \gls{rf} counterpart, it only needs a few gateways to achieve very high throughput \cite{barrios_rivoli-tn1:_2015}. This directly relates to a reduction in the cost of the ground-segment motivating the use of optical communications for feeder links. In addition, \gls{fso} communications benefit from the absence of frequency regulation constraints, small systems with lower power requirements and enhanced security.  

Optical links are impaired by several atmospheric phenomena like clouds, aerosols, turbulence etc. \cite{kaushal_optical_2016}. The two main categories of propagation impairments are 
\begin{itemize}
    \item {\bf Blockage Effects:} Cloud coverage constitutes the predominant fading mechanism, resulting in the blockage of the link \cite{kaushal_optical_2016}. This impairment is not localized but spread over a geographical extent. A cloud-blockage typically introduces significant attenuation on the link, potentially breaking the link. In order to maintain an optical link, ground system design involves choosing ground-based optical stations at places with a high cloud free line of sight (CFLOS) joint probability \cite{lyras_cloud_2017,perlot_model-oriented_2012}.
P    \item {\bf Turbulence and other small-scale fading effects:} Even under CFLOS conditions, the optical systems are severely affected by atmospheric turbulence.  This phenomenon leads to small-scale fading and impacts the link budget  \cite{perlot_optical_2012}. The estimation of this phenomenon taking also into account the beam wander, beam spread and amplitude scintillation is of critical importance  \cite{dimitrov_digital_2015}. In addition to turbulence,  aerosols, cirrus clouds impact the signal amplitude \cite{smith_millimeter_1993}.
\end{itemize}
Fade mitigation techniques are considered to mitigate the aforementioned impairments. These are categorized as:
\begin{itemize}
    \item {\bf Macro-Diversity:}  For cloud coverage,  multiple Optical Ground Stations (OGS) constituting a network are employed \cite{lyras_cloud_2017}. These stations are separated by hundreds of miles, so that a certain desired CFLOS probability of the whole network is achieved. Unfortunately, this requires  several ground stations increasing the cost of ground segment. 
    \item {\bf Micro-diversity} The mitigation techniques for turbulence are termed as micro-scopic diversity techniques. For the optical feeder uplink,  multiple apertures are placed in a distance higher than the coherence length of turbulence; this configuration, termed as transmitter diversity \cite{giggenbach_high-throughput_2015},  is used to combat turbulence. While several works have focused on exploiting the diversity gain achievable from \gls{mimo} optical setups e.g., \cite{mesleh_optical_2011,ozbilgin_optical_2015,simon_alamouti-type_2005}, they are typically considered for terrestrial optical networks and have certain shortcomings for \gls{fso}. The Repetition Coding (RC), considered for example in,  \cite{safari_we_2008,roy_optical_2015,giggenbach_high-throughput_2015}, where identical information is transmitted over multiple transmitters from different wavelengths.
\end{itemize}
The design of the optical feeder link depends on the on-board processing capabilities. Fully regenerative payloads offer the best performance due to their additional processing  partly because of its ability to include a strong \gls{fec} to enhance the optical link. However, the complexity of such payloads is rather high and would be considered in later generations of satellites. On the other hand, transparent processing offers a simple, yet effective, solution to enable \gls{fso}. Two architectures have been considered in the literature for transparent satellites  \cite{barrios_rivoli-tn1:_2015}; these are,
\begin{itemize}
    \item Analog Transparent: In this architecture, the \gls{rf} signal is used (after appropriate biasing) to modulate the intensity of the optical source. It offers a very simple modulation onto the optical carrier and demodulation on-board the satellites. However, it offers no protection to the optical link and can exhibit poor performance.
    \item Digital Transparent: Herein, the baseband radio signal is sufficiently oversampled (both the I/ Q channels), quantized and the resulting sequence of bits modulates an optical source digitally, e.g., Pulse position modulation or On-Off keying. This architecture offers the possibility to include \gls{fec} to mitigate impairments on the optical channel; however, it suffers from bandwidth expansion, additional noise injection and higher complexity
\end{itemize}
A comprehensive study of optical feeder links has been pursued in \cite{ONSET}. Herein, the nuances of the optical and \gls{rf} links are modelled and included in an end-to-end simulator with optical feeder links and \gls{rf} user links. Both micro and macro diversities are considered and performance studied for modelled channels as well as measurements. The results provide directions on the development of future \gls{fso} systems.

\subsubsection{Satellite swarms and synchronization}

Some implementations of distributed space systems such as constellations and satellite trains are relatively well established, whereas satellite swarms containing tens to even thousands of small spacecraft are still in an active research and development phase \cite{juan_Radhakrishnan2016a}. The use of very small satellites has been gaining popularity thanks to recent advances in electronics miniaturization and the decrease of cost promoted by the scale-economy and mass production \cite{juan_swarm_Edmonson2015, juan_Manchester2015}. 

The revolutionary strength of satellite swarms is in their enormous size and complete flexibility; they are envisioned to contain from tens to even thousands of individual spacecraft operating together to achieve their objectives resembling the behaviour of animal swarms. The set of spacecraft can be nano-satellites and even femto-satellites with a mass of a few grams, with restricted capabilities but the complete swarm spacecraft can potentially produce a very capable system addressing complex problems that could not be solved with monolithic missions. As an example, the implementation of Synthetic Aperture Radar missions from higher orbits (MEO or GEO) can only have a realistic power budget using swarm multi-static configurations \cite{juan_MontiGuarnieri2015,juan_Yu2009}.  Similarly, many  other applications can be enabled by satellite swarms missions, such as the characterization planetary atmospheres, the estimate the composition of asteroids, the deep-space exploration, the investigation on Earth’s ionosphere \cite{juan_swarm_Manchester2011}.  All these applications, in the remote sensing area, have been evaluated for its implementation in swarms. In these applications the designers can make use of sensor fusion and offline processing. However, the implementation of data link communications using satellite swarms have not been so popular, due to the stringent technical requirements, and in particular the synchronization requirements. Synchronization in terms of absolute phase, frequency and time for clock or local oscillator signals is essential for  distributed and collaborative beamforming\cite{juan_swarm_Merlano-Duncan2019}. 

\subsubsection{Deep Space Comms}
\label{DeepSpaceComm}

Because of their very specific nature, Deep Space Comms pose specific telecommunication challenges that required specific solutions. 
The first and most important cause of challenges in Deep Space Communications is the huge distance between the spacecraft and the Earth. According to the ITU definition, in fact, it is possible to talk about Deep Space Communications when the spacecraft is at least 2 Millions km away from the Earth. The first challenge poses by such a huge distance, is the very low available SNR. Just to give some number let us consider only the free space loss degradation. The increase of the FSPL with respect to the case of a GEO satellite for different object of our solar system is reported in Table \ref{FSPL}.

\begin{table}[!t]
\caption{Additional Free Space Loss and transmission delay for different location in the solar system with respect to a GEO satellite}
\begin{center}
\begin{tabular}{ |c|c|c| } 
 \hline
 Place & FSPL (wrt GEO sat) & Delay \\ 
 \hline
 Moon & +20.9151dB & 1.2 sec\\ 
 Mars & +78.4164dB & 12.5 min\\ 
 Jupiter & +86.9357dB & 44 min\\
 Pluto & +102.8534dB & 4h 37min\\
 \hline
\end{tabular}
\label{FSPL}
\end{center}
\end{table}

This limitation is particularly challenging for the downlink (from the spacecraft to Earth). While, in fact, in the uplink the signal is generated on Earth with basically no limitation on the available transmitted power, the situation is totally different for the downlink where the transmitted power is strongly limited by the power that the spacecraft is able to generate. Power generation is very difficult for a spacecraft far from the sun. Using solar panel for power generation, we have to keep in mind that the solar flux goes down by a factor of four each time the distance from the Sun doubles, so a solar panel at Jupiter can only generate a billionth the power as at Earth. A more efficient alternative is to generate the on-board power through a radioisotope thermoelectric generator (RTG). An RTG uses the fact that radioactive materials (such as plutonium) generate heat as they decay into non-radioactive materials. The heat is converted into electricity by an array of thermocouples which then power the spacecraft. While this power generation method is very effective from a technical standpoint, nuclear-based generators are expensive (due to the limited amount of nuclear material available) and politically sensitive and it poses a security problem. If, in fact, an accident happens to the rocket during the launch of the spacecraft the radioactive material can be spread in the atmosphere.

\begin{figure*}[!t]
\centering
\includegraphics[width=160mm,keepaspectratio]{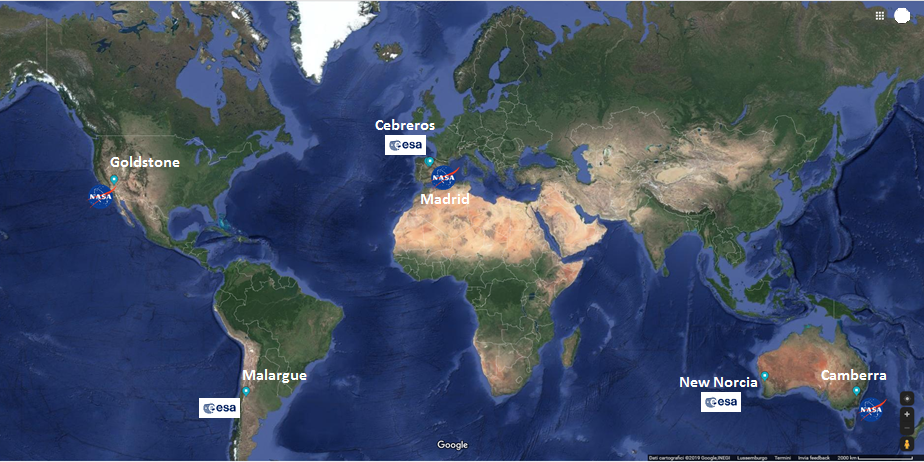}
\caption{Locations' of NASA Deep Space Network and ESA ESTRACK sites}
\label{DSN_Loc}
\end{figure*}

To overcome the limitation in terms of SNR, the space agencies have dedicated transmitting/receiving sites \cite{NASA_DSN} \cite{ESA_DSN}. These sites constitute what is generally known as Deep Space Network. In the Deep Space network sites, huge antennas (e.g. 35 and 70 meters diameter dish) are combined with cryogenic cooled antenna feeds: \cite{JPLCryo}, \cite{NASA_Cryo1} and \cite{NASA_Cryo2}. As shown in Fig. \ref{DSN_Loc} for the ESA and NASA Deep Space Network, these sites are separated on the Earth surface by approximately 120 degrees, in order to guarantee 24/7 coverage, despite the relative position between the Earth and the spacecraft. In addition, these locations have been selected in order to guarantee as limited as possible amount of interference and rain fading. 

Another telecommunication challenge related to the huge distance that we deal with in case of Deep Space Communication, is related to delay. As for the previous challenge, we report some numbers for different locations in our solar system in Table \ref{FSPL}.
Because of the huge transmission delay, it is evident that the spacecraft cannot be operated in real-time. On the contrary spacecraft are usually “sequenced”, meaning that a long list of commands is prepared in a program that is then transmitted to the spacecraft well in advance, in order to operate the spacecraft for long periods without commands from Earth. It is evident that the huge delay prevent the usage of any ARQ mechanism, so the transmission scheme must be very reliable in order to guarantee that the transmitted message is correctly received. This high-reliability level is accomplished through the use of very powerful error-correcting code and low order modulations as specified in CCDSD standard for both the downlink (aka Telemetry link) \cite{CCSDS_TM} and uplink (aka Telecommand link) \cite{CCSDS_TC}.

A potentially dangerous effect on communications between Earth and space probes is due to scintillation, due to propagation through the solar corona, when the signal encounters solar conjunction. This event might, in fact, cause error rate degradation and eventually residual carrier unlock. Accordingly, the amount of scintillation is due to charged particles of the solar corona and depends on the solar elongation (i.e. minimum distance of the signal ray path from the sun), solar cycle and sub-solar latitude of the signal path.

\begin{figure}[!t]
\centering
\includegraphics[width=45mm,keepaspectratio]{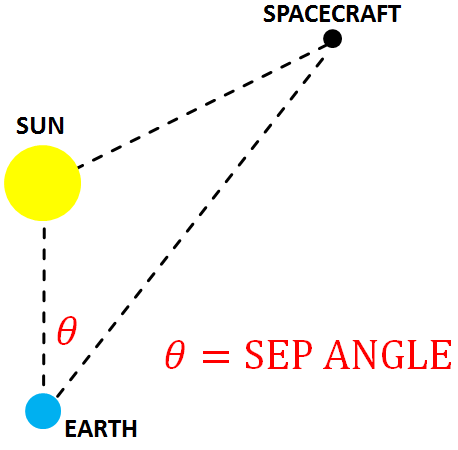}
\caption{Solar conjunction geometry}
\label{SolarConjGeometry}
\end{figure}

Several statistical models describe the effects of a scattering medium on radio communications. For solar scintillation, each coronal in-homogeneity can be modelled as a scattering center for the impinging electromagnetic wave. At the receiver the electric fields of the scattered waves add up, producing a time-varying interference pattern that may lead to fading. Considering the receiver far away from the scattering medium, it is reasonable to assume that the received electric field is the sum of a large number of statistically independent waves scattered from different regions within the medium. Application of the central limit theorem leads to a complex-valued received signal with independent Gaussian real and imaginary parts. Assuming the real and imaginary random components have the same variance we may thus model the scintillation channel as a multipath fading channel with a Rice distribution.

The Rician statistics depends on the carrier frequency as well as the geometry of the Sun, Earth and Probe i.e. the SEP angle shown in Fig. \ref{SolarConjGeometry}. Usually, the Rician fading distribution is specified in terms of the scintillation index, noted by $m$, which is the ratio of the standard deviation of the received signal power to its mean.

This topic has been recently investigated in the context of a research project funded by the ESA. During this project the several solutions have been proposed for this scenario, the details can be found in the following scientific publications: \cite{RESCUE3} \cite{RESCUE2} and \cite{RESCUE1}.


\section{Medium Access Control: Enablers \& Topics} \label{sec6}

The aim of this section is to address some fundamental aspects related to the Medium Access Control (MAC) layer of a satellite communication system. This layer is responsible for controlling channel access mechanisms to enable several terminals or network nodes to communicate in a network.

\subsection{MAC Protocols for UHTS}

\subsubsection{Forward Link Scheduling} 

Forward packet scheduling has been studied since the birth of DVB-S2 standard (developed in 2003) in order to fully exploit its new features. The air interface suggested in DVB-S2 is able to adapt the Code and Modulation (ACM) to the propagation conditions so that the spectral efficiency maximized. This is done by providing to each user with the most suitable modulation and code (ModCod) value according to the measured SINR. Therefore, DVB-S2 standard permits the management of different services guaranteeing a certain Quality of Service (QoS).

Packet scheduling mechanisms particularly play a key role to guarantee an efficient resource management, since they can play with the time dimension to distribute satellite resources among different beams and receivers based on the channel conditions and QoS requirements. In this context, the adaptation loop, which comprises the set of operations starting by the channel estimation at the satellite terminal and ending with reception of the information encoded/modulated according to the reported channel status, plays a fundamental role.
On the other hand, the packet traffic in broadband services is bursty (i.e., the data rate needed to support the different services is not constant). Therefore, the goal of the forward link satellite scheduler is to optimize bandwidth (capacity) utilization and QoS, in the presence of traffic flows generated by services with different requirements.

In general, the satellite scheduler can consider the following parameters for the design of the scheduling strategy:
\begin{itemize}
	\item\textbf{Channel status:} The channel status information reported by the satellite terminals is essential to combine packets in a single frame according to the propagation conditions. This includes changes in the link quality experienced by each terminal due to the weather conditions, mobility, jamming, and other factors.
	\item\textbf{Packet priority:} Lower priority packets can be delayed (or even dropped) in favor of high priority packets. For instance, emergency real-time packets, including emergency medical communications, rescue and natural disaster management related services, should be served with high priority. 
	\item\textbf{QoS requirements:} Give priority to packets with high QoS requirements.
	\item\textbf{Buffer occupation:} Scheduling algorithms are strictly related to buffer management problem. Give priority to packets allocated in highly congested buffers.
\end{itemize}

We can distinguish two scheduling cases, which are detailed in the following sections:
\begin{itemize}
	\item\textbf{Unicast Scheduling}: One user (per beam) is scheduled within each frame
	\item\textbf{Multicast Scheduling}: Multiple users (per beam) are scheduled within each frame
\end{itemize}

\paragraph{Unicast scheduling}

There are two design aspects to be taken into account:
\begin{itemize}
	\item	\textbf{Demand satisfaction}: We do not need to schedule a user which has an empty queue. We need to schedule users which have large pending data volume first. But this depends on the Service-Level Agreement (SLA) each user has signed with the satellite operator. SLA can be min rate over time, max rate over time, average rate over time, latency, etc.
    \item \textbf{Interference avoidance}: Whenever frequency is reused across beams, interference appears. Users scheduled in adjacent synchronous frames, should be far located one from the other to minimize interference. ``far distance'' can be translated as ``different channels'', or sometime called ``channel vectors that are as orthogonal as possible''. This means that the users serves simultaneously over different beams should have orthogonal (ideally) channel vectors. This is essentially the basis of the semi-orthogonality criteria originally proposed in \cite{yoo}.
\end{itemize}

\paragraph{Multicast scheduling}

Serving a single user within a single frame is not a practical assumption as it rarely happens in real systems. Before we observed that minimizing the inter-beam interference can be achieved by scheduling users within adjacent synchronous frames according to orthogonal channel conditions. When considering multiple users within a frame, another design constraint applies. Since all packets in a frame are served using the MODCOD imposed by the worst user contained in that frame, significant performance gains are expected from a scheduler that groups the terminals according to similar propagation conditions.

Clearly, the combination of throughput requirements (PHY layer) and service requirements (NET layer) claim for a cross-layer scheduling design, where the packets are queued according to QoS-class and channel conditions.

The satellite traffic scheduling has been widely addressed in the literature \cite{ParragaNiebla2005,Tropea2011,4114275,RendonMorales2011,Neely2003,NeelyPhd2003,Du2009,Fairhurst2008}. In \cite{ParragaNiebla2005,Tropea2011,4114275}, a new scheduling approach suitable for DVB-S2 systems characterized by a two level architecture is proposed. This structure is able to take into account QoS requirements (i.e. buffer congestion, buffer size, dropped packets, queue waiting time)and MODCOD parameters. In \cite{RendonMorales2011}, a scheduler for DVB-S2 based on the Weighted Round Robin (WRR) mechanism is proposed, whose weighting takes into account the traffic class as well as available capacity. In \cite{Neely2003,NeelyPhd2003}, the capacity region a multi-beam satellite with N time-varying downlink channels and N on-board output queues is established. In \cite{Du2009}, the Satellite Digital Multimedia Broadcasting (SDMB) is investigated. Given the unidirectional nature of the SDMB system and the point-to-multipoint services it provides, the authors in \cite{Du2009} propose a novel adaptive multidimensional QoS-based (AMQ) packet scheduling scheme for provisioning heterogeneous services to provide better QoS guarantee while achieving more efficient resource utilization via an adaptive service prioritization algorithm. In \cite{Fairhurst2008}, a novel channel-aware scheduler scheme compliant with DVB-S2 is proposed which also considers the expedited delivery requirements for delay-sensitive packets. 

Previous works have focused in cross-layer scheduling without considering aggressive frequency reuse. Introducing precoding techniques, the functionalities of PHY, MAC and NET layers become even more intertwined. The main reason is that the achieved user rates at PHY are dependent on the packet scheduling due to the non-orthogonal access of the medium \cite{joroughi,taricco,christopoulos,guidotti,lagunas3}.

The works in \cite{joroughi,taricco,christopoulos}, they all assume that the number of users to be grouped into the same frame is fixed and constant across the beams. In addition, \cite{joroughi,taricco,christopoulos} follow a two-step approach where first a single user is classified in each group, and next the rest of the users are classified. In particular, \cite{joroughi,taricco} they randomly chose a user as a reference and then define the remaining group members associated with that user, while \cite{christopoulos} selects the first user per group according to the semi-orthogonality criteria originally proposed in \cite{yoo}.
The works in \cite{guidotti,lagunas3} try to avoid the two-step approach and perform the user-per-group classification at once. The work in \cite{guidotti} makes use of a geographical strategy, by sectorizing the beam. The work in \cite{lagunas3} considers a graph-based partitioning approach using conventional spectral clustering. While \cite{lagunas3} assumes a fix number of users per frame, \cite{guidotti} does not impose any constraint on that. On the other hand, \cite{lagunas3} proposes a second step to orthognalize as much as possible the adjacent synchronous beam transmissions.

\subsubsection{Return Link Scheduling} 

\begin{figure}[t]
\centering
\centerline{\includegraphics[scale=0.7]{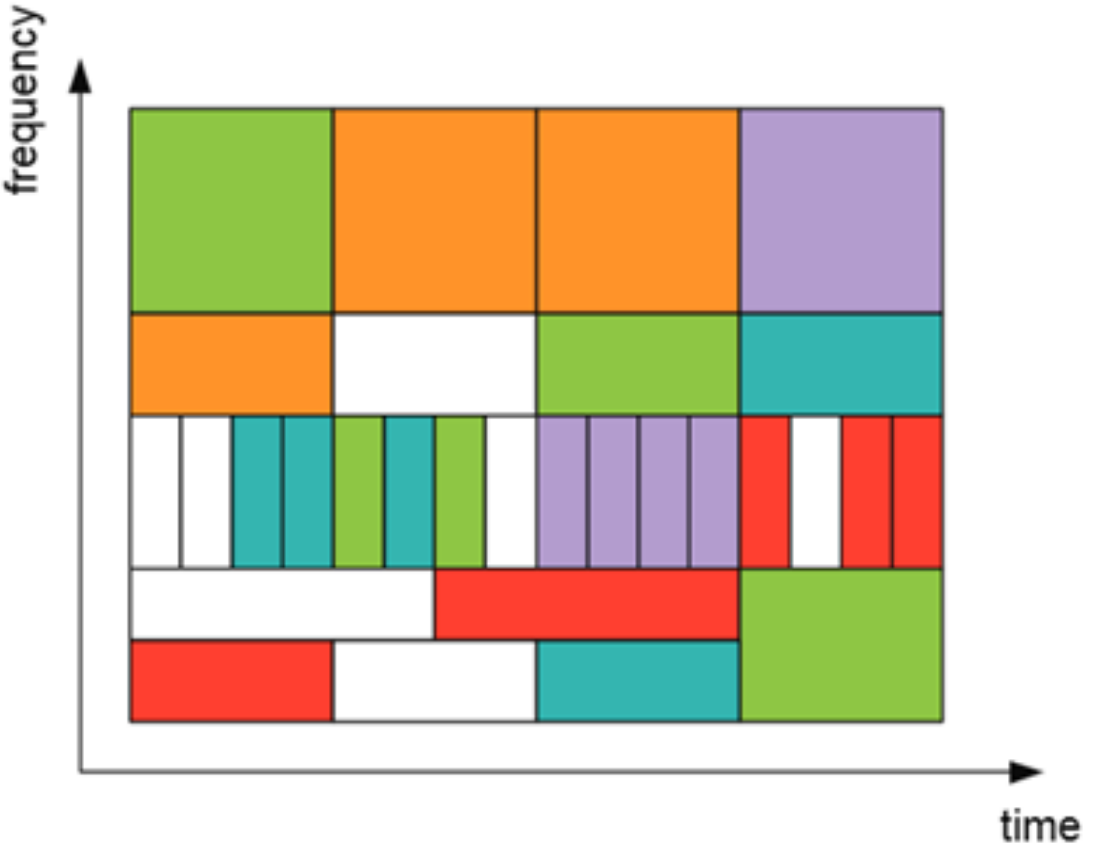}}
\caption{MF-TDMA scheme used in satellite uplink}
\label{MFTDMA_Fig}
\end{figure}

In current satellite systems, the Network Control Center (NCC) is the entity that collects the traffic demands of the Return Channel Satellite Terminals (RCSTs) and distributes the available resources accordingly. The return link access is based on the Multi-Frequency Time Division Multiple Access (MF-TDMA) scheme, which provides high bandwidth efficiency for multiple users. Fig. \ref{MFTDMA_Fig} shows the frequency-time distribution of a sample MF-TDMA scheme. MF-TDMA is a system of access control to a set of digitally modulated carriers whereby the RCSTs are capable of frequency hopping among those carriers for the purposes of transmitting short bursts of data within assigned time slots.  It is noteworthy that the return link can optionally use a continuous carrier (CC) instead of MF-TDMA. The advantage of this scheme is the more efficient adaptation to widely varying transmission requirements, typical of multimedia, at the expense of slightly more complex RCSTs. The NCC periodically broadcasts a signaling frame, the TBTP (Terminal Burst Time Plan), which updates the timeslot allocation within a super-frame between every competing RCTS. 

However, the MF-TDMA proposed in DVB standard has been shown to not perform optimally for bradband satellite systems \cite{5286351}. In situations where the traffic is bursty, fixed assignment mechanisms lead to inefficient use of the resources. Random access protocols are an interesting alternative. In random access, data packets are instantly transmitted, independent of other nodes activities. There is no coordination which translated into possible packet collisions. Unlike DVB-RCS, DVB-RCS2 optionally supports RA to return link. For more details on uplink scheduling and RA, the reader is referred to Section \ref{maciot_section}.


\begin{figure*}[t]
\centering
\includegraphics[width=180mm,keepaspectratio]{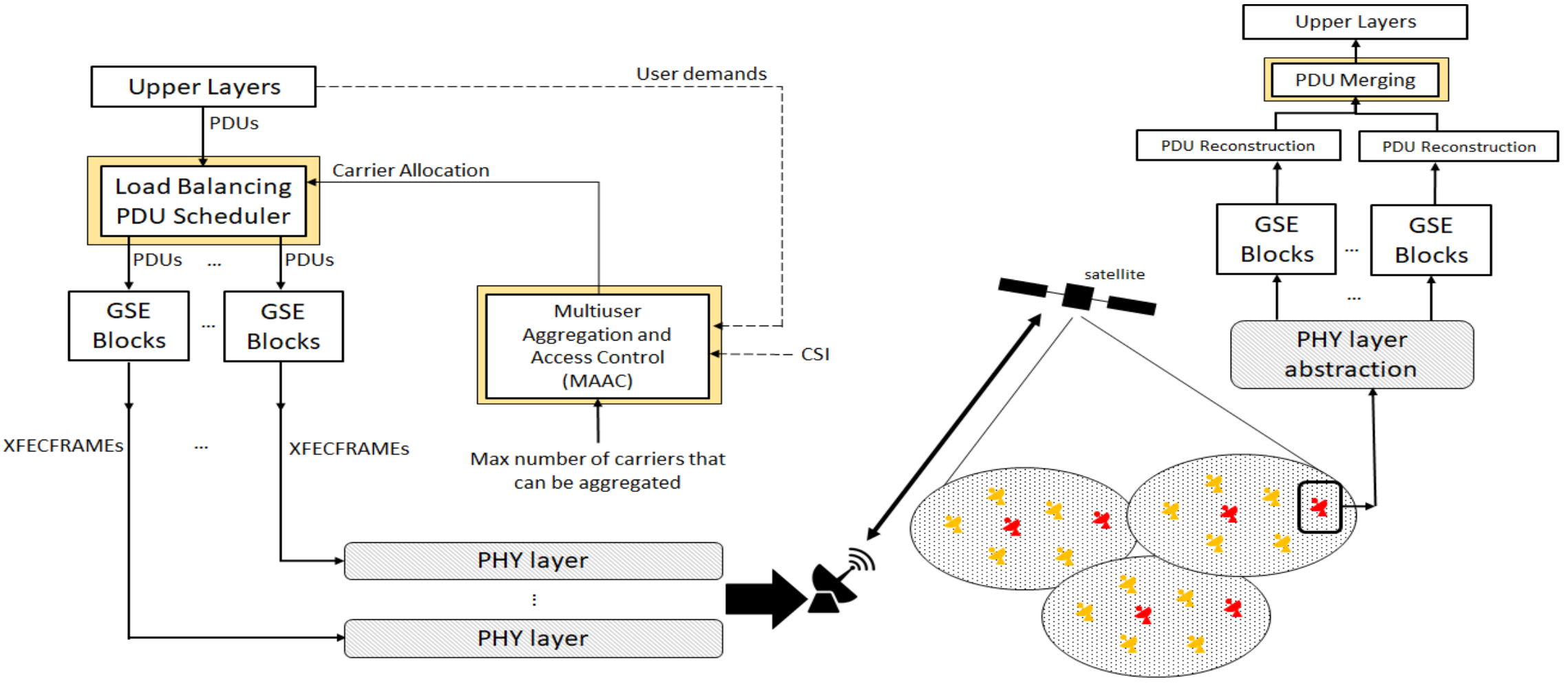}
\caption{CA Architecture as proposed in \cite{cadsat}.}
\label{CA_Fig}
\end{figure*}

\subsubsection{Resource allocation}

Satellite resources are expensive and thus it is necessary to optimize and time-share these precious resources. In this section, we review the current state-of-the-art of resource management solutions in multi-beam satellite systems.

\paragraph{Power Assignment}

Power is a huge concern for the satellite, as the available power on-board is limited and should be used wisely. In \cite{choi}, a power allocation and packet scheduling technique based on traffic demands and channel conditions is proposed. However, interbeam interference is neglected by assuming non-adjacent active beams. Interbeam interference is also overlooked in \cite{lei}. Interbeam interference is dependent on the power allocated to each beam and therefore, affects the total system performance. If not considered, it limits the flexibility of the system and can be a problem when a hot-spot requires coverage from multiple adjacent active beams. The benefits of power allocation are explored in \cite{aravanis}, where a sub-optimal solution is proposed providing some insights about the relation between assigned power and offered capacity. However, the complexity of the solution in \cite{aravanis} limits its applicability. 

With the advent of DTP payloads (see Section \ref{digital_Payloads_Section}) where each carrier can be independently power controlled on board through the digital channelization, power can be flexible moved from one beam to another. Power assignment is considered the first level of flexibility (and the easiest to implement). Sometimes, however, power flexibility is not enough and the channelization (e.g. bandwidth and frequency) should also be adapted to provide another degree of flexibility.

\paragraph{Channelization: Carrier and Bandwidth Assignment}

Dynamic bandwidth allocation techniques can be classified into three groups depending on the amount of spectrum that is shared: (i) Orthogonal but asymmetric carrier assignment across beams, with no inter-beam interference, (ii) Semi-orthogonal asymmetric carrier assignment, where certain overlap between spectrum of beams is allowed, and (iii) Complete full frequency reuse, where all beams share the total spectrum resource. While (i) seems to not provide enough capacity, (ii) and (iii) have been identified as most promising. The semi-orthogonal scenario has been considered in \cite{lei,choi,aravanis,cocco,alberti}. In \cite{alberti}, a very simple sub-optimal iterative bandwidth assignment is considered to deal with the demand-matching problem. More computationally expensive algorithms have been proposed in \cite{lei,aravanis,cocco} with similar objective. However, scenario (ii) and (iii) together with precoding have been overlooked as the introduction of makes the problem much more challenging, but at the same time with very high potential in terms of system performance. While in (iii) precoding is mandatory, in (ii) one can design which carriers to be precoded and which not, depending on the requested demand 

\subsubsection{Beamhopping}

In conventional broadband multibeam HTS system, all satellite beams are constantly illuminated, even if there is no demand to be satisfied. It is widely accepted that the beam data demand is not homogenous, shifting from beam to beam over the course of a day or seasonally. Clearly, such uneven beam traffic patterns claim for a more efficient resource allocation mechanism. This has given rise to the beam hopping concept, a novel beam-illumination technique able to flexibly allocate on--board resources over the service coverage \cite{freedman}. With beam hopping, all the available satellite resources are employed to provide service to a certain subset of beams, which is active for some portion of time, dwelling just long enough to fill the demand in each beam. The set of illuminated beams changes in each time--slot based on a time--space transmission pattern that is periodically repeated. By modulating the period and duration that each of the beams is illuminated, different offered capacity values can be achieved in different beams.

The beam hopping procedure on one hand allows higher frequency reuse schemes by placing inactive beams as barriers for the co--channel interference and on the other hand allows the use of a reduced number of on--board power amplifiers at each time slot. Beam hopping uncovers entirely new problems that were never considered before in satellite communications: the challenge of designing an illumination pattern able to perfectly match the demands \cite{eva_patent,ginesi_patent}, acquisition and synchronization of bursty transmitted data \cite{giraud}, the exploitation of extra degrees of freedom provided by the fact that certain regions of the coverage area are inactive.

In addition, in certain scenarios (like the high throughput full frequency reuses scenario) the performance of Beam Hopping is heavily degraded by the self--interference generated by the system, particularly when neighboring co--channel beams are activated at the same time \cite{kibria}. 

\subsubsection{Carrier Aggregation}

Carrier Aggregation (CA) is an integral part of current LTE terrestrial networks. Its ability to enhance the peak data
rate, to efficiently utilize the limited available spectrum resources and to satisfy the demand for data-hungry applications has drawn large attention from the satellite communications community. In particular, the application of CA in satellite communications has received interest within \cite{cadsat}, where several potential scenarios have been discussed and analyzed based on market, business and technical feasibility. The CA architecture is illustrated in Fig. \ref{CA_Fig}.

CA represents an improved version of Channel Bonding (CB). Accordng to the DVB-S2X standard CB combines multiple adjacent channels to constitute larger transmission bandwidths, while CA can aggregate both contiguous and non-contiguous carriers in different spectrum bands \cite{khan}. Most important, CB is primarily designed for broadcast applications and employs constant coding and modulation, while CA is tailored to the emerging broadband traffic and is compatible with the ACM functionality \cite{kibria2}.

CA represents a new level of flexibility by means of smart carrier assignment which, in case of carrier aggregation, it can be seen as flexible bandwidth assignment. Using CA technology has the following advantages: 
\begin{itemize}
	\item More efficient match of capacity demand distribution over satellite coverage,
    \item More users can be accommodated on a satellite,
    \item Commercial potential with higher revenues for broadband satellite operators.
\end{itemize}

Compared to conventional non-CA systems, CA is implemented by means of 3 main blocks. From the GW side, there is a block called ``Multiuser Aggregation and Access Control'' (MAAC) which represents the main intelligence of the system and which is in charge of designing the carrier allocation strategy for all the user terminals of the system as well as the multiplexing of each carrier. Then, a ``Load balancing and PDU scheduler'' module is in charge of implementing the decisions of the MAAC by distributing the incoming protocol data units (PDUs) across the available carriers. The ``Load balancing and PDU scheduler'' block needs to carefully design such that the PDUs are distributed across the selected carriers based on the link capacities so that, at the receiver side, the PDU disordering is minimized. At the receiver side, the most important block is the ``Traffic merging block'', which takes as input the PDU streams of the aggregated carriers and converts them into a single stream of received PDUs.

In \cite{kibria2}, most of the implementation effort is assigned to  the gateway side, so that the user terminal is as simple as possible with the minimum required changes to support CA. Following this approach, the ``Traffic merging block'' will consist on a simple First-In First-Out (FIFO) system. Therefore, it is of extremely importance that the ``Load balancing and PDU scheduler'' module at the gateway side makes sure to schedule the PDUs in a proper way such that they can be easily merged in a single stream with a simple FIFO buffer.

\subsection{MAC protocols for Satellite IoT}
\label{maciot_section}

Designing a MAC protocol for IoT communications is a crucial and challenging aspect, mainly driven by the low-complexity requirement and the need to support an enormous number of IoT devices generating a sporadic traffic to the network. These exist two main groups of MAC protocols in the literature for satellite-IoT applications as shown below.

\subsubsection{Fixed assignment based}

Protocols in this category ensure that each device in the network has separate resources in time, frequency or both for data transmission, hence avoiding data packet collision. A leading IoT technology which uses fixed assignment based protocol is NB-IoT. More specifically, in the downlink transmission OFDMA is used whereas the uplink is based on SC-FDMA. In an OFDMA (SC-FDMA) system, the time-frequency resources allocated to the users are different. Therefore, even in the case that many nodes transmit at the same time, data packet collision does not happen. Of course, in order to achieve this time-frequency separation, the users should be apriori informed on the resourced to use for data transmission. It is also worth highlighting here that a system based on OFDMA (SC-FDMA) requires a strict synchronization in order to maintain the orthogonality both in time and frequency in order to avoid inter-channel interference. The higher RTT delay in the satellite channel and the increased Doppler effects, especially in the LEO orbit, impose a significant challenge from the MAC layer perspective of such systems. As a matter of fact, authors in \cite{sat_iot_mac12} and \cite{sat_iot_mac13} study the impact of the Doppler effects in a satellite-based NB-IoT system and come up with a new resource allocation approach to handle this problem, without modifying the existing fixed assignment based MAC protocol.

\subsubsection{Random access based}
Random access (RA) based protocols are a natural solution for IoT over satellite communications since they match well to the traffic demand characteristics coming from the IoT devices. It is shown by authors in \cite{sat_iot_mac9} and \cite{sat_iot_mac10} that the traditionally used demand assignment multiple access (DAMA) protocol for the satellite return link does not perform well under sporadic IoT traffic with low duty-cycles and very short packet length. In the case of RA protocols, the devices transmit the data using the same channel without prior coordination. Due to the fact that the allocation of resources is random, possibly many devices will use the same resources for data transmission, hence causing packet collisions. The most representative and well-known RA protocol is Aloha. Even though it is quite old, leading IoT technologies such as LoRa and SigFox use a variation of this protocol \cite{sat_iot_mac1}. An Aloha based protocol, named time frequency ALOHA (TFA) is also proposed for the NB-IoT by the authors in \cite{sat_iot_mac14}.
Basically, when the nodes have some data to transmit, they do it without prior coordination. In case an acknowledgment (ACK) is not received from the network, the device goes to sleep and tries again to retransmit the same packet after a random time.  Despite being a simple protocol and performing well at very modest traffic, the increased propagation delay in the satellite channel creates potential network stability issues, making it an unattractive solution for modern IoT satellite applications \cite{sat_iot_mac2}. In the last decade, there has been an effort in investigating more advanced RA schemes for satellite IoT  and a survey can be found in \cite{sat_iot_mac3}. 
A comparative study of RA techniques for satellite-IoT \cite{sat_iot_mac4} shows that the most attractive ones in terms of spectral and energy efficiency are Enhanced Spread-Spectrum ALOHA (E-SSA) \cite{sat_iot_mac6}, Contention Resolution Diversity ALOHA (CRDSA) \cite{sat_iot_mac5}, and Asynchronous Contention Resolution Diversity
ALOHA (ACRDA) \cite{sat_iot_mac7}. The above mentioned best-performing techniques adopt iterative successive interference cancellation to increase the detection probability of the received packets. The authors in \cite{sat_iot_mac2} further investigate the performance of single-frequency and multifrequency CRDSA and ACRDA \cite{sat_iot_mac8} under realistic parameters and for a number of system scenarios of practical
interest. In \cite{sat_iot_mac11} the phase noise impact on the performance of CRDSA is analyzed. 

\subsection{System Coexistence}
\label{syscoex_section}

One of the promising solutions to address the spectrum scarcity problem caused due to spectrum segmentation and the dedicated assignment of available usable radio spectrum is to enable the spectral coexistence of two or more wireless systems over the same set radio frequencies. The spectral coexistence of heterogeneous wireless networks, i.e., coexistence of satellite and terrestrial networks \cite{Sharma2014ASMS} or the coexistence of two satellite networks \cite{Sharma2013IA, Freqpacking2013VTC} is challenging due to several aspects such as the underlying interference links and no prior coordination between primary and secondary systems. In spectral coexistence scenarios, there may be multiple secondary users trying to access the same portion of the radio spectrum. In this situation, the network access should be coordinated in a way that multiple cognitive users do not seek the same portion of the radio spectrum. 

The effective sharing of available radio spectrum among two or more wireless systems can be obtained by utilizing suitable Dynamic Spectrum Sharing (DSS) techniques, which can be divided into coordinated or uncoordinated based on whether the primary and secondary systems exchange the spectrum usage information, i.e., TV WhiteSpace database, or they can operate without any coordination between them, i.e., spectrum sensing. Also, the DSS models can be broadly classified into three types, namely, commons model, exclusive-use model and shared-use model \cite{Hassan2017exclusive}.  In the first model, i.e., spectrum commons model, all the unlicensed or secondary users can access the spectrum with equal rights while in the exclusive-user model, the secondary users acquire the exclusive rights of using the radio spectrum either by providing a cooperation award from the primary system or by purchasing a certain portion of the radio spectrum from spectrum licensees or primary service providers, also known as spectrum trading. On the other hand, the shared-use DSS model utilizes the underutilized or vacant spectrum either in an underlay (interference-avoidance) or interweave (opportunistic) manner \cite{Sharma2015CR}.  Furthermore, several advanced mechanisms which can be employed to enable the spectrum sharing of heterogeneous networks include Licensed Shared Access (LSA), Licensed Assisted Access (LAA), Carrier Aggregation (CA) and Channel Bonding (CB) and Spectrum Access System (SAS) \cite{SharmaFD2018,Yang2016advanced}. 

\subsubsection{Coordinated}
Two multibeam satellites may coexist in the same orbital position by utilizing different architectures, namely, conventional frequency splitting, cooperation, coordination and cognition \cite{Christdualsat}. In the first approach, the total available bandwidth in the forward link is divided into two equal portions, with each segment assigned to one satellite system.   In the second approach, two satellites having multibeam communications payloads with the aggressive frequency reuse are interconnected and synchronized with a high-speed link between the gateways. With the help of advanced signal processing techniques such as precoding, two transmitters located in two different satellites will behave like a large satellite with the equivalent payloads of two satellites. Two interconnected gateways have to exchange the channel state information and data reliably to enable the implementation of precoding techniques.  The main challenge in this architecture is to meet the stringent synchronization demand between two physically separated satellites. 
To reduce the overhead of data exchange and to lower the system complexity, instead of full coordination, partial cooperation between the two coexisting transmitters can be employed.  Such coordination will require the exchange of smaller amount of information, i.e., CSI and does not need to perform symbol level synchronization, thus leading to a reduced system complexity. Although intra-satellite multiuser interference in the coordinated dual satellite architecture can be completely mitigated by employing the precoding techniques, interference arising from the adjacent satellite limits the system performance \cite{Christdualsat}.

\subsubsection{Uncoordinated} 
Another approach for the system coexistence is cognition via high-speed links between the satellite gateways on the Earth. Two satellite systems operating in the same or different orbits may operate over the same set of radio frequencies, with one satellite system being primary and another as secondary by utilizing various techniques such as cognitive interference alignment and cognitive beamhopping. In the cognitive interference alignment approach \cite{Sharma2013IA},  the secondary terminals can employ precoding in away that the received secondary signals at the primary receiver becomes aligned across the alignment vector, which can then be filtered by sacrificing some part of the desired received energy at the primary receiver. Based on the level of coordination between primary and secondary systems, the IA techniques can be of static, uncoordinated and coordinated. 

In the cognitive beamhopping approach \cite{Sharmacogbeamhop}, the secondary satellite having smaller beams can adapt its beamhopping pattern based on the prior knowledge of the beamhopping pattern of the primary satellite in way that the operation of primary (incumbent) satellite does not gets impacted. To enable this, beamhopping pattern of the primary satellite as well as the timing information can be shared to the secondary satellite via a high-speed signaling link between their gateways.  

For the coexistence of NGSO and GSO satellites, inline interference, which arises when an Non-Geostationary (NGSO) satellite passes through a line of sight path between an earth station and a Geostationary (GSO) satellite, may become problematic \cite{SharmaInline2014}. To this end, ITU-R recommendations ITU-R S.1431 \cite{ITURS1431} and ITU-R S.1325 \cite{ITURS1325} provide recommendations for various static and uncoordinated solutions to mitigate inline interference including the following. 
\begin{enumerate}
\item	Satellite diversity: The traffic of the impacted satellite can be switched to an alternative satellite to avoid the main beam to the main beam interference whenever inline events occur.
\item Transmission Ceassation: The link budget design can be designed to accept some outage without switching to another satellite.
\item GSO arc avoidance based on the latitude: With this approach, the coupling of the main beam of NGSO satellites and the main beam of GSO earth station can be avoided by providing sufficient angular separation with respect to the equatorial plane.
\item GSO arc avoidance based on discrimination angle: By switching off the beams when the point of interest in the Earth observes an angular separation (between an NGSO satellite’s main beam and GSO arc) less than a predefined angle.
\item Sidelobe design of NGSO satellite and terminal antennas: The amount of harmful interference from/to satellites and GSO terminals can be minimized by designing the low side-lobe antennas on the terminals and NGSO satellite, respectively. 
\item Satellite selection strategies: Interference scenarios can be avoided by selecting  a satellite that has the highest angular discrimination with respect to other GSO and NGSO and GSO satellites.
\item Frequency channelization: The carrier-to-interference levels can be enhanced by dividing the frequency band into smaller sub-bands and assigning these sub-bands to a distinct beams. 
\end{enumerate}


\section{Networking: Enablers \& Upper-Layer Integration} \label{sec7}

The aim of this section is to cover the main technical advances related to networking and upper-layer integration of SatComs with 5G network. 

\subsection{Software Defined Networking and Network Function Virtualization}

During the last decade, the networking community has witnessed a paradigm shift towards more open architectures based on Software Defined Networking (SDN) in a quest for improved agility, flexibility and cost reduction, in the deployment and operation of networks. The General reference of SDN architectures have been specified by the Open Networking Foundation (ONF) and Internet Engineering Task Force (IETF) in \cite{onftr1,haleplidis1} respectively, reflecting the key principles of SDN: (1) separation of data plane resources (e.g. data forwarding functions) from control and management functions; (2) centralization of the management-control functions and; (3) programmabillity of network functionality through device-neutral and vendor-neutral abstractions and Application Programming Interfaces (APIs). 

The first efforts for implementing the SDN principles on computer networks can go back to more than 2 decades \cite{feamster1}, for example, some works studied the introduction of programmability of some network functions \cite{tennenhouse1,calvert1}, the separation of the control and user plane \cite{van1,smith1}, or the centralization of management-control functions \cite{feamster2,caesar1}, nevertheless, these efforts did not have a practical impact on the network community. However, in the mid-2000s the emergence of a series of works such as the definition of the set of architectural, modeling, associated terminology and protocol requirements to logically separate the control and data forwarding planes of an IP networking devices by IETF \cite{salim1,khosravi1}, or the subsequent development of the first open interface between the control and data planes by the Forwarding and Control Element Separation (ForCES) IETF group \cite{yang1}, gradually unleashed a greater interest on the part of the academy. Subsequently, in the late of 2000’s
emerged some network designs that included a practical deployment and operation such as the developed by Ethane Project \cite{casado1} and the development of the standardized Openflow (OF) API interface \cite{mcKeown1}, that finally triggered a general interest and its adoption. Thenceforth, we have witnessed the appearance of several controller platforms, applications as well as a wide variety options of commercial OF switches, promoting a virtuous circle in favor of its development and adoption. 

Mobile networks also have been progressively embracing SDN concepts and technologies to decouple the control plane from the user plane. In this regard, a variety of proposals for adopting SDN concepts in mobile network architectures have been presented \cite{sama1,bojic1}, likewise, some standardization works as the so called Control and User Plane Separation (CUPS) architecture has been developed as an enhancement of the 4G/LTE standards to fully split control and user plane functions within the Evolved Packet Core (EPC) \cite{3gpp1}, or the new 5G Core Network (5CN) specifications that have consolidated this separation as a key design principle \cite{3gpp2}. In the other hand, while until recently the SDN scope has been focused on the packet-oriented Layers 2 and 3 (e.g. Ethernet, IP/MPLS), different extensions are underway for covering the abstractions necessary for the applicability of SDN in mobile networks \cite{sama1}, for example, for the management of optical transmission (Transport SDN \cite{onf1}) or wireless transport devices (Wireless Transport Networks \cite{onf2}). 

Regarding satellite networks, as they are expected to be an integral part of 5G service deployment \cite{netWorld1}-\cite{giambene1}, the evolution of satellite technology must also follow the guideline towards more open architectures based on SDN technology that are being consolidated within the 5G landscape, not only to bring to satellite technology the SDN benefits, but also to greatly facilitate the seamless integration for combined satellite and terrestrial networks \cite{vital1}-\cite{rossi1}. In this context, satellite networks must also be outfitted with a set of control and management functions and interfaces (API and/or network protocols) compatible with the mainstream SDN architectures and technologies in order to realize a full End-to-End (E2E) networking concept where the whole satellite-terrestrial network behavior can be programmed in a consistent and interoperable manner \cite{mendoza1}. In this regard, although satellite technology has not adopted the SDN concepts at the pace that terrestrial communications networks have done, important advances have been carried out in the recent years regarding the analysis of the potential use cases, requirements, and definition of functional frameworks for the exploitation of SDN technologies in satellite networks. It should be noted that one of the most notable use cases of SDN applied to satellite networks has been Network Function Virtualization (NFV). While the concept of Network Virtualization, that is generaly regarded as an abstraction of the physical network in terms of a logical network \cite{feamster1} is independent of SDN, the key principles of SDN have positioned it as a technological enabler for network virtualization \cite{feamster1}. In this regard, some of the first works were presented in \cite{bertaux1}\cite{bao1}. In \cite{bertaux1}, authors investigated the advantages of introducing network programmability and virtualization using SDN and/or NFV by analyzing four use cases as well as their impacts on a typical satellite system architecture while in \cite{bao1}, authors presented a satellite network architecture based on the idea of decoupling data plane and control plane to gain high efficiency, fine-grained control and flexibility. Subsequently, a variety of works have been presented, some aimed at the research of benefits and technical challenges brought by introducing SDN/NFV into the satellite networks, detailing a set of use cases, opportunities, scenarios and research challenges, but especially, identifying the SDN as a promising enabler in the evolution of service delivery over the integrated satellite-terrestrial networks \cite{rossi1}, \cite{Kapovits1}-\cite{xu1}. Other works, more aimed at the development of platforms and architectures have been presented in \cite{vital1}, \cite{bao1}, \cite{akyildiz1}-\cite{ahmed2}. For example, in \cite{vital1}, authors presented a generic functional architecture for satellite ground segment systems embracing SDN/NFV technologies, detailing the interaction of the SDN controller with the satellite network control plane functions of the satellite network (e.g., network control centre [NCC] functions), as well as the characteristics of both externally exposed and internal interfaces, including a study of the pros and cons of several interfaces and data models that could be leveraged. Likewise, other works aimed at investigating SDN and its integration into satellite networks through several applications have been presented in \cite{bertaux1}, \cite{ferrus4}-\cite{artes1}. For example, the applicability of the functional architecture in a combined satellite-terrestrial backhauling scenario 
presented in \cite{vital1} was further developed in \cite{ferrus4}\cite{mendoza2} with a focus on the use of SDN technologies for the realization of end-to-end Traffic Engineering (TE) applications across the terrestrial and satellite segments. The benefits of such architecture were assessed in \cite{mendoza3} in terms of improved network resource efficiency achieved through the centralized and more fine-grained control of traffic routing enabled by the SDN-based TE applications. In this context, other research works has further progressing in this research area presenting some experimental proof of concepts (PoC) and testbeds for validations on the use of SDN technologies, as will be discussed in detail below (see section VIII.B). Furtermore, other relevant research projects coping with the applicability of SDN/NFV technologies are currently on-going in \cite{Sat5G}\cite{artes1}. In this respect, an overview of the current 5G initiatives and projects followed by a proposed architecture for 5G satellite networks, where the SDN/NFV approach facilitates the integration with the 5G terrestrial system is provided in \cite{giambene1} which also analyses a novel technique based on network coding for the joint exploitation of multiple paths in integrated satellite-terrestrial systems. 

SDN has managed to establish itself as a powerful tool for the solution of several networking problems. In the field of satellite communications the opinion is not different, seen as a key facilitator to enhance the delivery of satellite communications services and to achieve a better integration of the satellite component within the 5G ecosystem. However, SDN is still an emerging technology and its development and maturity are still in process. In the field of satellite communications, while important progress has been achieved so far on network architectural and functional aspects, as well as on the assessment of their benefits mainly via mathematical modelling and more or less sophisticated simulation environments, further research is still warranted towards the practical implementation of integrated satellite-terrestrial solutions and their assessment under more realistic conditions.

\subsection{Caching over Satellite}

One of the challenges in the edge caching is how to effectively prefetch the popular content to the caches considering the high volume of data \cite{multi_cache_2001}. In order to overcome this issue, satellite backhauling has attracted much attention as a promising solution for cache placement phase to exploit the large coverage of the satellite beams. Satellite systems have the ability to provide high throughput links and to operate in multi/broad-cast modes for immense area coverage. 

Due to their multi-hop unicast architecture, the cached content via terrestrial networks has to go through multiple links and has to be transmitted individually towards each base station (BS). On the other hand, with wide area coverage, the satellite backhaul can broadcast content to all BSs or multi-cast contents to multiple groups of BSs. Therefore, bringing these two technologies together can further off-load the network. The main idea is to integrate the satellite and terrestrial telecommunication systems in order to create a hybrid federated content delivery network, which can improve the user experience. The joint deployment of satellite and terrestrial networks can be found in \cite{Brinton13,Kalan17,VuKa,VuSAT19}. The application of satellite communications in feeding several network caches at the same time using broad/multi-cast is investigated in~\cite{sat_wire_2005,Satellite_sate_2000,Brinton13}. The work of~\cite{Satellite_sate_2000} proposes using the broad/multi-cast ability of the satellite to send the requested contents to the caches located at the user side. Online satellite-assisted caching is studied in~\cite{Brinton13}. In this work, satellite broadcast is used to help placing the files in the caches located in the proxy servers. Each server uses the local and global file popularity to update the cache.

Recent works on caching over satellite are presented in  \cite{CS1,CS2,CS3,CS4,CS5,CS6,CS7}.
A two-layer caching algorithm is studied in \cite{CS2} in which cache on the satellite is the first caching layer and the cache in the ground station is the second one. The joint cache optimal is carried out via generic algorithm of the original mixed integer linear programming. In \cite{CS3}, a service model is proposed for hybrid terrestrial/satellite networks  in order to identify viable alternatives to deploy converged satellite-terrestrial services. Two caching policies, namely pull-based and push-based, are studied.
In \cite{CS4}, a back-tracing partition directed on-path
caching mechanism is proposed for hybrid LEO constellation and terrestrial network. By reducing intermittent connectivity as much as possible, it is shown that the redundant transmissions of data access for different users can be largely reduced since the requested files are favorably fetched from intermediate caching nodes, instead of directly from the source.
The authors in \cite{CS5} propose a resource allocation strategy for cache filling in hybrid optimal-satellite networks. It is shown that the placement time can be notably reduced in a hybrid terrestrial-satellite backhaul network, particularly in case of bad weather that impacts the data rate of the wireless optical links. 
The authors in \cite{CS6} propose a novel caching algorithm for optimizing content placement in LEO satellite networks based on many-to-many matching game. 
In \cite{CS7}, the authors investigate the performance of hybrid satellite-terrestrial relay network (HSTRN) under different caching policies. Analytical closed-forms are derived for the outage probability under the most popular uniform content based caching schemes.

By equipped with some computation capabilities in addition to storage capacity, satellite communications have shown potential applications in mobile edge computing (MEC). Thanks to the wide coverage, satellite can be used for task off-loading from mobile users which are out of range from terrestrial MEC servers. It is shown in \cite{mec_sat1, mec_sat2} that with a proper network virtualization algorithm, satellite MEC can significantly reduce latency and improve the energy efficiency compared to the stand-alone 5G terrestrial.


\section{Testbeds \& Prototyping} \label{sec8}

This section focuses on communication testbeds which have been developed for different communication layers in order to practically demonstrate some of the advanced SatCom concepts.

\subsection{PHY \& MAC: SDR Based}

The phenomenal advances of the electronics industry create a trend toward ever
smaller, smarter, cheaper, and more capable devices, from sensors to computers to radios suitable for use in spacecraft applications. Their availability has led to the ongoing revolution in small and medium sized satellites and also in ground based consumer equipment. The community is pushing the effort toward re-configurable SDR SoC (System-on-a-Chip) ground receivers and to the ultimate extreme of satellite-on-a-chip. \cite{juan_Manchester2015}.
The SDR technologies have become popular in the last decade, with plenty of demonstrations for terrestrial wireless communications \cite{juan_Maheshwarappa2017}. 

Multi-standard and adaptive communication systems can be implemented in easily using software-defined radio (SDR) techniques. It consists in that must of signal processing is performed in the digital domain by an appropriate digital signal processing (DSP) device \cite{juan_Kozlowski2018}. An SDR platform consists of a hardware radio frequency (RF) front-end and a DSP unit implemented in signal processors, field programmable gate arrays (FPGA), or GPUs. These platforms are designed to be highly flexible, where all the receiver and transmitter functionalities can be updated by a simple modification of the software code of the DSP devices \cite{juan_8566051}. 

However, there are still not many testbeds in academic papers published on SATCOMs, where the tendency to use SDR technologies has only been seen in recent years, particularly in the small satellites community, where a universal programmable hardware is desirable.
which intensifies the interest in software-defined radio (SDR) in recent years. 

Few works have been published for SDR testbeds for GEO orbits. Those research work focus on interference mitigation in multibeam satellite systems. The popularization of the de-facto Cubesat standard fostered the use of SDR platforms. In the beginning, those platforms were custom designed by the universities and research centers, and later around the beginning to the decade of 2010 by some specialized companies like GomSpace \cite{juan_GOMspace}, Nanoavionics \cite{juan_nanoavionics}, Alen Space \cite{juan_alen}, ISIS Space \cite{juan_isis}, and Tyvak \cite{juan_tyvak}, among others. These companies provide SDR based payloads for Cubesat platforms. The SDR platforms are mainly based on SoCs (System-on-a-Chip) implementing signal acquisition and signal processing using programmable logic (PL) fabric (FPGAs, CPLDs, etc) and processing system (PS) units. Most of these SDR platforms use the popular Zynq 7000 hybrid ARM/FPGA SoCs. There are some examples of communication system payloads using SDR techniques; one example is the GomSpace DVB-S2 compliant transmitters.

Research and academic institutions have been very active in the development of nano-satellites to micro-satellites \cite{juan_7811224} for communication and earth observation applications. However, there a few works using experimental communication testbeds using middle sized or bigger platforms. Some of the published work provide some specific techniques applicable to satellite communications and verified using SDR prototyping. 

Many of the specific SDR implementations are focused on channel coding aspects such as in  \cite{juan_sdr_fec1,juan_sdr_fec2, juan_sdr_fec3}. Low-Density Parity-Check (LDPC) codes, which are the core of the FEC functionalities in the  DBV-S2 and DVB-S2X standards, gained must of the interest in the community because its outstanding BER performance, close to Shannon limit, at relatively low complexity and latency. The implementation of such type of coding, was a technological challenge that was solved with the help of the flexibility of SDR approaches.

There are also several works regarding the waveform design and synchronization aspects for satellite communications. One example of SDR prototyping for pulse shaping optimization and multi-component signaling (MSC) is found in \cite{juan_sdr_mcs}. References \cite{juan_sdr_dvb, juan_sdr_dvbs2} show implementations of DVB-S2 transmitter and receivers.

There are other aspects that had grab some attention from the research community on top of the waveform design and channel coding. These aspects are the interference mitigation and MU-MIMO schemes in multi-beam satellite systems. For the prototyping of such complex systems, in emulation environments, the SDR techniques were the only alternative. 

The reference \cite{juan_6865984} describes an emulation system of geostationary satellite channels by means of software-defined radio techniques. The emulator is based on the National Instruments Universal Software Radio Peripheral (USRP) \cite{juan_ni}, and use LabVIEW programming environment. The emulator is able to process a 1.6 MBaud signal stream in real-time, while adding thermal noise, phase noise, and propagation delay according to the system which is modelled.

LASSENA group \cite{juan_lassena}, at University of Quebec, has been developing an SDR infrastructure for interference mitigation in satellite communications. The objective of these developments aims to create a technical framework for the detection, measurement and mitigation of RFI to resolve satellite link interference issues and increase the global robustness. The infrastructure uses several devices in a hybrid approach. The GEO channel emulation is based on the commercially available single-link satellite channel emulator RT Logic T400CS \cite{juan_kratos}. The payload emulation is based on the BEECube BEE4 SDR platform, which uses multiple FPGA for high-bandwidth real-time signal processing. Finally, the transmitters and receivers are implemented by means of the SDR platforms provided by the company Nutaq Inc., in particular, the Nutaq PicoSDR, and, the Nutaq ZeptoSDR \cite{juan_nutaq}. This SDR infrastructure has been used to test radio frequency interference excision schemes \cite{juan_sdr_Getu2017, juan_sdr_Getu2018}, and also the evaluation of scenarios for airplane connectivity \cite{juan_sdr_Zhang2017}. 

\begin{figure}
\centering
 \includegraphics[width= \columnwidth]{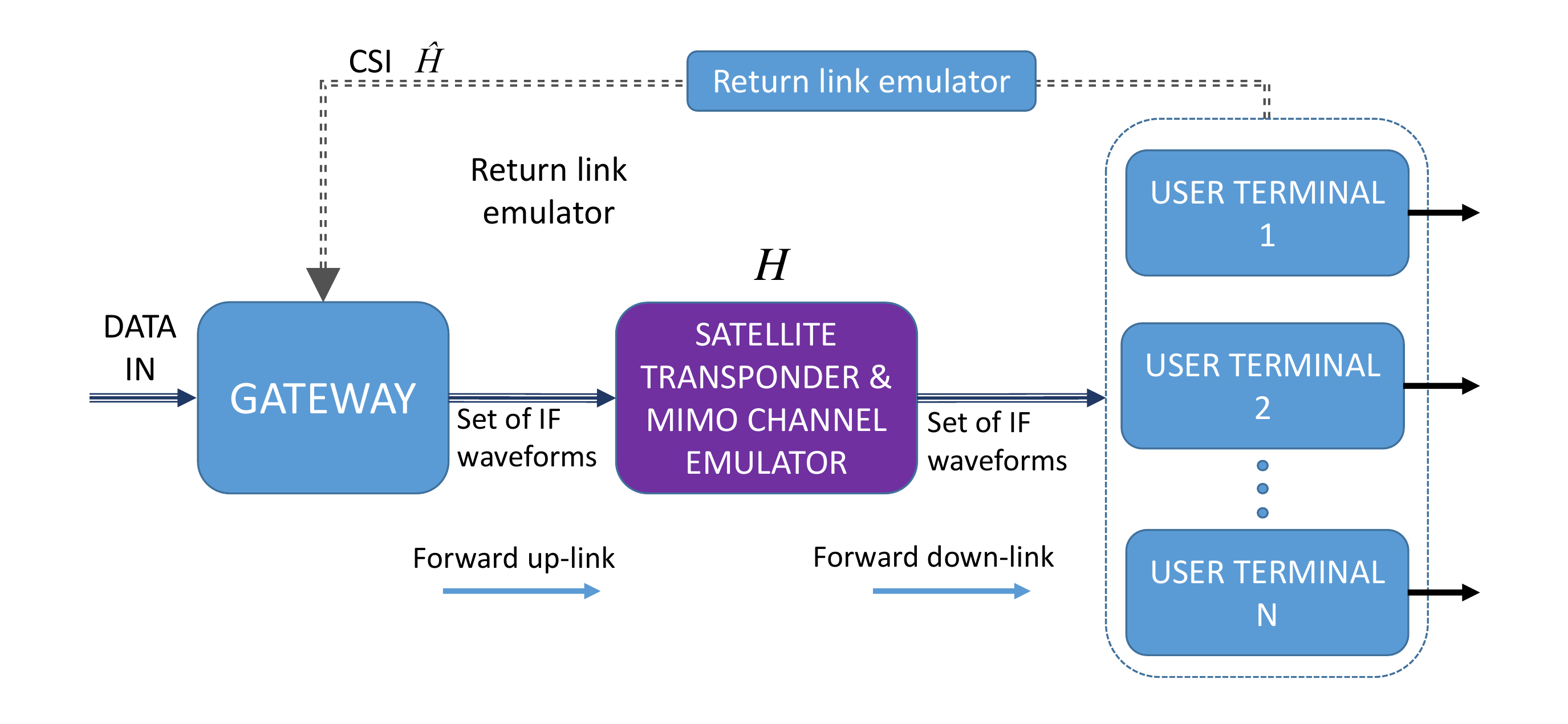} 
\caption{General diagram of end-to-end satellite forward link hardware testbed}
\label{fig:serenade_main}
\end{figure}

\begin{figure}
\centering
 \includegraphics[width= \columnwidth]{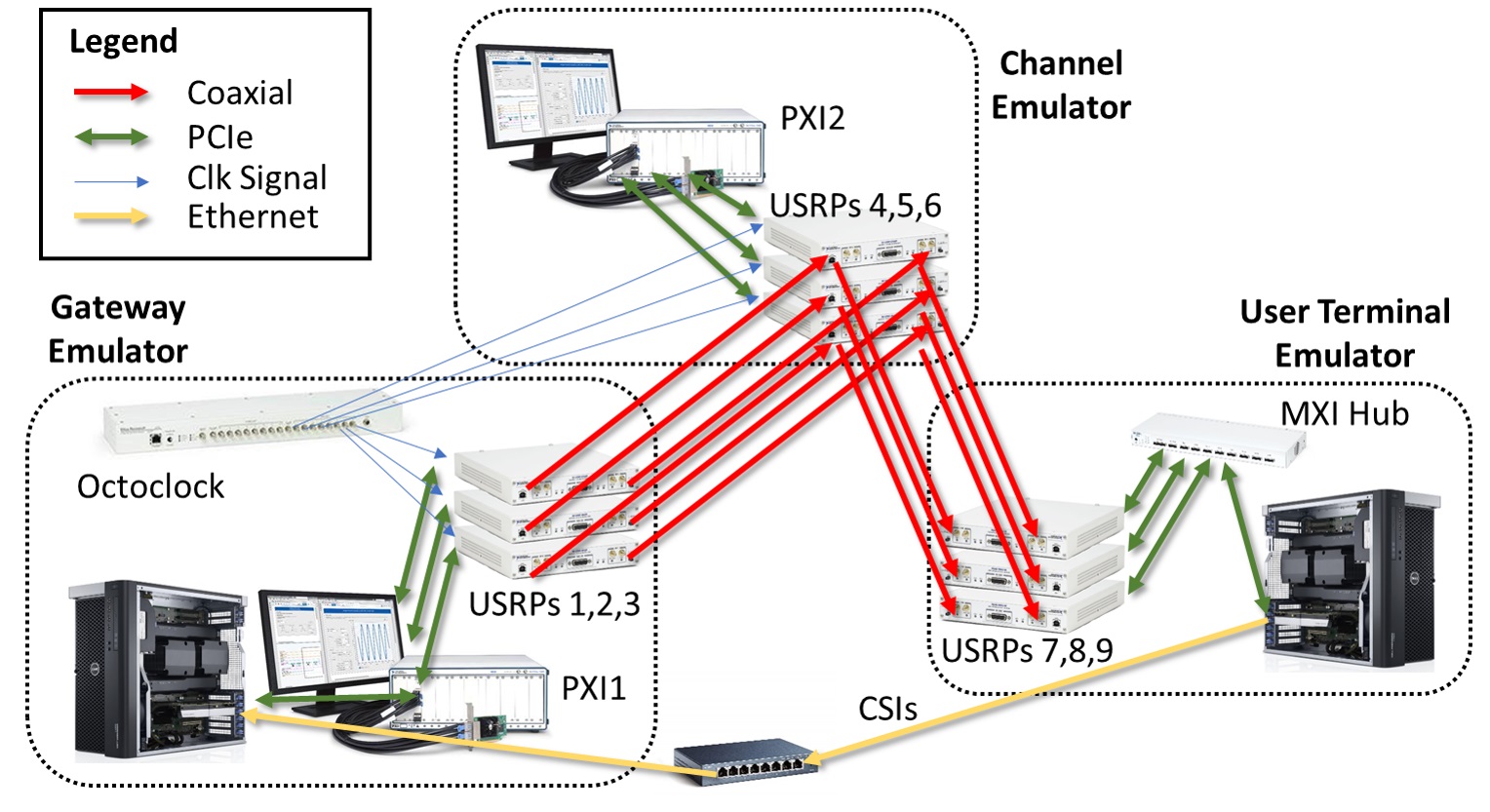} 
\caption{SDR infrastructure of end-to-end satellite forward link hardware testbed}
\label{fig:serenade_DEMO}
\end{figure}

\begin{figure}
\centering
 \includegraphics[width= \columnwidth]{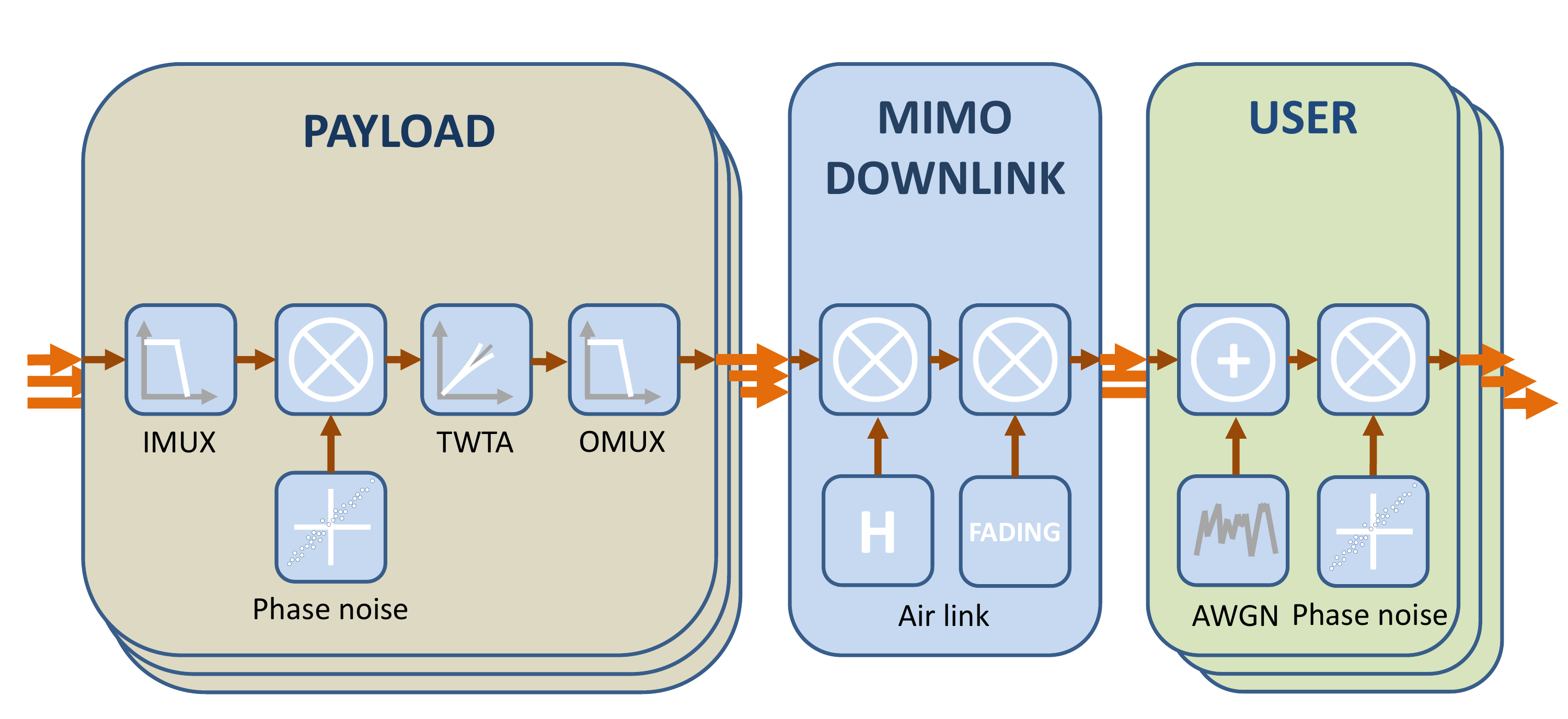} 
\caption{Satellite payload and MIMO downlink channel emulator}
\label{fig:serenade_CE}
\end{figure}

The work in \cite{juan_serenade7, juan_serenade6, juan_serenade5} describes an end-to-end multi-beam satellite communication system emulator, called SERENADE, which is used for the evaluation of precoding and interference mitigation techniques \cite{juan_8340048}. The full testbed is based in National instrument infrastructure, which uses the NI USRP. The end-to-end testbed is hosted in the University of Luxembourg, and emulates the complete forward link for different of transponder orbits from LEO to GEO. The forward link includes a multi-beam satellite ground-based gateway transmitter using the DVB-S2X standard, a multi-beam channel and transponder emulator, and a set of User Terminals (UT) receivers. Fig. \ref{fig:serenade_main} shows a functional block diagram of the SERENADE forward satellite link hardware testbed emulator. 
Fig. \ref{fig:serenade_DEMO} shows a generic description of the SERENADE SDR infrastructure,
which is flexible and scalable for different number of channels.
The infrastructure consists of the NI PXI (PCI EXtension for Instruments) 1085 chassis, which allow centralized connection of the set of USRPs, and FPGA processing units. The FPGA (Field-Programmable Gate Array) processing units, model NI FlexRIO 7976R, are inserted in the PXI chassis slots to increase to real-time processing capabilities, and consist of the Xilinx FPGA Kintex-7 410T. The complete testbed can be configured to have MIMO sizes of up to 16x16 using a modular satellite payload and channel emulator as the one shown in Fig. \ref{fig:serenade_main}. 
A detailed functional diagram of the payload and MIMO channel emulator is shown in Fig. \ref{fig:serenade_CE}.  The channel emulator receives the transmitted signals, applies the payload impairments, and applies the MIMO linear interference pattern to generate the signal provided towards the users. The payload impairments include:
\begin{itemize}
    \item IMUX and OMUX frequency response.
     \item Phase noise emulation. This includes the phase and frequency drifts over time, and can be controlled independently at each of the transponder channels \cite{juan_serenade8}.
     \item Amplifier non-linearities, with re-configurable parameters.
\end{itemize}
The MIMO downlink applies the channel matrix, a fading pattern and a re-configurable delay, and finally the user emulators apply Gaussian noise and the phase noise of a typical UT hardware.

\begin{figure*}[!t]
\begin{centering}
\subfigure a){\includegraphics[width=80mm,keepaspectratio]{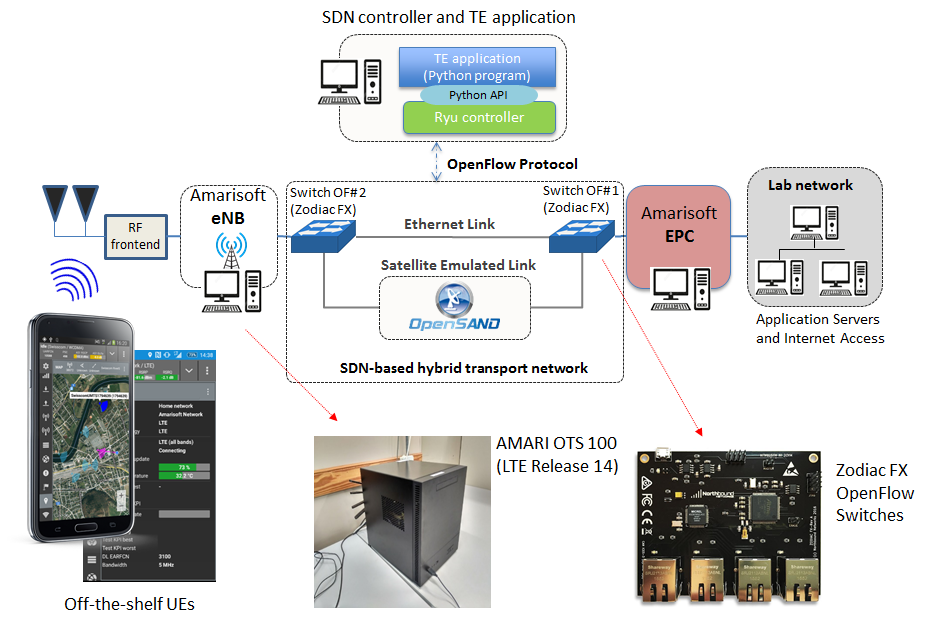}}
\subfigure b){\includegraphics[width=94mm,keepaspectratio]{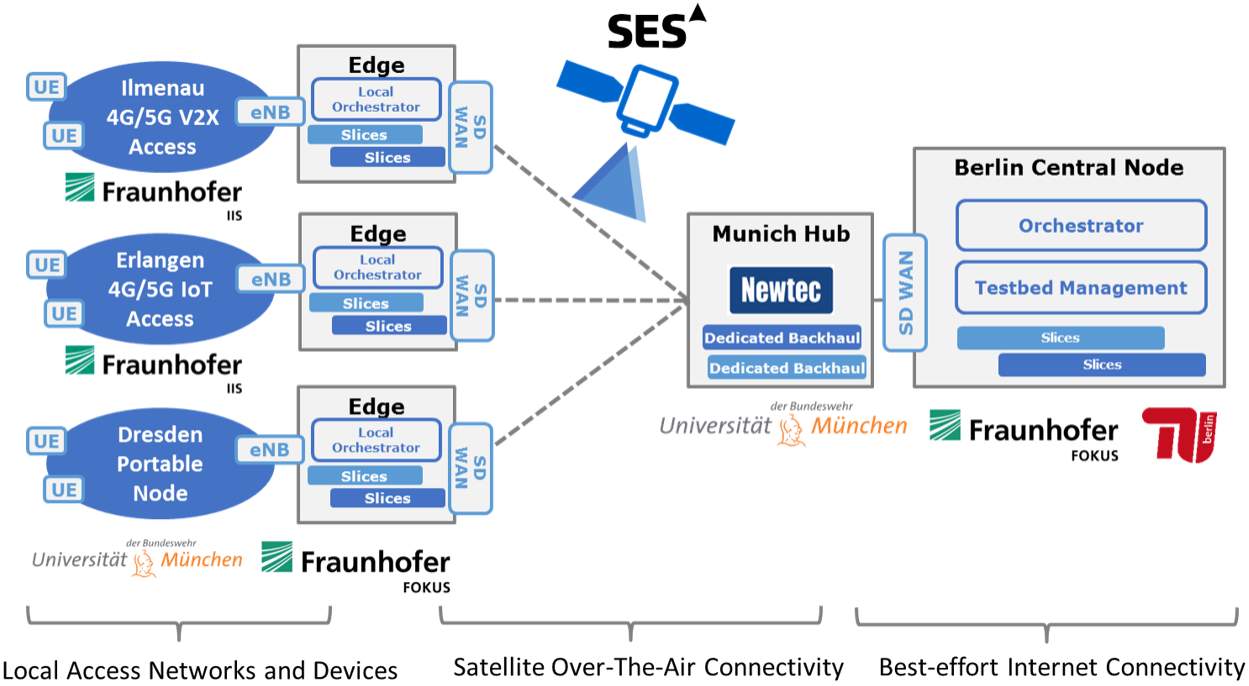}}
\caption{High level view of the experimental test bed components a)  SDN-based traffic engineering solution PoC \cite{mendoza1}; b) SATis5 \cite{satis1}.}
\label{pocs}
\par\end{centering}
\end{figure*}

The UT emulators receive the output signal of the channel emulator, and perform synchronization, channel state estimation, and decode the information stream. The channel state information (CSI) is feedback to the Gateway using a return link emulator over an Ethernet link. The transmitter uses this CSI s to compute the precoding signals. This infrastructure has been used to experiment with novel optimized precoding techniques, know and Symbol Level Precoding (SLP), that optimizes the precoding vectors per every modulated set of input symbols \cite{juan_serenade4, juan_serenade3, juan_serenade2, juan_serenade1 }. 

\subsection{Network Testbeds}

Due to the difficult access to satellite systems and its high costs, tools such as satellite system simulators/emulators, which in turn have a vital role in the development of PoCs/testbeds are becoming increasingly relevant in satellite technology research. In this regard, PoCs/testbeds can play an important role in conducting demonstrations and rigorous evaluation of feasibility, performance, manageability, etc. for new architectures, strategies, protocols, algorithms, etc., under low cost schemes and realistic and reproducible network scenarios. One of the main attributes of a simulator/emulator is the precision among the reproduced models and the systems to be evaluated. In satellite domain there are three most representative elements to be reproduced; (1) Network Topology: e.g. modelling star-mesh structures, satellite types (transparent and regenerative), multispot and multigateway configurations, as well as the representation of network elements as the Gateway (GW), Satellite Terminal (ST), Satellite (SAT), etc.; (2) Physical Layer: e.g. the RF propagation characteristics as channel attenuation models (separated forward/return, uplink/downlink channels) which in turn will determine essential performance parameters (e.g. bit error rate, availability, channel capacity, etc.), and; (3) Delay: the transmission and propagation delay for different satellite orbits (GEO, MEO, HEO/LEO) and ISLs. In case of DVB-S2/RCS systems, emulators/simulators must also reproduce at least important features of these systems as some Network access and Radio Resource Management (RRM) functions (e.g. the adaptability of channel conditions by modulation and coding schemes, etc). Furthermore, satellite simulators/emulators must have other types of features that are very important
for a successful implementation and analysis, such as the performace, interfaces, system interconnections capabilities with real equipments and applications, performance analysis tools, ease of operation/configuration, etc. There is a variety of satellite system simulators/emulators on the market as iTrinegy’s Network Emulators \cite{itrinegy1} or DataSoft Satellite Network Simulator \cite{datasoft1}, among others. Likewise, there are some OpenSource options among which we can highlight OpenSAND \cite{opensand1} initially developed by Thales Alenia Space, the Satellite Network Simulator 3 (SNS3) \cite{sns1} initially developed by Magister Solutions Ltd in the frame of ESA ARTES projects, etc., and others that are still under development as the Real-Time Satellite Network Emulator \cite{artes2} by the European Space Agency (ESA).

As mentioned before, the applicability of PoCs/testbeds developments can cover a broad spectrum of scientific research. One of these examples can be represented by developments focused on the current crucial issue of satellite integration in 5G networks, some of them presented in \cite{mendoza1,satis1,allstar1,Sat5G}. For example, in \cite{mendoza1}, following with the outcomes delivered by the VITAL project \cite{vital1}, the authors presented an experimental proof of concept (PoC) and validation based on the use of SDN technologies for the realization of E2E TE applications in integrated hybrid terrestrial-satellite backhaul mobile scenarios (Fig. \ref{pocs}a). Other three remarkable examples can be found in the still in progress projects SATis5, 5G-ALLSTAR and SAT5G, presented in \cite{satis1,allstar1,sat5g1}, respectively. The SATis5 project \cite{satis1} aims to build a large-scale real-time live end-to-end 5G integrated satellite terrestrial network proof-of-concept testbed (Fig. \ref{pocs}b) in order to implement, deploy and evaluate an integrated satellite-terrestrial 5G network, showcasing the benefits of the satellite integration with the terrestrial infrastructures as part of a comprehensive communication system. The 5G-ALLSTAR project \cite{allstar1}, is aimed to develop selected technologies targeting a set of PoCs to validate and demonstrate in the following heterogeneous real setup: new radio based feasibility of satellite access for providing broadband and reliable 5G services; multi-connectivity support based on cellular and satellite access; spectrum sharing between cellular and satellite access; etc. Finally, the SAT5g project \cite{sat5g1}, is focused in the validation of technical challenges for cost effective satcom solutions for 5G as: virtualisation of satcom network functions to ensure compatibility with the 5G Software Defined Networking (SDN) and Network Functions Virtualisation (NFV) architecture; cellular network management system to control satcoms radio resources and service; Link aggregation scheme for small cell connectivity mitigating Quality of Service (QoS) and latency imbalance between satellite and cellular access; Leveraging 5G features/technologies in satcoms; Optimising/harmonising key management and authentication methods between cellular and satellite access technologies; etc.

As an important tool in the development and innovation of satellite technology by academia and industry, the development of satellite systems simulators/emulators as well as the development of PoCs/testbeds have taken on great relevance in the recent years. However, as justified in \cite{artes2}, the development of such tools must include new and better capabilities such as a highly configurable real-time network (e.g. time-varying topology and link characteristics in satellite constellation networks) and highly accurate models at low-cost equipment, allowing fast developments and simplicity in design. Furthermore, as SDN is also seen as a key facilitator to enhance the delivery of satellite communications services and achieve a better integration of the satellite component within the 5G ecosystem (see section VII.A), the new developments require the introduction of the additional SDN components as key enabling technologies.


\section{Future \& Open Topics} \label{sec9}

\subsection{Digital twins for satellite systems}
Digital twin represents the digital replica of physical objects, places, system, people and devices, which can be utilized for various objectives with the help of sensor/IoT and data analytic technologies. It reflects the involved elements and dynamics of the process by which IoT devices/sensors gather data from the environment, operate and live throughout the life cycles of final products \cite{Digitaltwin1}. Various technologies including machine-to-machine interactions, natural language processing, machine learning, video processing and data analytics can be used to extract and understand the dynamics of the environment, and the extracted knowledge can be subsequently utilized to dynamically recalibrate the environment, leading to the significant impact on the design, build and operational phases of a particular device/product. 

In the domain of satellite systems, the existing methods utilized for fleet management, system design and certification, which are mainly based on heuristic design principles, physical testing and statistical distributions of physical device properties, are not suitable for future generation of satellites which demand for lighter mass with the capability of handling higher loads and the requirements of operating with extreme service conditions over longer duration \cite{Digitaltwin2}. To address these drawbacks, digital twin is expected to play a crucial role in integrating historical and fleet data, maintenance history and sensor data from the satellite on-board integrated health management system to enhance the safety and reliability of satellites/space vehicles. By analyzing all the available information, digital twin helps to forecast different attributes such as response to critical events, the health of a satellite/vehicle system, probability of mission success and remaining useful life, and to activate self-healing mechanisms whenever needed.

Another promising future application of digital twin is to enable space-based monitoring and communication services. The cost and time needed to provide space-based services can be drastically reduced by utilizing software defined components in the satellites, which can be remotely configured from the Earth \cite{Digitaltwin3}. Also, digital twin can enable the creation of autonomous swarms in the satellites by incorporating the intelligent sensing and communication capabilities to the satellite systems. Furthermore, digital twin at the satellites seems promising to enable the global sharing of services and skills by dynamically creating new services, i.e., supply chain in the space, and generating a sharing-based economy in the space.

One crucial aspect to be addressed with regard to the commonly accessible digital twin is to protect the privacy of individual entities and to prevent the information misuse without acquiring the permission of concerned entities. In this direction, one promising enabler could be block-chain technology, which can opt out the records that should not be shared among others. Another issue in digital twin enabled nanosatellite systems is to properly track, control and decommission nano-satellites in order prevent any threats to the ground or other satellites \cite{Digitaltwin3}. Other future issues include how to manage the space debris and pollution by removing the failed or inoperative satellites and how to regulate the digital twin-enabled infrastructure in terms of preventing data misuse by the governments, criminals or terrorist bodies. 

\subsection{Cooperative satellite swarms and clusters enabled by inter-satellite links}

In contrast to monolithic conventional satellite missions, the NewSpace methodology proposes novel distributed architectures promising a paradigm shift in the space industry. Distributed Space Systems (DSS), and in spacial, satellite swarms and clusters will provide improved re-configurability, flexibility, upgrade-ability, responsiveness, and adaptability to structural and functional changes. Large satellite swarms, based on small spacecraft, can also enlarge the autonomy of the mission by upgrading or replacing defective units while the mission remain under operation.
Clusters and swarms are implementation of DSS consisting of an array of autonomous satellites which share the same mission goals, and require communication and cooperation to achieve those goals. In clusters the satellites fly in close formation, and require an accurate observability and controllability of the satellite positions and attitudes for coordinate their operations. This accuracy can be in orders of micrometers, and even, picometers, and is only achievable with powerful propulsion and actuator systems. Usually clusters arrays contain a few to tens of satellites. 
In contrast, swarms, which may consist of tens to even thousands of satellites aiming for the same mission goals, but without keeping a tight relative position in the array formation. This configuration allows the use of cheaper and smaller spacecraft and to enlarge the number of array elements. Recently, the concept of cohesive swarms was coined \cite{juan_swarm_Merlano-Duncan2019} to describe the swarm implementations that do not require to have an accurate controllability of the array relative positions and attitudes but has an accurate monitoring of these parameters. The cohesive swarm use the parameters observations to compensate them in the signals transmitted or received by the whole array. 
In general, the implementation of satellite swarms is in an active research and development phase, and is envisioned to be applied in different space mission that would be impractical, and even impossible, with current monolithic or multi-satellite missions. 

The synchronization of the swarm nodes is a very challenging task due to the dynamic characteristic of the transmission channel between the nodes, and the limited accuracy of the time and frequency references available at the small satellites \cite{juan_Radhakrishnan2016a,juan_swarm_EGlennon2018a}. In order to achieve a proper synchronization, the nodes must implement ISLs \cite{juan_8412572, juan_swarm_Edmonson2015} (ISL) to obtain an accurate reference from an external source. 

With the current state of the technology, establishing a direct data transfer between the flying units in a swarm is regarded as economically unfeasible due to the high payload costs. The implementation of the ISL requires additional transceivers which add to the weight and power consumption in each of the satellites \cite{juan_Engelen2014, juan_ISL2_6496947}.  For this reason, the development of space missions from swarms have not been contemplated in the past for data communications applications, but for science missions where the nature of the distributed space system is crucial or strictly required to fulfil the mission objectives. Some examples, from proposed concept to actual missions, can be found in the remote sensing literature \cite{juan_swarm_Merlano-Duncan2019}. One example of synchronization and formation flying is performed in the Tandem-X mission from the German Aerospace Center. This mission consists of two SAR satellites following an orbit in close formation, with a variable distance between them of few hundreds of meters \cite{juan_TANDEMX_Fiedler2008}. The spacecraft in here are not precisely nanosatellites, however this mission pioneered the formation flying concept. Another good example is the OLFAR project. The objective of this mission in to perform as a distributed radio telescope with satellites spreading in a cloud with a diameter of 100 km. The satellites will share their captured astronomical data of at least 6 Mbit/s/satellite \cite{juan_olfar_Bentum2009, juan_olfar_Rajan2011, juan_olfar_Willink-Castro2012, juan_olfar_82db56c8ed40472eb74af00b0512cb42}. Another example is the ongoing QB50 mission project,  QB50 is an international network of CubeSats for multi-point, in-situ measurements in the lower thermosphere and re-entry research \cite{juan_QB50, juan_qb50_e592ba6fe65043a2a4bd954d3a2b7b1e }.

Going in this direction, for swarms in general, the enabling factor is the capability of performing data exchange and distributed processing. In order to exchange information between satellites, RF and optical and ISLs have been proposed. Additionally, some LEO and MEO systems use RF links to improve availability and ensure a good quality of service \cite{juan_iSL1}. 

As an example, the satellites in the Iridium constellation use radio systems around 22 GHz to route traffic via the intra-plane and inter-plane neighboring satellites \cite{juan_iridium_5340513}. However, this type of solution, is not feasible, for the nanosatellites used in swarm missions, for the reasons mentioned before. 
The implementation of such ISLs is still an open research topic since his implementation represent an increase in system complexity and power consumption to the total system.

\subsection{ Hierarchical Aerial Networks}

Hierarchical area networks with multiple types of flying layers are promising to provide extended coverage and improve secured communications to some specific areas and events in the new space era. In this architect, multiple types of flying layers will cooperate to improve the space-to-ground link reliability and capacity \cite{Ahamadi17}. The unmanned aerial vehicle (UAV) will serve the ground users at low and medium layer, while high-altitude-pseudo satellites (HAPS) will serve both UAVs and ground users from high altitude and act as relaying nodes from the satellites when necessary. However, due to the difference in the height and velocity, link connections between the HAPS and UAVs are disconnected frequently. Therefore, how to harmonize the flight of the UAVs and HAPS to maintain reliable connections is of great importance as the current routing protocol is not applicable to vertical space networks. One should note that the desired routing protocol for vertical area networks should take into account the heterogeneous connects between the links, e.g., free space optical among the HAPS, hybrid radio frequency/free space optical between the HAPS and UAVs. Another open problem is how to efficiently deploy the hierarchical area network \cite{haps-teleo}. A joint design of communications and HAPS/UAV flights is expected to achieve the global performance. This will include not only UAVs and HAPs placement design but also trajectory optimizations. 

\subsection{Internet of Space Things/ Planetary Communications}

\begin{figure}[!t]
\centering
\includegraphics[width=85mm,keepaspectratio]{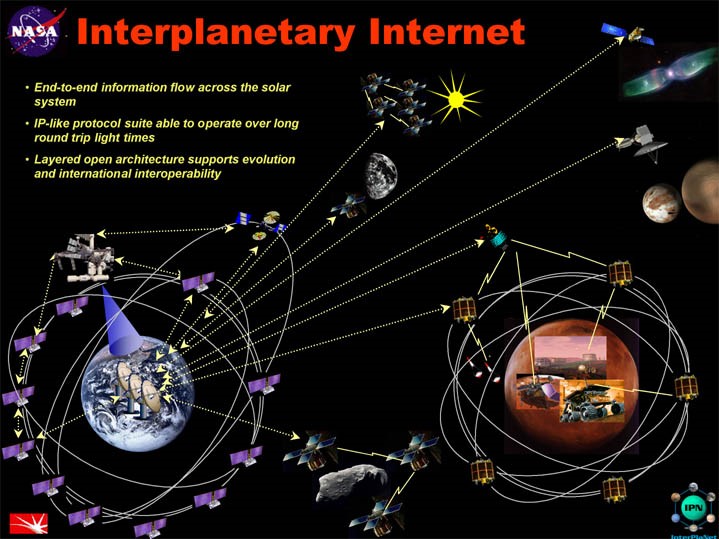}
\caption{Interplanetary Internet Network Concept. Credits: NASA}
\label{SWW}
\end{figure}

Space communication technology has steadily evolved from expensive, one-of-a-kind point-to-point architectures, to the re-use of technology on successive missions, to the development of standard protocols agreed upon by space agencies of many countries. This last phase has gone on since 1982 through the efforts of the Consultative Committee for Space Data Systems (CCSDS). With the current rate of astronomy and space exploration, it is clear that a Space Wide Web network  will spread to all over the solar system in the near future. 
As depicted in Fig. \ref{SWW} the vision of NASA \cite{NASAinternet} for the future of space communication is a huge network of communications nodes so that messages can hope between different intermediate nodes to reach their final destination. This architecture completely matches the architecture used on Earth for the World Wide Web and this is the reason why this new path for space communication is commonly referred to as Space Wide Web. 
While IP-like network layer protocols are feasible for short hops, such as ground station to orbiter, rover to lander, lander to orbiter, and so on, delay-tolerant networking is needed to get information from one region of the Solar System to another. 
Delay-tolerant networking (DTN) has the target to enable standardized communications over long distances and through time delays. At its core is something called the Bundle Protocol (BP), which is similar to the Internet Protocol, or IP, that serves as the heart of the Internet here on Earth. 
Several research groups are currently working on the development of such a new network layer protocol for the Space Internet. 

\subsection{Onboard regeneration / Flying Base Stations}
Several developments have been considered towards catering to the emerging challenge of handling different types of mobile traffic originating from multitude of devices supporting various use cases. To realize the flexibility and scalability in this context, micro-cell and small-cells with certain operational autonomy and ease of deployment have been considered. These deployments involve static base-stations. A next step in achieving flexibility in this direction is the use of mobile infrastructure to provide necessary services; in this context, the Flying base-stations have been considered \cite{David_FBS}. Flying base stations  are mounted on general-purpose unmanned aerial vehicles (UAVs) and these UAVs are further integrated into wireless network. It has been shown in \cite{David_FBS}, that integration of such systems into mobile networks can be an efficient alternative to ultra-dense small cell deployment, especially in scenarios with users moving in crowds.

With the envisaged deployment of low latency LEO satellites supporting terrestrial communication waveforms, an additional design flexibility can be incorporated in the system by developing base-station capabilities into the satellites. The extensive use of on-board processing in the emerging satellite systems paves way for the realization of a flying base-station on-board a satellite. On-board regeneration is essential for implementing this capability; the processing is not only restricted to PHY, but MAC and NET layer functionalities need to be added as well. Key aspects include serving a terminal using multiple satellites (CoMP scenario) and appropriate routing of the packets over ISLs. 

\subsection{Aggressive frequency reuse and dynamic spectrum management for both GSO and NGSO}
The term aggressive frequency reuse in multibeam satellite systems refers to the use of very low reuse factors in assigning user link bandwidth across multiple beams of a satellite. The reuse factor of one implies the maximum possible frequency reuse with all available bandwidth being allocated to all the beams. However, this results in a very high level of co-channel interference, leading to the need of suitable co-channel interference mitigation strategies. One promising approach in this regard is to utilize advanced signal processing techniques such as precoding by exploiting the spatial degrees of freedom provided obtained from the multibeam antenna. 

As compared to the Geostationary (GSO) satellites, NGSO satellites enjoy the benefits of less free space attenuation, small propagation delay and the reduced in-orbit injection cost per satellite \cite{Vatalaroanalysis}. Due to these advantages, the trend of deploying NGSO satellites is increasing over the recent years but the available usable radio spectrum is limited and is costly for the satellite operators. This has led to the need of spectrum coexistence of LEO/MEO satellites with the already existing GSO satellites and/or the spectral coexistence between different NGSO satellites \cite{SharmaInline2014,Pourmoghadas2017}. The interference analysis between GSO and NGSO systems operating  over the same set of radio frequencies becomes challenging as the relative position of the co-channel spots changes over time in NGSO systems \cite{Vatalaroanalysis,Park2010interference}. 

Regarding the frequency allocation in the Ka-band (uplink: 27.5 – 30 GHz, downlink: 17.7 – 20.2 GHz), the frequency bands 29.5 – 30 GHz and 19.7 – 20.2 GHz have been exclusively assigned to the Fixed Satellite Services (FSS) satellites. On the other hand, 17.7-19.7 GHz and 27.5-29.5 GHz bands are allocated to the terrestrial Fixed Service (FS) links on the primary basis \cite{SharmaInline2014}. These bands can also be utilized by the FSS satellites by providing sufficient protection to the existing FS links.  Furthermore, in the following sub-bands of the Ka-band, GSO and NGSO satellites have equal right: (i) 28.6 – 29.1 GHz (uplink), 18.8 – 19.3 GHz (downlink), and (ii) 29.1 – 29.5 GHz (uplink), 19.3 – 19.7 GHz (downlink). In these bands, ITU RR No. 9.11 A specifies that GSO and NGSO satellites must coordinate with the previously filed GSO and NGSO networks, and also with the existing other primary services in these bands. According to ITU-R footnote 5.523A, in the bands 17.8–19.3/28.6–29.1 GHz, such coordination should be based on the date of filing.  In the rest of the frequency bands, the limits on the Effective Power Flux Density (EPFD) mentioned in the RR Article 22 must be respected while coordinating with the already existing satellite systems. The EPFD specifies the maximum permissible interference that the NGSO FSS systems can cause to the GSO FSS systems and there arises no need of coordination with the GSO networks if these limits are respected but the coordination with other NGSO needs to be considered.  

As highlighted earlier, in-line interference, may be challenging for the coexistence of GSO and NGSO satellites, mainly in the equatorial region. In such a scenario, an earth station that is in-line with GEO and NGEO satellites may create and receive interference through its main beam. Furthermore, in the uplink, besides the inline interference from the main lobe of the NGSO terminal, the aggregate interference caused by the side-lobe gains of the beampatterns of many NGSO terminals in the ground may cause harmful interference to the GSO satellite. Also, if there exist multiple NGSO satellites in different  equatorial orbits over the same spectrum, the GSO terminals located at the equatorial region in the earth may receive aggregate interference from the side-lobes of multiple NGSO satellite  beampatterns \cite{SharmaInline2014}. 

Although the inline interference event can be predetermined and avoided by using proper planning considering the constellation geometries and utilizing the aforementioned static methods suggested by ITU-R recommendation S.1431 \cite{ITURS1431} and ITU-R S.1325 \cite{ITURS1325}, the performance of the primary system (GSO or NGSO depending on the coexistence scenario) may be impacted due to limited dynamicity of these methods. Also, the QoS of the secondary NGSO system may not be guaranteed while utilizing these static approaches.  In this regard, there arises the need to investigate more dynamic/flexible approaches for the real-time mitigation of inline interference events, which may occur while operating GSO-NGSO or NGSO-NGSO satellites over the same frequency band. 

One of the promising flexible approaches for interference mitigation could be to employ beamhopping principle at the secondary satellite so that the interference to the primary GSO or NGSO satellite can be avoided by adapting the beamhopping patterns in the real-time by utilizing the principle of cognitive beamhopping framework proposed in \cite{Sharmacogbeamhop}. Another promising solution could be to employ adaptive power control mechanisms \cite{Pourmoghadas2017} at the NGSO terminal to mitigate harmful interference towards the GSO satellite in the uplink coexistence scenario, and at the NGSO satellite to mitigate harmful interference towards the GSO terminal in the downlink coexistence scenario. Furthermore, another dynamic approach is to incorporate sensing mechanisms with the help of intelligent sensors at the NGSO terminals in a way that the inline interference can be detected during the reception mode. Moreover, terminal-side beamforming \cite{Terminalside2016} can be employed at the secondary NGSO terminal to mitigate harmful interference towards/from the primary GSO or NGSO satellite.  

\subsection{Satellite Network Automation}
The upcoming integrated 5G-satellite networks will largely increase in size and complexity due to the wide adoption of heterogeneous mobile devices and wireless access, which poses increased degrees of freedom in the network management process.
In many use cases, optimal solutions for terrestrial-satellite network management can be difficult to model due to the complex environment and the presence of too many uncertainties. 
Developing fast and high-quality heuristics or closed-form analytical models are not always a viable option for such use cases. 
As a result, network performance could be degraded, and the capital expenditure (CAPEX) and operational expenditure (OPEX) overheads can increase. 
With the growing complexity and reliability requirements, the conventional test-and-verification methods for network management will be challenging. This is because network operators are not comfortable deploying live traffic on untested/unoptimized configurations. The concepts of self-optimization and self-organization network (synonymous with network automation) are highly suited for such complex problems.

A promising architecture and implementation comes from SDN where networks can be dynamically programmed through centralized control points and from NFV enabling the cost-efficient deployment and runtime of network functions as software only. 
Based on NFV and SDN, network slicing (NS) is a service-oriented construct providing ``Network as a Service'' to concurrent applications. The slices will deliver different SLAs based on a unified pool of resources. 
The envisioned satellite-terrestrial network will be capable to support end-to-end services (and their management) across heterogeneous environments by means of a single (converged) common network. 
Through this paradigm, the specific services can be highly customized, enabling the seamless integration of heterogeneous networks in a 5G-satellite ecosystem.
Unlike the conventional one-type-fits-all network, the network slicing presents not just a cutting-edge technique, but opens new horizons for efficient and intelligent resource configuration for integrated terrestrial-satellite systems.

In this context,  the combination of terrestrial and non-terrestrial links, e.g., satellite, in transport networks has introduced new dimensions of network heterogeneity and dynamicity. Several open issues have to be addressed.
Firstly, 
one of the main challenges is to devise network-slicing algorithms, e.g., slicing configuration, virtual resource isolation, that can efficiently and autonomously configure the large number of parameters present in a virtualized dynamic graph representing an integrated satellite-terrestrial transport network.
Secondly,  most of the works on virtual network embedding (VNE) are based on a static design, i.e., based on a snapshot of a deterministic network graph. However, a realistic integrated NGSO satellite-terrestrial network is highly dynamic, resulting in fast variations of the virtual network topology over time.  Dealing with the graph dynamics in the context of online network-slice management is an essential challenge.
Thirdly,  the de facto standard protocol between the data and control planes, i.e., OpenFlow, in SDN/NFV networks has to be extended and compatible to satellite-terrestrial networks by considering satellite characteristics, e.g., LEO and MEO satellites' motion, available on-board energy, storage capacity, and computational power.

\subsection{Advanced Satellite Resource Orchestration}

As already mentioned, one of the concepts that is revolutioning the infrastructure of current communication systems is the so-called SDR technology. In short, SDR refers to a radio communication system where the major part of its functionality is implemented by means of software. The advances in this software disruptive paradigm is currently reinventing future network architectures, accelerating service deployment, and facilitating infrastructure management. Satellite communications are not an exception.

The main advantage of SDR is the capacity of adaptation which has been identified as a crucial characteristic of the future broadband satellite systems. By replacing as much hardware with software, the satellite payload becomes much more flexible and allows to deliver cost-competitive connectivity in response to evolving consumer demand and price expectations. Software defined payloads are less dependent on hardware and becomes more flexible and automatically reactive, able to face the dynamicity envisaged in the forthcoming wireless traffic.
The ability to reprogram beam pattern, frequency and power allocation dynamically in at anytime during the satellite mission, makes SDR technology very attractive in the forthcoming day where the data markets are more uncertain. The aforementioned capabilities open a door to advance resource management strategies for satellite communications, but at the same time bring new research challenges. In particular, the new on-board processing capabilities combined with the emerging role of active antenna systems, requires advanced resource management techniques capable of maximizing the satellite resource utilization while maintaining QoS guarantees, and dynamically matching the distribution of the satellite capacity on ground to the geographic distribution of the traffic demand and following its variations in time.

SDR-based satellite systems bring important improvements from a network management point of view, by allowing a better orchestration of the satellite resources. Unavoidably, “softwarization” will expand to the whole satellite ecosystem, replacing the custom hardware solutions, resulting in a more flexible and dynamic system with overall better performance and efficiency.

\subsection{Quantum Key Distribution through Optical Satcom}
The RSA protocol has been the cornerstone of cryptographic systems due to computational power needed to break it. However, with the advent of significantly large increase in computing power, alternative options whose performance is not vulnerable to computing power have been considered. In this context, Quantum key distribution (QKD), first proposed in \cite{QKD1}, involves establishing a private encryption key between two parties. QKD is inherently an optical technology, and has the ability to deliver encryption keys between any two points that share an optical link automatically. However, use of QKD over the mature optical fibre networks for long-range, long-scale applications is limited by the transmission losses that increase exponentially with distance.  In this context, QKD over satellite is being increasingly considered  with a project to develop such a space-based waveform already underway \cite{QKD3}.

Key to the success of QKD over satellite is the ability to set-up stable optical links by overcoming the various impediments in transmission. The links should ensure certain minimum quantum bit error rate (QBER), which is the QKD counterpart of signal-to-noise ratio, is met. This requires appropriate selection of optical frequencies, components and mechanisms for pointing, acquisition and tracking. Also of significant interest is the transmit and receive processing to ensure high fidelity link while satisfying constraints on size (e.g., on-board receiving lens cannot be large), power (e.g., constraints imposed  not to harm existing links/ equipment etc) and possibly computational power. Thus, in addition to its consideration for solving spectrum crunch, optical satellite communications will enable the QKD in the coming years; this motivates further investigations into optical satellite communications focussing on QKD scenarios in future. 

\subsection{Machine Learning Applications}
Machine Learning (ML) techniques in the literature can be broadly categorized into supervised, unsupervised and Reinforcement Learning (RL) \cite{SharmamassiveMTC}. Out of these, supervised learning requires the labelled training data-set while the unsupervised learning does not require the labelled data-sets. In contrast to these approaches which require training data-sets, the RL does not need a training data-set and enables a learning agent to learn from the prior experience. 

In the context of satellite systems, the application of ML has been already being explored in several scenarios including opportunistic weather monitoring, earth observation applications, satellite operations and sensor fusion for navigation. Furthermore, with the growing trend of investigating the applicability of ML in wireless communications, investigating its applications in the satellite communications has recently received increasing attention from the academia as well as SatCom industries/agencies. The ML/AI techniques can find potential applications in addressing various issues in satellite communications including interference mitigation to enable the coexistence of satellite systems with terrestrial systems, optimization of radio resources (spectrum, power),optimization of SatCom network operation, and management of large satellite constellations. 

In the above context, some promising use-cases to investigate the applications of ML techniques include: (i) adaptive allocation of carrier/power for the hybrid satellite-terrestrial scenarios, (ii) adaptive beamforming to enhance the performance of multibeam satellites with non-uniform demand, (iii) scheduling and precoding to mitigate interference in multibeam satellites, (iv) beamhopping and resource scheduling in multi-beam satellite systems with heterogeneous traffic demand per beam, and (v) detection of spectrum events in spectrum monitoring applications \cite{Raj2019unsupervised}.

\section{Conclusion} \label{sec10}

Satellite communications have recently entered in a crucial phase of their evolution, mainly motivated by the explosive growth of various Interned-based applications and services, which have triggered an ever increasing demand for broadband high-speed, heterogeneous, ultra-reliable and low latency communications. Due to their unique features and technical advances in the field, satellites can be a cornerstone in satisfying this demand, either as a stand-alone solution, or as an integrated satellite-terrestrial network. 

To ths end, this paper has captured the latest technical advances in scientific, industrial and standardisation analyses in the domain of satellite communications. In particular, the most important applications and use cases under the current focus of SatCom research have been highlighted. Moreover, an in-depth literature review has been provided covering the latest SatCom contributions in terms of system aspects, air interface, medium access control techniques and networking. The communication testbeds which have been developed in order to practically demonstrate some of the advanced SatCom concepts are shown. Finally, some important future challenges and their respective open research topics have been described.


%



\ifCLASSOPTIONcaptionsoff
  \newpage
\fi



%

\bibliographystyle{IEEEtran}
\bibliography{IEEEabrv,references}

\begin{thebibliography}{100}
\providecommand{\url}[1]{#1}
\csname url@samestyle\endcsname
\providecommand{\newblock}{\relax}
\providecommand{\bibinfo}[2]{#2}
\providecommand{\BIBentrySTDinterwordspacing}{\spaceskip=0pt\relax}
\providecommand{\BIBentryALTinterwordstretchfactor}{4}
\providecommand{\BIBentryALTinterwordspacing}{\spaceskip=\fontdimen2\font plus
\BIBentryALTinterwordstretchfactor\fontdimen3\font minus
  \fontdimen4\font\relax}
\providecommand{\BIBforeignlanguage}[2]{{%
\expandafter\ifx\csname l@#1\endcsname\relax
\typeout{** WARNING: IEEEtran.bst: No hyphenation pattern has been}%
\typeout{** loaded for the language `#1'. Using the pattern for}%
\typeout{** the default language instead.}%
\else
\language=\csname l@#1\endcsname
\fi
#2}}
\providecommand{\BIBdecl}{\relax}
\BIBdecl

\bibitem{teledesic2}
M.~Sturza. (1995) The teledesic satellite system: Overview and design trades.

\bibitem{starlink}
\BIBentryALTinterwordspacing
 [Online]. Available: \url{https://en.wikipedia.org/wiki/SpaceX_Starlink}
\BIBentrySTDinterwordspacing

\bibitem{sesref}
\BIBentryALTinterwordspacing
{SES and Thales Reach Record Speed and Enhanced Coverage via Integrated GEO/MEO
  Network}. [Online]. Available: \url{https://www.ses.com/press-release/}
\BIBentrySTDinterwordspacing

\bibitem{edrs}
\BIBentryALTinterwordspacing
{European Data Relay Satellite}. [Online]. Available:
  \url{https://artes.esa.int/edrs-global}
\BIBentrySTDinterwordspacing

\bibitem{audacy}
\BIBentryALTinterwordspacing
 [Online]. Available: \url{https://audacy.space/}
\BIBentrySTDinterwordspacing

\bibitem{5gppref1}
\BIBentryALTinterwordspacing
5GPPP. (2016) {5G} empowering vertical industries. [Online]. Available:
  \url{https://5g-ppp.eu/roadmaps/}
\BIBentrySTDinterwordspacing

\bibitem{5gppref2}
\BIBentryALTinterwordspacing
------. (2017) {5G} innovations for new business opportunities. [Online].
  Available: \url{https://5g-ppp.eu/roadmaps/}
\BIBentrySTDinterwordspacing

\bibitem{ITUref}
\BIBentryALTinterwordspacing
ITU-R. {IMT vision—framework and overall objectives of the future deployment
  of IMT for 2020 and beyond}. [Online]. Available:
  \url{https://www.itu.int/dms\_pubrec/itu-r/rec/m/R-REC-M.2083-0-201509-I!!PDF-E.pdf.}
\BIBentrySTDinterwordspacing

\bibitem{3gppref1}
3GPP, ``{Technical Specification Group Radio Access Network; Study on New Radio
  (NR) to support non terrestrial networks; (Release 15)},'' {3rd Generation
  Partnership Project (3GPP)}, Technical Report (TR) 38.811, 09-2019, version
  15.2.0.

\bibitem{3gppref2}
------, ``{Technical Specification Group Radio Access Network; Solutions for NR
  to support non-terrestrial networks; (Release 16)},'' {3rd Generation
  Partnership Project (3GPP)}, Technical Report (TR) 38.821, 09-2019, version
  0.9.0.

\bibitem{3gppnews}
\BIBentryALTinterwordspacing
------. {5G in Release 17 – strong radio evolution}. [Online]. Available:
  \url{https://www.3gpp.org/news-events/2098-5g-in-release-17-\%E2\%80\%93-strong-radio-evolution}
\BIBentrySTDinterwordspacing

\bibitem{3gppref3}
------, ``{Technical Specification Group Services and System Aspects; Study on
  using Satellite Access in 5G; (Release 16)},'' {3rd Generation Partnership
  Project (3GPP)}, Technical Report (TR) 38.822, 06-2018, version 16.0.0.

\bibitem{usecases1}
\BIBentryALTinterwordspacing
K.~Liolis, A.~Geurtz, R.~Sperber, D.~Schulz, S.~Watts, G.~Poziopoulou,
  B.~Evans, N.~Wang, O.~Vidal, B.~Tiomela~Jou, M.~Fitch, S.~Diaz~Sendra,
  P.~Sayyad~Khodashenas, and N.~Chuberre, ``{Use cases and scenarios of 5G
  integrated satellite-terrestrial networks for enhanced mobile broadband: The
  SaT5G approach},'' \emph{International Journal of Satellite Communications
  and Networking}, vol.~37, no.~2, pp. 91--112, 2019. [Online]. Available:
  \url{https://onlinelibrary.wiley.com/doi/abs/10.1002/sat.1245}
\BIBentrySTDinterwordspacing

\bibitem{VLEO-comms}
J.~V. Llop, P.~K. Roberts, Z.~Hao, L.~Tomas, and V.~Beauplet, ``Very low earth
  orbit mission concepts for earth observation: Benefits and challenges.'' in
  \emph{Reinventing Space Conference}, 2014.

\bibitem{aerial-comms}
\BIBentryALTinterwordspacing
L.~Reynaud and T.~Rasheed, ``Deployable aerial communication networks:
  Challenges for futuristic applications,'' in \emph{Proceedings of the 9th ACM
  Symposium on Performance Evaluation of Wireless Ad Hoc, Sensor, and
  Ubiquitous Networks}, ser. PE-WASUN '12.\hskip 1em plus 0.5em minus
  0.4em\relax New York, NY, USA: ACM, 2012, pp. 9--16. [Online]. Available:
  \url{http://doi.acm.org/10.1145/2387027.2387030}
\BIBentrySTDinterwordspacing

\bibitem{LEO-MSS-HAPS}
Y.~{Li}, H.~{Wei}, L.~{Li}, Y.~{Han}, J.~{Zhou}, and W.~{Zhou}, ``An extensible
  multi-layer architecture model based on leo-mss and performance analysis,''
  in \emph{2019 IEEE 90th Vehicular Technology Conference (VTC2019-Fall)}, Sep.
  2019, pp. 1--6.

\bibitem{VLEO_attitude}
\BIBentryALTinterwordspacing
J.~Virgili-Llop, H.~C. Polat, and M.~Romano, ``Attitude stabilization of
  spacecraft in very low earth orbit by center-of-mass shifting,''
  \emph{Frontiers in Robotics and AI}, vol.~6, p.~7, 2019. [Online]. Available:
  \url{https://www.frontiersin.org/article/10.3389/frobt.2019.00007}
\BIBentrySTDinterwordspacing

\bibitem{VLEO-comms2}
S.~M. {Dakka}, ``Vleo satellites — a new earth observation space systems
  commercial and business model,'' in \emph{36th International Communications
  Satellite Systems Conference (ICSSC 2018)}, Oct 2018, pp. 1--11.

\bibitem{happiest}
\BIBentryALTinterwordspacing
``Happiest: High altitude pseudo-satellites: Proposal of initiatives to enhance
  satellite telecommunications,'' Tech. Rep., June 2018. [Online]. Available:
  \url{http://uleia.unileon.es/index.php/en/research-projects/aerospace-system-engineering1/unconventional-unmanned-aircrafts/16-research-projects/aerospace-engineering/high-altitude-pseudo-satellites-haps/52-stratospheric-platforms-happiest}
\BIBentrySTDinterwordspacing

\bibitem{haps-teleo}
\BIBentryALTinterwordspacing
``Haps-teleo: High-altitude pseudo-satellites for telecommunication and
  complementary space applications,'' Tech. Rep., July 2018. [Online].
  Available:
  \url{https://nebula.esa.int/sites/default/files/neb\_study/2470/C4000118800ExS.pdf}
\BIBentrySTDinterwordspacing

\bibitem{Loon1}
\BIBentryALTinterwordspacing
M.~Koziol, ``Loon’s balloons deliver emergency internet service to peru
  following 8.0 earthquake,'' \emph{IEEE Spectrum}. [Online]. Available:
  \url{https://spectrum.ieee.org/tech-talk/telecom/wireless/loons-balloons-deliver-emergency-service-to-peru-following-80-earthquake}
\BIBentrySTDinterwordspacing

\bibitem{Loon2}
{Iskandar}, M.~A. {Wibisono}, S.~{Priatna}, T.~{Juhana}, {Hendrawan}, and
  N.~{Rachmana}, ``On the design and development of flying bts system using
  balloon for remote area communication,'' in \emph{2017 11th International
  Conference on Telecommunication Systems Services and Applications (TSSA)},
  Oct 2017, pp. 1--5.

\bibitem{LAP-example}
A.~Qiantori, A.~B. Sutiono, H.~Hariyanto, H.~Suwa, and T.~Ohta, ``An emergency
  medical communications system by low altitude platform at the early stages of
  a natural disaster in indonesia,'' \emph{Journal of Medical Systems},
  vol.~36, pp. 41--52, 2010.

\bibitem{UAV-comms}
B.~{Li}, Z.~{Fei}, and Y.~{Zhang}, ``Uav communications for 5g and beyond:
  Recent advances and future trends,'' \emph{IEEE Internet of Things Journal},
  vol.~6, no.~2, pp. 2241--2263, April 2019.

\bibitem{UAV-comms2}
\BIBentryALTinterwordspacing
M.~Marchese, A.~Moheddine, and F.~Patrone, ``Iot and uav integration in 5g
  hybrid terrestrial-satellite networks,'' \emph{Sensors}, vol.~19, no.~17,
  2019. [Online]. Available: \url{https://www.mdpi.com/1424-8220/19/17/3704}
\BIBentrySTDinterwordspacing

\bibitem{5G-Sky}
\BIBentryALTinterwordspacing
``5g-sky: Interconnecting the sky in 5g and beyond - a joint communication and
  control approach,'' Tech. Rep., 2019. [Online]. Available:
  \url{https://wwwfr.uni.lu/snt/research/sigcom/projects/5g\_sky\_interconnecting\_the\_sky\_in\_5g\_and\_beyond\_a\_joint\_communication\_and\_control\_approach}
\BIBentrySTDinterwordspacing

\bibitem{HAPs_Further1}
X.~{Cao}, P.~{Yang}, M.~{Alzenad}, X.~{Xi}, D.~{Wu}, and H.~{Yanikomeroglu},
  ``Airborne communication networks: A survey,'' \emph{IEEE Journal on Selected
  Areas in Communications}, vol.~36, no.~9, pp. 1907--1926, Sep. 2018.

\bibitem{HAPs_Further2}
S.~{Karapantazis} and F.~{Pavlidou}, ``Broadband communications via
  high-altitude platforms: A survey,'' \emph{IEEE Communications Surveys
  Tutorials}, vol.~7, no.~1, pp. 2--31, First 2005.

\bibitem{InternationalAirTransportAssociation2018}
\BIBentryALTinterwordspacing
{International Air Transport Association}, ``{IATA Forecast Predicts 8.2
  billion Air Travelers in 2037},'' 2018. [Online]. Available:
  \url{https://www.iata.org/pressroom/pr/Pages/2018-10-24-02.aspx}
\BIBentrySTDinterwordspacing

\bibitem{Strohmeier2014}
M.~Strohmeier, M.~Sch{\"{a}}fer, V.~Lenders, and I.~Martinovic, ``{Realities
  and challenges of nextgen air traffic management: The case of ADS-B},''
  \emph{IEEE Communications Magazine}, vol.~52, no.~5, pp. 111--118, 2014.

\bibitem{Francis2011}
R.~Francis, R.~Vincent, J.~M. No{\"{e}}l, P.~Tremblay, D.~Desjardins,
  A.~Cushley, and M.~Wallace, ``{The flying laboratory for the observation of
  ADS-B signals},'' \emph{International Journal of Navigation and Observation},
  no. August, 2011.

\bibitem{Knudsen2014}
B.~G. Knudsen, M.~Jensen, A.~Birklykke, P.~Koch, J.~Christiansen, K.~Laursen,
  L.~Alminde, and Y.~{Le Moullec}, ``{ADS-B in space: Decoder implementation
  and first results from the GATOSS mission},'' \emph{Proceedings of the
  Biennial Baltic Electronics Conference, BEC}, vol. 2015-Novem, pp. 57--60,
  2014.

\bibitem{juan_RTCA-DO-144A}
\BIBentryALTinterwordspacing
{RTCA Special Committee 209}, ``{Minimum Operational Characteristics for Air
  Traffic Control Radar Beacon System (ATCRBS) Airborne Equipment},'' 2008.
  [Online]. Available: \url{https://standards.globalspec.com/std/1390603/RTCA
  DO-181}
\BIBentrySTDinterwordspacing

\bibitem{SpireWeb}
``Spire website,'' \url{https://www.spire.com/en}, accessed: 2019-11-22.

\bibitem{juan_adsb_aireon}
\BIBentryALTinterwordspacing
``{Aireon Space Based ADS-B.}'' [Online]. Available: \url{https://aireon.com/}
\BIBentrySTDinterwordspacing

\bibitem{juan_IMO2018}
\BIBentryALTinterwordspacing
IMO, ``{About IMO},'' 2018. [Online]. Available: \url{http://www.imo.org}
\BIBentrySTDinterwordspacing

\bibitem{juan_Yang2019}
\BIBentryALTinterwordspacing
D.~Yang, L.~Wu, S.~Wang, H.~Jia, and K.~X. Li, ``{How big data enriches
  maritime research–a critical review of Automatic Identification System
  (AIS) data applications},'' \emph{Transport Reviews}, vol.~39, no.~6, pp.
  755--773, nov 2019. [Online]. Available:
  \url{https://www.tandfonline.com/doi/full/10.1080/01441647.2019.1649315}
\BIBentrySTDinterwordspacing

\bibitem{juan_Winther2014}
M.~Winther, J.~H. Christensen, M.~S. Plejdrup, E.~S. Ravn, {\'{O}}.~F.
  Eriksson, and H.~O. Kristensen, ``{Emission inventories for ships in the
  arctic based on satellite sampled AIS data},'' \emph{Atmospheric
  Environment}, vol.~91, pp. 1--14, 2014.

\bibitem{juan_FernandezArguedas2018}
V.~{Fernandez Arguedas}, G.~Pallotta, and M.~Vespe, ``{Maritime Traffic
  Networks: From Historical Positioning Data to Unsupervised Maritime Traffic
  Monitoring},'' \emph{IEEE Transactions on Intelligent Transportation
  Systems}, vol.~19, no.~3, pp. 722--732, mar 2018.

\bibitem{juan_Kaluza2010}
P.~Kaluza, A.~K{\"{o}}lzsch, M.~T. Gastner, and B.~Blasius, ``{The complex
  network of global cargo ship movements},'' pp. 1093--1103, jul 2010.

\bibitem{juan_7947788}
G.~{Sahay}, P.~{Meghana}, V.~V. {Sravani}, T.~P. {Venkatesh}, and V.~{Karna},
  ``Sdr based single channel s-ais receiver for satellites using system
  generator,'' in \emph{2016 IEEE International Conference on Advanced Networks
  and Telecommunications Systems (ANTS)}, Nov 2016, pp. 1--6.

\bibitem{juan_ais_Sahay2017}
G.~Sahay, P.~Meghana, V.~V. Sravani, T.~P. Venkatesh, and V.~Karna, ``{SDR
  based single channel S-AIS receiver for satellites using system generator},''
  in \emph{2016 IEEE International Conference on Advanced Networks and
  Telecommunications Systems, ANTS 2016}.\hskip 1em plus 0.5em minus
  0.4em\relax Institute of Electrical and Electronics Engineers Inc., jun 2017.

\bibitem{PlanetWeb}
``Planet website,'' \url{https://www.planet.com/}, accessed: 2019-11-22.

\bibitem{ESAmoonVILLAGE}
``Esa moon village project webpage,''
  \url{https://www.esa.int/About\_Us/Ministerial\_Council\_2016/Moon\_Village},
  accessed: 2019-11-22.

\bibitem{PlanRes}
``Consensys webpage,'' \url{https://consensys.space/}, accessed: 2019-11-22.

\bibitem{monitor}
M.~Baldi, G.~Cerri, F.~Chiaraluce, L.~Eusebi, and P.~Russo, ``Non-invasive uwb
  sensing of astronauts' breathing activity,'' \emph{Sensors (Basel,
  Switzerland)}, vol.~15, pp. 565--91, 12 2014.

\bibitem{Elbert_book}
B.~R. Elbert, ``The satellite communication applications handbook.''\hskip 1em
  plus 0.5em minus 0.4em\relax Artech House, Inc., 2003.

\bibitem{ground_segment_book}
B.~Elbert, \emph{The Satellite Communication Ground Segment and Earth Station
  Handbook}.\hskip 1em plus 0.5em minus 0.4em\relax Artech House Space
  Technology and Applications Library, 2000.

\bibitem{5g_ntn_arch}
A.~{Guidotti}, A.~{Vanelli-Coralli}, M.~{Conti}, S.~{Andrenacci},
  S.~{Chatzinotas}, N.~{Maturo}, B.~{Evans}, A.~{Awoseyila}, A.~{Ugolini},
  T.~{Foggi}, L.~{Gaudio}, N.~{Alagha}, and S.~{Cioni}, ``Architectures and key
  technical challenges for 5g systems incorporating satellites,'' \emph{IEEE
  Transactions on Vehicular Technology}, vol.~68, no.~3, pp. 2624--2639, March
  2019.

\bibitem{5g_ntn_ch}
O.~{Kodheli}, A.~{Guidotti}, and A.~{Vanelli-Coralli}, ``Integration of
  satellites in 5g through leo constellations,'' in \emph{GLOBECOM 2017 - 2017
  IEEE Global Communications Conference}, Dec 2017, pp. 1--6.

\bibitem{5g_ntn_ch2}
M.~Höyhtyä, M.~Corici, S.~Covaci, and M.~Guta, ``5g and beyond for new space:
  Vision and research challenges,'' 10 2019.

\bibitem{aws}
\BIBentryALTinterwordspacing
{AWS Ground Station: Easily control satellites and ingest data with fully
  managed Ground Station as a Service}. [Online]. Available:
  \url{https://aws.amazon.com/ground-station/}
\BIBentrySTDinterwordspacing

\bibitem{esoa}
{ESOA - EMEA Satellite Operator Association}, ``{Satellite Spectrum},''
  \url{https://www.esoa.net/spectrum/satellite-spectrum.asp}.

\bibitem{6689909}
Y.~{Li}, Z.~{Wang}, and W.~{Tan}, ``The frequency spectrum management for
  aerospace tt c system,'' in \emph{2013 5th IEEE International Symposium on
  Microwave, Antenna, Propagation and EMC Technologies for Wireless
  Communications}, Oct 2013, pp. 595--600.

\bibitem{8766193}
Y.~{Guan}, F.~{Geng}, and J.~H. {Saleh}, ``Review of high throughput
  satellites: Market disruptions, affordability-throughput map, and the cost
  per bit/second decision tree,'' \emph{IEEE Aerospace and Electronic Systems
  Magazine}, vol.~34, no.~5, pp. 64--80, May 2019.

\bibitem{Perez2019}
A.~I. {Perez-Neira}, M.~A. {Vazquez}, M.~R.~B. {Shankar}, S.~{Maleki}, and
  S.~{Chatzinotas}, ``Signal processing for high-throughput satellites:
  Challenges in new interference-limited scenarios,'' \emph{IEEE Signal
  Processing Magazine}, vol.~36, no.~4, pp. 112--131, July 2019.

\bibitem{gupta}
A.~{Gupta} and R.~K. {Jha}, ``A survey of 5g network: Architecture and emerging
  technologies,'' \emph{IEEE Access}, vol.~3, pp. 1206--1232, 2015.

\bibitem{5GPPP}
{5G Infrastructure Association (5G-PPP)}, ``{The need for 5G spectrum},''
  \url{https://5g-ppp.eu/wp-content/uploads/2017/03/2\_EC\_Stantchev\_5G-WS\_5GPPP\_June7.pdf}.

\bibitem{cbandA}
{C-Band Alliance}, ``{5G For Everyone},'' \url{https://c-bandalliance.com/}.

\bibitem{kyrgiazos}
A.~Kyrgiazos, B.~Evans, and P.~Thompson, ``{High Throughput Satellite Gateway
  Feeder Link (Q/V/W bands)},'' \emph{Ka and Broadband Communications,
  Navigation and Earth Observation Conference (KaConf)}, 2015.

\bibitem{6856179}
A.~{Kyrgiazos}, B.~G. {Evans}, and P.~{Thompson}, ``On the gateway diversity
  for high throughput broadband satellite systems,'' \emph{IEEE Transactions on
  Wireless Communications}, vol.~13, no.~10, pp. 5411--5426, Oct 2014.

\bibitem{6666239}
A.~{Gharanjik}, B.~S. M.~R. {Rao}, P.~{Arapoglou}, and B.~{Ottersten}, ``Large
  scale transmit diversity in q/v band feeder link with multiple gateways,'' in
  \emph{2013 IEEE 24th Annual International Symposium on Personal, Indoor, and
  Mobile Radio Communications (PIMRC)}, Sep. 2013, pp. 766--770.

\bibitem{braun}
C.~Braun, A.~Voicu, L.~Simic, and P.~Mahonen, \emph{{}IEEE International
  Symposium on Dynamic Spectrum Access Networks (DySPAN)}.

\bibitem{ITURS1325}
ITU-R, ``Simulation methodologies for determining statistics of short-term
  interference between co-frequency, codirectional non-geostationary-satellite
  orbit fixed-satellite service systems in circular orbits and other
  non-geostationary fixed-satellite service systems in circular orbits or
  geostationary-satellite orbit fixed-satellite service networks,'' Tech. Rep.
  Rec. ITU-R S.1325-3, 2003.

\bibitem{ITURS1431}
------, ``Methods to enhance sharing between non-{GSO} {FSS} systems in
  frequency bands between 10-30 {GHz},'' Tech. Rep. Rec. ITU-R S.1431, 2000.

\bibitem{lagunas}
E.~{Lagunas}, S.~K. {Sharma}, S.~{Maleki}, S.~{Chatzinotas}, and
  B.~{Ottersten}, ``Resource allocation for cognitive satellite communications
  with incumbent terrestrial networks,'' \emph{IEEE Transactions on Cognitive
  Communications and Networking}, vol.~1, no.~3, pp. 305--317, Sep. 2015.

\bibitem{chatzinotas}
\BIBentryALTinterwordspacing
S.~Chatzinotas, B.~Evans, A.~Guidotti, V.~Icolari, E.~Lagunas, S.~Maleki, S.~K.
  Sharma, D.~Tarchi, P.~Thompson, and A.~Vanelli-Coralli, ``Cognitive
  approaches to enhance spectrum availability for satellite systems,''
  \emph{International Journal of Satellite Communications and Networking},
  vol.~35, no.~5, pp. 407--442, 2017. [Online]. Available:
  \url{https://onlinelibrary.wiley.com/doi/abs/10.1002/sat.1197}
\BIBentrySTDinterwordspacing

\bibitem{sansa}
{H2020 SANSA project}, ``{Shared Access Terrestrial-Satellite Backhaul Network
  enabled by Smart Antennas},'' \url{http://sansa-h2020.eu/}.

\bibitem{viasat}
{ViaSatellite magazine}, ``{Cellular Backhaul: The Ever Growing Opportunity for
  Satellite},''
  \url{http://www.satellitetoday.com/publications/2015/02/12/cellular-backhaul-the-ever-growing-opportunity-for-satellite/}.

\bibitem{Sat5G}
{5G-PPP project Sat5G}, ``{Satellite and Terrestrial Network for 5G},''
  \url{https://5g-ppp.eu/sat5g/}.

\bibitem{lagunas2}
E.~{Lagunas}, L.~{Lei}, S.~{Chatzinotas}, and B.~{Ottersten}, ``Satellite links
  integrated in 5g sdn-enabled backhaul networks: An iterative joint power and
  flow assignment,'' in \emph{2019 27th European Signal Processing Conference
  (EUSIPCO)}, Sep. 2019, pp. 1--5.

\bibitem{artiga}
X.~{Artiga}, A.~{Perez-Neira}, J.~{Baranda}, E.~{Lagunas}, S.~{Chatzinotas},
  R.~{Zetik}, P.~{Gorski}, K.~{Ntougias}, D.~{Perez}, and G.~{Ziaragkas},
  ``Shared access satellite-terrestrial reconfigurable backhaul network enabled
  by smart antennas at mmwave band,'' \emph{IEEE Network}, vol.~32, no.~5, pp.
  46--53, Sep. 2018.

\bibitem{shaat}
M.~{Shaat}, E.~{Lagunas}, A.~I. {Perez-Neira}, and S.~{Chatzinotas},
  ``Integrated terrestrial-satellite wireless backhauling: Resource management
  and benefits for 5g,'' \emph{IEEE Vehicular Technology Magazine}, vol.~13,
  no.~3, pp. 39--47, Sep. 2018.

\bibitem{DVB}
``{DVB} website,'' \url{https://www.dvb.org/}.

\bibitem{CCSDS01}
``{CCSDS} website,'' \url{http://www.ccsds.org/}.

\bibitem{PDA1}
{Pantelis-Daniel Arapoglou}, K.~Liolis, M.~Bertinelli, A.~Panagopoulos,
  P.~Cottis, and R.~D. Gaudenzi, ``{MIMO} over satellite: A review,''
  \emph{IEEE Communications Surveys and Tutorials}, vol.~13, no.~1, pp.
  722--732, 2011.

\bibitem{Bundeswehr}
A.~C. Hofmann, K.-U. Storek, R.~T. Schwarz, and A.~Knopp, ``Spatial mimo over
  satellite. : A proof of concept.'' in \emph{Proc. 2016 IEEE International
  Conference on Communications ({ICC})}, 2016.

\bibitem{RD15}
I.-R. P.681-6. (2003) {Propagation Data Required for the Design of Earth Space
  Land Mobile Telecommunication Systems}.

\bibitem{RD17}
R.~P. Cerdeira, F.~P. Fontán, P.~Burzigotti, A.~B. Alamañac, and
  I.~Sanchez-Lago, ``Versatile two-state land mobile satellite channel model
  with first application to dvb-sh analysis,'' \emph{International Journal of
  Satellite Communications and Networking}, June 2010.

\bibitem{RD19}
I.-R. M.1225. {Guidelines for Evaluation of Radio Transmission Technologies for
  IMT-2000}.

\bibitem{gouss_1}
S.~K. {Rao}, ``Advanced antenna technologies for satellite communications
  payloads,'' \emph{IEEE Transactions on Antennas and Propagation}, vol.~63,
  no.~4, pp. 1205--1217, April 2015.

\bibitem{gouss_2}
B.~{Palacin}, N.~J.~G. {Fonseca}, M.~{Romier}, R.~{Contreres}, J.~{Angevain},
  G.~{Toso}, and C.~{Mangenot}, ``Multibeam antennas for very high throughput
  satellites in europe: Technologies and trends,'' in \emph{2017 11th European
  Conference on Antennas and Propagation (EUCAP)}, March 2017, pp. 2413--2417.

\bibitem{gouss_3}
H.~{Fenech}, S.~{Amos}, and T.~{Waterfield}, ``The role of array antennas in
  commercial telecommunication satellites,'' in \emph{2016 10th European
  Conference on Antennas and Propagation (EuCAP)}, April 2016, pp. 1--4.

\bibitem{gouss_4}
J.~{Hill}, Y.~{Demers}, A.~{Liang}, E.~{Amyotte}, K.~{Glatre}, and
  S.~{Richard}, ``Multibeam antenna architectures for flexible capacity
  allocation,'' in \emph{39th ESA Antenna Workshop on Innovative Antenna
  Systems and Technologies for Future Space Missions}, October 2018.

\bibitem{gouss_4a}
S.~{Egami} and M.~{Kawai}, ``An adaptive multiple beam system concept,''
  \emph{IEEE Journal on Selected Areas in Communications}, vol.~5, no.~4, pp.
  630--636, May 1987.

\bibitem{gouss_4b}
``Eutelsat 172b satellite: On the road to kourou, march 2017,''
  \url{https://news.eutelsat.com/pressreleases/eutelsat-172b-satellite-on-the-road-to-kourou-1857558},
  accessed: 2019-12.

\bibitem{gouss_4c}
\BIBentryALTinterwordspacing
T.~K. Henriksen and C.~Mangenot, ``Large deployable antennas,'' \emph{CEAS
  Space Journal}, vol.~5, no.~3, pp. 87--88, Dec 2013. [Online]. Available:
  \url{https://doi.org/10.1007/s12567-013-0055-4}
\BIBentrySTDinterwordspacing

\bibitem{gouss_5}
S.~{Amos}, G.~{Thomas}, S.~{McLaren}, and A.~{Warburton}, ``Answering the
  market pull with active antennas and the technology challenges,'' in
  \emph{39th ESA Antenna Workshop on Innovative Antenna Systems and
  Technologies for Future Space Missions}, October 2018.

\bibitem{gouss_6}
M.~{Cooley}, ``Phased array fed reflector (pafr) antenna architectures for
  space-based sensors,'' in \emph{2015 IEEE Aerospace Conference}, March 2015,
  pp. 1--11.

\bibitem{gouss_12}
W.~{Tang}, D.~{Bresciani}, H.~{Legay}, G.~{Goussetis}, and N.~J.~G. {Fonseca},
  ``Circularly polarised multiple beam antenna for satellite applications,'' in
  \emph{2017 11th European Conference on Antennas and Propagation (EUCAP)},
  March 2017, pp. 2628--2630.

\bibitem{gouss_6a}
E.~{Doumanis}, G.~{Goussetis}, W.~{Steffe}, D.~{Maiarelli}, and S.~A.
  {Kosmopoulos}, ``Helical resonator filters with improved power handling
  capabilities for space applications,'' \emph{IEEE Microwave and Wireless
  Components Letters}, vol.~20, no.~11, pp. 598--600, Nov 2010.

\bibitem{gouss_6ai}
J.~Moron, R.~Leblanc, P.~Frijlink, F.~Lecourt, M.~Sigler, G.~Goussetis,
  G.~Amendola, G.~Codispoti, G.~Valente, and G.~Parca,
  ``\BIBforeignlanguage{English}{A novel high-performance v-band gan mmic hpa
  for the qv-lift project},'' in \emph{\BIBforeignlanguage{English}{Ka and
  Broadband Communications Conference 2018}}, ser. Ka and Broadband
  Communications Conference.\hskip 1em plus 0.5em minus 0.4em\relax Ka and
  Broadband Communications, 2018.

\bibitem{gouss_6b}
C.~A. Balanis, \emph{Antenna Theory: Analysis and Design, 4th Edition}.\hskip
  1em plus 0.5em minus 0.4em\relax Wiley, 2016.

\bibitem{gouss_7}
G.~{Toso}, C.~{Mangenot}, and P.~{Angeletti}, ``Recent advances on space
  multibeam antennas based on a single aperture,'' in \emph{2013 7th European
  Conference on Antennas and Propagation (EuCAP)}, April 2013, pp. 454--458.

\bibitem{gouss_8}
G.~{Toso}, P.~{Angeletti}, and C.~{Mangenot}, ``Multibeam antennas based on
  phased arrays: An overview on recent esa developments,'' in \emph{The 8th
  European Conference on Antennas and Propagation (EuCAP 2014)}, April 2014,
  pp. 178--181.

\bibitem{gouss_9}
W.~{Rotman} and R.~{Turner}, ``Wide-angle microwave lens for line source
  applications,'' \emph{IEEE Transactions on Antennas and Propagation},
  vol.~11, no.~6, pp. 623--632, November 1963.

\bibitem{gouss_10}
P.~Jankovic, P.~Tapia, J.~Galdeano, G.~Toso, and P.~Angeletti, ``The
  multi-functional payload based on rotman-lens beamforming network,'' 03 2019.

\bibitem{gouss_11}
F.~{Doucet}, N.~J.~G. {Fonseca}, E.~{Girard}, H.~{Legay}, and R.~{Sauleau},
  ``Analytical model and study of continuous parallel plate waveguide lens-like
  multiple-beam antennas,'' \emph{IEEE Transactions on Antennas and
  Propagation}, vol.~66, no.~9, pp. 4426--4436, Sep. 2018.

\bibitem{gouss_13}
F.~{Vidal}, H.~{Legay}, G.~{Goussetis}, M.~{Garcia Vigueras}, S.~{Tubau}, and
  J.~D. {Gayrard}, ``Development of a modular simulation tool for flexible
  payload design and benchmark of antenna architectures applied to a ka-band
  broadband leo mission,'' in \emph{4th ESA Workshop on Advanced Flexible
  Telecom Payloads}, March 2019.

\bibitem{gouss_14}
S.~{Tubau}, F.~{Vidal}, H.~{Legay}, B.~{Palacin}, and G.~{Goussetis}, ``Novel
  multiple beam antenna farms for megaconstellations,'' in \emph{40th ESA
  Antenna Workshop, ESA-ESTEC, Nordwijk, The Netherlands}, October 2019.

\bibitem{gouss_15}
``O3b corporate presentation – itu,''
  \url{https://www.itu.int/dms\_pub/itu-r/md/12/iturka.band/c/R12-ITURKA.BAND-C-0010!!PDF-E.pdf},
  accessed: 2019-12.

\bibitem{gouss_16}
S.~{Gao}, Y.~{Rahmat-Samii}, R.~E. {Hodges}, and X.~{Yang}, ``Advanced antennas
  for small satellites,'' \emph{Proceedings of the IEEE}, vol. 106, no.~3, pp.
  391--403, March 2018.

\bibitem{gouss_17}
\emph{Flat Panel Satellite Antennas, 4th Edition}.\hskip 1em plus 0.5em minus
  0.4em\relax NSR, 2019.

\bibitem{gouss_17c}
A.~{Rao}, A.~{Chatterjee}, J.~{Payne}, and J.~Trujillo, ``Multi-beam/multi-band
  aeronautical antenna: Opportunities and challenges,'' in \emph{29th ESA
  Antenna Workshop}, October 2018.

\bibitem{gouss_18}
\url{https://www.thinkom.com/}, accessed: 2019-12.

\bibitem{gouss_20}
E.~{Doumanis}, G.~{Goussetis}, R.~{Dickie}, R.~{Cahill}, P.~{Baine}, M.~{Bain},
  V.~{Fusco}, J.~A. {Encinar}, and G.~{Toso}, ``Electronically reconfigurable
  liquid crystal based mm-wave polarization converter,'' \emph{IEEE
  Transactions on Antennas and Propagation}, vol.~62, no.~4, pp. 2302--2307,
  April 2014.

\bibitem{SPM}
A.~I. {Pérez-Neira}, M.~. {Vázquez}, M.~R. {Bhavani~ Shankar}, S.~{Maleki},
  and S.~{Chatzinotas}, ``Signal processing for high throughput satellites:
  Challenges in new interference-limited scenarios,'' \emph{IEEE Signal
  Processing Magazine}, vol.~36, no.~4, pp. 112--131, 2019.

\bibitem{ref_9}
T.~M. Braun, in \emph{Satellite Communications : Payload and System}.\hskip 1em
  plus 0.5em minus 0.4em\relax Wiley-IEEE Press, 2012.

\bibitem{ref_10}
P.~Angeletti, R.~{de~Gaudenzi}, and M.~Lisi, ``From bent pipes to software
  defined payloads: evolution and trends of satellite communications systems,''
  in \emph{Proc. 26th AIAA Int. Commun. Satellite Systems Conf.}, 2008.

\bibitem{ref_11}
P.~Angeletti, A.~B. {Alamanac}, F.~Coromina, and {et. al}, ``Satcoms 2020 r\&d
  challenges: Part ii: Mobile communications,'' in \emph{Proc. 27th AIAA Int.
  Commun. Satellite Systems Conf.}, 2009.

\bibitem{ref_13}
\BIBentryALTinterwordspacing
SES. (2019) {SES-12}. [Online]. Available:
  \url{https://www.ses.com/press-release/ses-12-goes-operational-serve-asia-pacific-and-middle-east}
\BIBentrySTDinterwordspacing

\bibitem{TAES_DTP}
V.~Sulli, D.~Giancristofaro, F.~Santucci, and M.~Faccio, ``Performance of
  satellite digital transparent processors through equivalent noise,''
  \emph{IEEE Transactions on Aerospace and Electronic Systems}, vol.~54, no.~6,
  pp. 2643 -- 2661, 2018.

\bibitem{Nicolo_DTP}
N.~Mazzali, M.~R.~B. Shankar, and B.~Ottersten, ``On-board signal predistortion
  for digital transparent satellites,'' in \emph{Proc. 16th IEEE International
  Workshop on Signal Processing Advances in Wireless Communications {(SPAWC)}},
  2015.

\bibitem{Christos_DTP}
C.~Politis, S.~Maleki, C.~Tsinos, S.~Chatzinotas, and B.~Ottersten, ``Weak
  interference detection with signal cancellation in satellite
  communications,'' in \emph{Proc. 42 IEEE International Conference on
  Acoustics, Speech and Signal Processing {(ICASSP)}}, 2017.

\bibitem{5674689}
S.~P. {Lee}, J.~H. {Jo}, {Moon Hee You}, and J.~S. {Choi}, ``Integration,
  testing and in orbit validation of ka-band communication payload of coms,''
  in \emph{2010 International Conference on Information and Communication
  Technology Convergence (ICTC)}, Nov 2010, pp. 309--313.

\bibitem{8761654}
N.~{Mazzali}, S.~{Andrenacci}, and S.~{Chatzinotas}, ``Non-intrusive flexible
  measurement for the amplitude response of wideband transponders,'' in
  \emph{ICC 2019 - 2019 IEEE International Conference on Communications (ICC)},
  May 2019, pp. 1--6.

\bibitem{mazalli}
M.~Mazalli, S.~Andrenacci, and S.~Chatzinotas, ``{Method and Device for
  Adaptive In-orbit Testing of a Satellite},'' \emph{{Luxembourg Patent Office,
  LU101119}}, Feb. 2019.

\bibitem{Mishra_2015}
G.~Mishra, A.~Mishra, and S.~Singh, ``{Simulation and Validation of Spread
  Spectrum Technique in In-Orbit Testing of the Satellite Transponder},''
  \emph{International Journal of Emerging Technology and Advanced Engineering},
  vol.~5, no.~4, Apr. 2015.

\bibitem{Roy1997}
\BIBentryALTinterwordspacing
R.~Roy and B.~Ottersten, ``Spatial division multiple access wireless
  communication systems,'' May 1996, {US Patent} 5,515,378. [Online].
  Available: \url{https://www.google.com/patents/US5515378}
\BIBentrySTDinterwordspacing

\bibitem{Liu2011}
Y.-F. Liu, Y.-H. Dai, and Z.-Q. Luo, ``{Coordinated beamforming for {MISO}
  interference channel: Complexity analysis and efficient algorithms},''
  \emph{IEEE Transactions on Signal Processing}, vol.~59, no.~3, pp.
  1142--1157, 2011.

\bibitem{Bjornson2014}
E.~Bj\"{o}rnson, M.~Bengtsson, and B.~Ottersten, ``Optimal multiuser transmit
  beamforming: A difficult problem with a simple solution structure [lecture
  notes],'' \emph{IEEE Signal Processing Magazine}, vol.~31, no.~4, pp.
  142--148, July 2014.

\bibitem{Gershman2010}
A.~B. Gershman, N.~D. Sidiropoulos, S.~Shahbazpanahi, M.~Bengtsson, and
  B.~Ottersten, ``Convex optimization-based beamforming,'' \emph{IEEE Signal
  Processing Magazine}, vol.~27, no.~3, pp. 62--75, May 2010.

\bibitem{Bengtsson2001}
M.~Bengtsson and B.~Ottersten, ``Optimal and suboptimal transmit beamforming,''
  in \emph{Handbook of Antennas in Wireless Communications}.\hskip 1em plus
  0.5em minus 0.4em\relax CRC Press, 2001.

\bibitem{Schubert2004}
M.~Schubert and H.~Boche, ``Solution of the multiuser downlink beamforming
  problem with individual {SINR} constraints,'' \emph{IEEE Transactions on
  Vehicular Technology}, vol.~53, no.~1, pp. 18--28, Jan 2004.

\bibitem{DVB_S2X}
{ETSI EN 302 307-2}, ``Digital video broadcasting ({DVB}); second generation
  framing structure, channel coding and modulation systems for broadcasting,
  interactive services, news gathering and other broadband satellite
  applications; part 2: {DVB-S2} extensions ({DVB-S2X}).''

\bibitem{Alodeh2018_tutorial}
M.~{Alodeh}, D.~{Spano}, A.~{Kalantari}, C.~G. {Tsinos}, D.~{Christopoulos},
  S.~{Chatzinotas}, and B.~{Ottersten}, ``Symbol-level and multicast precoding
  for multiuser multiantenna downlink: A state-of-the-art, classification, and
  challenges,'' \emph{IEEE Communications Surveys Tutorials}, vol.~20, no.~3,
  pp. 1733--1757, thirdquarter 2018.

\bibitem{Spano2018IJSC}
D.~Spano, S.~Chatzinotas, S.~Andrenacci, J.~Krause, and B.~Ottersten,
  ``Per-antenna power minimization in symbol-level precoding for the multi-beam
  satellite downlink,'' \emph{International Journal of Satellite Communications
  and Networking}, May 2018.

\bibitem{Sidiropoulos2006}
N.~Sidiropoulos, T.~Davidson, and Z.-Q. Luo, ``Transmit beamforming for
  physical-layer multicasting,'' \emph{{IEEE} Trans. Signal Process.}, vol.~54,
  no.~6, pp. 2239--2251, 2006.

\bibitem{Silva2009}
Y.~C.~B. Silva and A.~Klein, ``Linear transmit beamforming techniques for the
  multigroup multicast scenario,'' \emph{{IEEE} Trans. Veh. Technol.}, vol.~58,
  no.~8, pp. 4353--4367, 2009.

\bibitem{christopoulos}
D.~{Christopoulos}, S.~{Chatzinotas}, and B.~{Ottersten}, ``Multicast
  multigroup precoding and user scheduling for frame-based satellite
  communications,'' \emph{IEEE Transactions on Wireless Communications},
  vol.~14, no.~9, pp. 4695--4707, Sep. 2015.

\bibitem{Christopoulos2014}
D.~Christopoulos, S.~Chatzinotas, and B.~Ottersten, ``Weighted fair multicast
  multigroup beamforming under per-antenna power constraints,'' \emph{IEEE
  Transactions on Signal Processing}, vol.~62, no.~19, pp. 5132--5142, Oct
  2014.

\bibitem{Mohammed2013}
S.~K. Mohammed and E.~G. Larsson, ``Per-antenna constant envelope precoding for
  large multi-user {MIMO} systems,'' \emph{IEEE Transactions on
  Communications}, vol.~61, no.~3, pp. 1059--1071, March 2013.

\bibitem{Studer2013}
C.~Studer and E.~G. Larsson, ``{PAR}-aware large-scale multi-user {MIMO}-{OFDM}
  downlink,'' \emph{IEEE Journal on Selected Areas in Communications}, vol.~31,
  no.~2, pp. 303--313, February 2013.

\bibitem{Spano2018TSP}
D.~Spano, M.~Alodeh, S.~Chatzinotas, and B.~Ottersten, ``Symbol-level precoding
  for the nonlinear multiuser {MISO} downlink channel,'' \emph{IEEE
  Transactions on Signal Processing}, vol.~66, no.~5, pp. 1331--1345, March
  2018.

\bibitem{survey_noma}
S.~M.~R. {Islam}, N.~{Avazov}, O.~A. {Dobre}, and K.~{Kwak}, ``Power-domain
  non-orthogonal multiple access (noma) in 5g systems: Potentials and
  challenges,'' \emph{IEEE Communications Surveys Tutorials}, vol.~19, no.~2,
  pp. 721--742, Secondquarter 2017.

\bibitem{leitwc_noma}
L.~{Lei}, D.~{Yuan}, C.~K. {Ho}, and S.~{Sun}, ``Power and channel allocation
  for non-orthogonal multiple access in 5g systems: Tractability and
  computation,'' \emph{IEEE Transactions on Wireless Communications}, vol.~15,
  no.~12, pp. 8580--8594, Dec 2016.

\bibitem{leitvt_noma}
L.~{Lei}, L.~{You}, Y.~{Yang}, D.~{Yuan}, S.~{Chatzinotas}, and B.~{Ottersten},
  ``Load coupling and energy optimization in multi-cell and multi-carrier noma
  networks,'' \emph{IEEE Transactions on Vehicular Technology}, vol.~68,
  no.~11, pp. 11\,323--11\,337, Nov 2019.

\bibitem{7562275}
L.~{Lei}, D.~{Yuan}, and P.~{Värbrand}, ``On power minimization for
  non-orthogonal multiple access (noma),'' \emph{IEEE Communications Letters},
  vol.~20, no.~12, pp. 2458--2461, Dec 2016.

\bibitem{atsc_noma}
L.~{Fay}, L.~{Michael}, D.~{Gómez-Barquero}, N.~{Ammar}, and M.~W. {Caldwell},
  ``An overview of the atsc 3.0 physical layer specification,'' \emph{IEEE
  Transactions on Broadcasting}, vol.~62, no.~1, pp. 159--171, March 2016.

\bibitem{perezneira2018noma}
A.~I. Perez-Neira, M.~Caus, M.~A. Vazquez, and N.~Alagha, ``Noma schemes for
  multibeam satellite communications,'' 2018.

\bibitem{caus_noma}
M.~{Caus}, M.~. {Vázquez}, and A.~{Pérez-Neira}, ``Noma and interference
  limited satellite scenarios,'' in \emph{2016 50th Asilomar Conference on
  Signals, Systems and Computers}, Nov 2016, pp. 497--501.

\bibitem{anyueoctr_noma}
A.~{Wang}, L.~{Lei}, E.~{Lagunas}, A.~I. {Pérez Neira}, S.~{Chatzinotas}, and
  B.~{Ottersten}, ``On fairness optimization for noma-enabled multi-beam
  satellite systems,'' in \emph{2019 IEEE 30th Annual International Symposium
  on Personal, Indoor and Mobile Radio Communications (PIMRC)}, Sept. 2019, pp.
  1--6.

\bibitem{beigi2018interference}
N.~A.~K. {Beigi} and M.~R. {Soleymani}, ``Interference management using
  cooperative noma in multi-beam satellite systems,'' in \emph{2018 IEEE
  International Conference on Communications (ICC)}, May 2018, pp. 1--6.

\bibitem{jsacnoma2017}
X.~{Zhu}, C.~{Jiang}, L.~{Kuang}, N.~{Ge}, and J.~{Lu}, ``Non-orthogonal
  multiple access based integrated terrestrial-satellite networks,'' \emph{IEEE
  Journal on Selected Areas in Communications}, vol.~35, no.~10, pp.
  2253--2267, Oct 2017.

\bibitem{jsacsp2019_noma}
Z.~{Lin}, M.~{Lin}, J.~{Wang}, T.~{de Cola}, and J.~{Wang}, ``Joint beamforming
  and power allocation for satellite-terrestrial integrated networks with
  non-orthogonal multiple access,'' \emph{IEEE Journal of Selected Topics in
  Signal Processing}, vol.~13, no.~3, pp. 657--670, June 2019.

\bibitem{yan2018hybrid}
X.~{Yan}, H.~{Xiao}, K.~{An}, G.~{Zheng}, and W.~{Tao}, ``Hybrid satellite
  terrestrial relay networks with cooperative non-orthogonal multiple access,''
  \emph{IEEE Communications Letters}, vol.~22, no.~5, pp. 978--981, May 2018.

\bibitem{sat_iot_phy2}
U.~{Raza}, P.~{Kulkarni}, and M.~{Sooriyabandara}, ``Low power wide area
  networks: An overview,'' \emph{IEEE Communications Surveys Tutorials},
  vol.~19, no.~2, pp. 855--873, Secondquarter 2017.

\bibitem{sat_iot_phy1}
Y.~.~E. {Wang}, X.~{Lin}, A.~{Adhikary}, A.~{Grovlen}, Y.~{Sui},
  Y.~{Blankenship}, J.~{Bergman}, and H.~S. {Razaghi}, ``A primer on 3gpp
  narrowband internet of things,'' \emph{IEEE Communications Magazine},
  vol.~55, no.~3, pp. 117--123, March 2017.

\bibitem{sat_iot_phy3}
SigFox, ``{Technical Overview},'' {SigFox}, {White Paper}, 05-2017.

\bibitem{sat_iot_phy4}
L.~Vangelista, A.~Zanella, and M.~Zorzi, ``Long-range iot technologies: The
  dawn of lora{\texttrademark},'' in \emph{\textit{Future Access Enablers for
  Ubiquitous and Intelligent Infrastructures}}, V.~Atanasovski and
  A.~Leon-Garcia, Eds.\hskip 1em plus 0.5em minus 0.4em\relax Cham: Springer
  International Publishing, 2015, pp. 51--58.

\bibitem{sat_iot_mac13}
O.~{Kodheli}, S.~{Andrenacci}, N.~{Maturo}, S.~{Chatzinotas}, and F.~{Zimmer},
  ``Resource allocation approach for differential doppler reduction in nb-iot
  over leo satellite,'' in \emph{2018 9th Advanced Satellite Multimedia Systems
  Conference and the 15th Signal Processing for Space Communications Workshop
  (ASMS/SPSC)}, Sep. 2018, pp. 1--8.

\bibitem{sat_iot_phy7}
Y.~{Qian}, L.~{Ma}, and X.~{Liang}, ``Symmetry chirp spread spectrum modulation
  used in leo satellite internet of things,'' \emph{IEEE Communications
  Letters}, vol.~22, no.~11, pp. 2230--2233, Nov 2018.

\bibitem{sat_iot_phy5}
J.~{Doré} and V.~{Berg}, ``Turbo-fsk: A 5g nb-iot evolution for leo satellite
  networks,'' in \emph{2018 IEEE Global Conference on Signal and Information
  Processing (GlobalSIP)}, Nov 2018, pp. 1040--1044.

\bibitem{sat_iot_phy6}
Y.~{Roth}, J.~{Doré}, L.~{Ros}, and V.~{Berg}, ``Turbo-fsk: A new uplink
  scheme for low power wide area networks,'' in \emph{2015 IEEE 16th
  International Workshop on Signal Processing Advances in Wireless
  Communications (SPAWC)}, June 2015, pp. 81--85.

\bibitem{sat_iot_phy8}
Y.~{Qian}, L.~{Ma}, and X.~{Liang}, ``The acquisition method of symmetry chirp
  signal used in leo satellite internet of things,'' \emph{IEEE Communications
  Letters}, vol.~23, no.~9, pp. 1572--1575, Sep. 2019.

\bibitem{sat_iot_phy9}
C.~{Yang}, M.~{Wang}, L.~{Zheng}, and G.~{Zhou}, ``Folded chirp-rate shift
  keying modulation for leo satellite iot,'' \emph{IEEE Access}, vol.~7, pp.
  99\,451--99\,461, 2019.

\bibitem{sat_iot_phy10}
S.~{Scalise}, C.~P. {Niebla}, R.~{De Gaudenzi}, O.~{Del Rio Herrero},
  D.~{Finocchiaro}, and A.~{Arcidiacono}, ``S-mim: a novel radio interface for
  efficient messaging services over satellite,'' \emph{IEEE Communications
  Magazine}, vol.~51, no.~3, pp. 119--125, March 2013.

\bibitem{sat_iot_phy11}
C.~A. {Hofmann} and A.~{Knopp}, ``Direct access to satellites: An internet of
  remote things technology,'' in \emph{2019 IEEE 2nd 5G World Forum (5GWF)},
  Sep. 2019, pp. 578--583.

\bibitem{sat_iot_phy12}
------, ``Ultra-narrowband waveform for iot direct random multiple access to
  satellites,'' \emph{IEEE Internet of Things Journal}, pp. 1--1, 2019.

\bibitem{mdm5000}
\BIBentryALTinterwordspacing
``Product leaflet. newtec mdm5000 satellite modem,'' Tech. Rep. 3.2, June 2017.
  [Online]. Available:
  \url{http://www.newtec.eu/frontend/files/leaflet/newtec-mdm5000-satellite-modem-r3.2.pdf}
\BIBentrySTDinterwordspacing

\bibitem{hdrm}
\BIBentryALTinterwordspacing
``Rt logic hdrm high data rate modem,'' Tech. Rep. [Online]. Available:
  \url{http://www.rtlogic.com/products/modemsreceivers/hdrm-high-data-rate-modem}
\BIBentrySTDinterwordspacing

\bibitem{satix}
\BIBentryALTinterwordspacing
``Satixfy introducing wideband 500 mhz dvb-s2x sdr modem with record 2.4 gbps
  throughput,'' Tech. Rep., March 2018. [Online]. Available:
  \url{https://www.satixfy.com/news/ \\
  satixfy-introducing-wideband-500-mhz-dvb-s2x-sdr-modem-with \\
  -record-2-4-gbps-throughput/}
\BIBentrySTDinterwordspacing

\bibitem{Wideband_modem}
S.~Kisseleff, N.~Maturo, S.~Chatzinotas, H.~Fanebust, B.~Rislow, K.~Kansanen,
  M.~Arzel, and H.~Haugli, ``{}user terminal wideband modem for very high
  throughput satellites.''

\bibitem{gharanjik_large_2013}
A.~Gharanjik, B.~Rao, P.-D. Arapoglou, and B.~Ottersten, ``Large scale transmit
  diversity in {Q}/{V} band feeder link with multiple gateways,'' in \emph{2013
  {IEEE} 24th {International} {Symposium} on {Personal} {Indoor} and {Mobile}
  {Radio} {Communications} ({PIMRC})}, Sep. 2013, pp. 766--770.

\bibitem{kyrgiazos_gateway_2014}
A.~Kyrgiazos, B.~Evans, and P.~Thompson, ``On the {Gateway} {Diversity} for
  {High} {Throughput} {Broadband} {Satellite} {Systems},'' \emph{IEEE
  Transactions on Wireless Communications}, vol.~13, no.~10, pp. 5411--5426,
  Oct. 2014.

\bibitem{cowley_optical_2014}
W.~Cowley, D.~Giggenbach, and R.~Calvo, ``Optical transmission schemes for {}
  feeder links,'' in \emph{2014 {IEEE} {International} {Conference} on
  {Communications} ({ICC})}, Jun. 2014, pp. 4154--4159.

\bibitem{gharanjik_spatial_2014}
A.~Gharanjik, K.~Liolis, M.~Shankar, and B.~Ottersten, ``Spatial multiplexing
  in optical feeder links for high throughput satellites,'' in \emph{2014
  {IEEE} {Global} {Conference} on {Signal} and {Information} {Processing}
  ({GlobalSIP})}, Dec. 2014, pp. 1112--1116.

\bibitem{dimitrov_digital_2014}
S.~Dimitrov, B.~Matuz, G.~Liva, R.~Barrios, R.~Mata-Calvo, and D.~Giggenbach,
  ``Digital modulation and coding for satellite optical feeder links,'' in
  \emph{Advanced {Satellite} {Multimedia} {Systems} {Conference} and the 13th
  {Signal} {Processing} for {Space} {Communications} {Workshop}
  ({ASMS}/{SPSC}), 2014 7th}, Sep. 2014, pp. 150--157.

\bibitem{barrios_rivoli-tn1:_2015}
R.~Barrios, R.~Mata-Calvo, and D.~Giggenbach, ``{RivOLi}-{TN}1: {Review} of
  {RoFSO} techniques for {Optical} {GEO} {Feeder} {Links},'' Jan. 2015.

\bibitem{kaushal_optical_2016}
\BIBentryALTinterwordspacing
H.~Kaushal and G.~Kaddoum, ``Optical {Communication} in {Space}: {Challenges}
  and {Mitigation} {Techniques},'' \emph{IEEE Communications Surveys \&
  Tutorials}, pp. 1--1, 2016. [Online]. Available:
  \url{http://ieeexplore.ieee.org/document/7553489/}
\BIBentrySTDinterwordspacing

\bibitem{lyras_cloud_2017}
\BIBentryALTinterwordspacing
N.~K. Lyras, C.~I. Kourogiorgas, and A.~D. Panagopoulos, ``Cloud {Attenuation}
  {Statistics} {Prediction} {From} {Ka}-{Band} to {Optical} {Frequencies}:
  {Integrated} {Liquid} {Water} {Content} {Field} {Synthesizer},'' \emph{IEEE
  Transactions on Antennas and Propagation}, vol.~65, no.~1, pp. 319--328, Jan.
  2017. [Online]. Available: \url{http://ieeexplore.ieee.org/document/7747494/}
\BIBentrySTDinterwordspacing

\bibitem{perlot_model-oriented_2012}
\BIBentryALTinterwordspacing
N.~Perlot and J.~Perdigues-Armengol, ``Model-oriented availability analysis of
  optical {GEO}-ground links,'' Feb. 2012, p. 82460P. [Online]. Available:
  \url{http://proceedings.spiedigitallibrary.org/proceeding.aspx?doi=10.1117/12.908706}
\BIBentrySTDinterwordspacing

\bibitem{perlot_optical_2012}
N.~Perlot, T.~Dreischer, C.~Weinert, and J.~Perdigues, ``Optical {GEO} feeder
  link design,'' in \emph{Future {Network} {Mobile} {Summit} ({FutureNetw}),
  2012}, Jul. 2012, pp. 1--8.

\bibitem{dimitrov_digital_2015}
S.~Dimitrov, R.~Barrios, B.~Matuz, G.~Liva, R.~Mata-Calvo, and D.~Giggenbach,
  ``Digital modulation and coding for satellite optical feeder links with
  pre-distortion adaptive optics,'' \emph{Int. J. Satellite Commun. and
  Networking}, pp. 1--22, 2015.

\bibitem{smith_millimeter_1993}
\BIBentryALTinterwordspacing
J.~J. Degnan, ``\BIBforeignlanguage{en}{Millimeter accuracy satellite laser
  ranging: {A} review},'' in \emph{\BIBforeignlanguage{en}{Geodynamics
  {Series}}}, D.~E. Smith and D.~L. Turcotte, Eds.\hskip 1em plus 0.5em minus
  0.4em\relax Washington, D. C.: American Geophysical Union, 1993, vol.~25, pp.
  133--162, dOI: 10.1029/GD025p0133. [Online]. Available:
  \url{http://www.agu.org/books/gd/v025/GD025p0133/GD025p0133.shtml}
\BIBentrySTDinterwordspacing

\bibitem{giggenbach_high-throughput_2015}
D.~Giggenbach, E.~Lutz, J.~Poliak, R.~Mata-Calvo, and C.~Fuchs, ``A
  {High}-{Throughput} {Satellite} {System} for {Serving} whole {Europe} with
  {Fast} {Internet} {Service}, {Employing} {Optical} {Feeder} {Links},'' in
  \emph{Broadband {Coverage} in {Germany}. 9th {ITG} {Symposium}.
  {Proceedings}}, Apr. 2015, pp. 1--7.

\bibitem{mesleh_optical_2011}
R.~Mesleh, H.~Elgala, and H.~Haas, ``Optical {Spatial} {Modulation},''
  \emph{IEEE/OSA Journal of Optical Communications and Networking}, vol.~3,
  no.~3, pp. 234--244, Mar. 2011.

\bibitem{ozbilgin_optical_2015}
T.~Ozbilgin and M.~Koca, ``Optical {Spatial} {Modulation} {Over} {Atmospheric}
  {Turbulence} {Channels},'' \emph{Journal of Lightwave Technology}, vol.~33,
  no.~11, pp. 2313--2323, Jun. 2015.

\bibitem{simon_alamouti-type_2005}
M.~K. Simon and V.~Vilnrotter, ``Alamouti-type space-time coding for free-space
  optical communication with direct detection,'' \emph{IEEE Transactions on
  Wireless Communications}, vol.~4, no.~1, pp. 35--39, Jan. 2005.

\bibitem{safari_we_2008}
M.~Safari and M.~Uysal, ``Do {We} {Really} {Need} {OSTBCs} for {Free}-{Space}
  {Optical} {Communication} with {Direct} {Detection}?'' \emph{IEEE
  Transactions on Wireless Communications}, vol.~7, no.~11, pp. 4445--4448,
  Nov. 2008.

\bibitem{roy_optical_2015}
\BIBentryALTinterwordspacing
B.~Roy, S.~Poulenard, S.~Dimitrov, R.~Barrios, D.~Giggenbach, A.~L. Kernec, and
  M.~Sotom, ``Optical feeder links for high throughput satellites.''\hskip 1em
  plus 0.5em minus 0.4em\relax IEEE, Oct. 2015, pp. 1--6. [Online]. Available:
  \url{http://ieeexplore.ieee.org/document/7425074/}
\BIBentrySTDinterwordspacing

\bibitem{ONSET}
\BIBentryALTinterwordspacing
Luximpulse onset project: Optical feeder links study for satellite networks.
  [Online]. Available: \url{https://wwwfr.uni.lu/snt/research/sigcom/projects}
\BIBentrySTDinterwordspacing

\bibitem{juan_Radhakrishnan2016a}
R.~Radhakrishnan, W.~W. Edmonson, F.~Afghah, R.~M. Rodriguez-Osorio, F.~Pinto,
  and S.~C. Burleigh, ``{Survey of Inter-Satellite Communication for Small
  Satellite Systems: Physical Layer to Network Layer View},'' \emph{IEEE
  Communications Surveys and Tutorials}, vol.~18, no.~4, pp. 2442--2473, 2016.

\bibitem{juan_swarm_Edmonson2015}
W.~Edmonson, S.~Gebreyohannes, A.~Dillion, R.~Radhakrishnan, J.~Chenou,
  A.~Esterline, and F.~Afghah, ``{Systems engineering of inter-satellite
  communications for distributed systems of small satellites},'' in \emph{9th
  Annual IEEE International Systems Conference, SysCon 2015 -
  Proceedings}.\hskip 1em plus 0.5em minus 0.4em\relax Institute of Electrical
  and Electronics Engineers Inc., jun 2015, pp. 705--710.

\bibitem{juan_Manchester2015}
Z.~R. Manchester, ``{CENTIMETER-SCALE SPACECRAFT: DESIGN, FABRICATION, AND
  DEPLOYMENT},'' Ph.D. dissertation, 2015.

\bibitem{juan_MontiGuarnieri2015}
\BIBentryALTinterwordspacing
A.~{Monti Guarnieri}, A.~Broquetas, A.~Recchia, F.~Rocca, and J.~Ruiz-Rodon,
  ``{Advanced Radar Geosynchronous Observation System: ARGOS},'' \emph{IEEE
  Geoscience and Remote Sensing Letters}, vol.~12, no.~7, pp. 1406--1410, jul
  2015. [Online]. Available: \url{http://ieeexplore.ieee.org/document/7055288/}
\BIBentrySTDinterwordspacing

\bibitem{juan_Yu2009}
\BIBentryALTinterwordspacing
Z.~Yu, J.~Chen, C.~Li, Z.~Li, and Y.~Zhang, ``{Concepts, properties, and
  imaging technologies for GEO SAR},'' in \emph{MIPPR 2009: Multispectral Image
  Acquisition and Processing}, J.~K. Udupa, N.~Sang, L.~G. Nyul, and H.~Tong,
  Eds., vol. 7494, oct 2009, p. 749407. [Online]. Available:
  \url{http://proceedings.spiedigitallibrary.org/proceeding.aspx?doi=10.1117/12.832809}
\BIBentrySTDinterwordspacing

\bibitem{juan_swarm_Manchester2011}
\BIBentryALTinterwordspacing
Z.~R. Manchester and M.~A. Peck, ``{Stochastic space exploration with
  microscale spacecraft},'' in \emph{AIAA Guidance, Navigation, and Control
  Conference 2011}, 2011. [Online]. Available: \url{http://arc.aiaa.org}
\BIBentrySTDinterwordspacing

\bibitem{juan_swarm_Merlano-Duncan2019}
\BIBentryALTinterwordspacing
J.~C. Merlano-Duncan, J.~Querol, A.~Camps, S.~Chatzinotas, and B.~Ottersten,
  ``{Architectures and Synchronization Techniques for Coherent Distributed
  Remote Sensing Systems},'' \emph{IGARSS 2019 - 2019 IEEE International
  Geoscience and Remote Sensing Symposium}, pp. 8875--8878, jul 2019. [Online].
  Available: \url{https://ieeexplore.ieee.org/document/8898444/}
\BIBentrySTDinterwordspacing

\bibitem{NASA_DSN}
``{NASA} deep space network website,''
  \url{https://eyes.nasa.gov/dsn/dsn.html}.

\bibitem{ESA_DSN}
``{ESA} estrack website,'' \url{http://estracknow.esa.int}.

\bibitem{JPLCryo}
``Macgregor s. reid: "low-noise systems in the deep space network" nasa-jpl
  descanso book series, february 2008,''
  \url{https://descanso.jpl.nasa.gov/monograph/series10/Reid\_DESCANSO\_sml-110804.pdf}.

\bibitem{NASA_Cryo1}
F.~{Manshadi}, ``Nasa ultra low noise x-band microwave feeds for deep space
  communication,'' in \emph{15th International Conference on Microwaves, Radar
  and Wireless Communications (IEEE Cat. No.04EX824)}, vol.~2, May 2004, pp.
  733--736 Vol.2.

\bibitem{NASA_Cryo2}
J.~{Yang}, M.~{Pantaleev}, P.~{Kildal}, B.~{Klein}, Y.~{Karandikar},
  L.~{Helldner}, N.~{Wadefalk}, and C.~{Beaudoin}, ``Cryogenic 2–13 ghz
  eleven feed for reflector antennas in future wideband radio telescopes,''
  \emph{IEEE Transactions on Antennas and Propagation}, vol.~59, no.~6, pp.
  1918--1934, June 2011.

\bibitem{CCSDS_TM}
``{CCSDS} recommended standard 131.0-b-3: "tm synchronization and channel
  coding",'' \url{https://public.ccsds.org/Pubs/131x0b3e1.pdf}.

\bibitem{CCSDS_TC}
``{CCSDS} recommended standard 231.0-b-3 : "tc synchronization and channel
  coding",'' \url{https://public.ccsds.org/Pubs/231x0b3.pdf}.

\bibitem{RESCUE3}
M.~{Baldi}, F.~{Chiaraluce}, N.~{Maturo}, G.~{Ricciutelli}, R.~{Abello}, J.~{De
  Vicente}, M.~{Mercolino}, A.~{Ardito}, F.~{Barbaglio}, and S.~{Finocchiaro},
  ``Coding for space telemetry and telecommand transmissions in presence of
  solar scintillation,'' in \emph{2016 International Workshop on Tracking,
  Telemetry and Command Systems for Space Applications (TTC)}, Sep. 2016, pp.
  1--8.

\bibitem{RESCUE2}
S.~{Finocchiaro}, A.~{Ardito}, F.~{Barbaglio}, M.~{Baldi}, F.~{Chiaraluce},
  N.~{Maturo}, G.~{Ricciutelli}, L.~{Simone}, R.~{Abello}, J.~{de Vicente}, and
  M.~{Mercolino}, ``Improving deep space telecommunications during solar
  superior conjunctions,'' in \emph{2017 IEEE Aerospace Conference}, March
  2017, pp. 1--13.

\bibitem{RESCUE1}
M.~{Baldi}, F.~{Chiaraluce}, N.~{Maturo}, G.~{Ricciutelli}, A.~{Ardito},
  F.~{Barbaglio}, S.~{Finocchiaro}, L.~{Simone}, R.~{Abello}, J.~{De Vicente},
  and M.~{Mercolino}, ``Coded transmissions for space links affected by solar
  scintillation: Baseband analysis,'' vol.~37, no.~6, February 2019, pp.
  571--587.

\bibitem{yoo}
{Taesang Yoo} and A.~{Goldsmith}, ``On the optimality of multiantenna broadcast
  scheduling using zero-forcing beamforming,'' \emph{IEEE Journal on Selected
  Areas in Communications}, vol.~24, no.~3, pp. 528--541, March 2006.

\bibitem{ParragaNiebla2005}
C.~Parraga-Niebla and C.~Kissling, ``Design and performance evaluation of
  efficient scheduling techniques for second generation dvb-s systems,''
  \emph{AIAA Int. Commun. Satellite Systems Conf. (ICSSC), Rome , Italy}, Sep.
  2005.

\bibitem{Tropea2011}
M.~Tropea, F.~Veltri, F.~De~Rango, A.~Santamaria, and L.~Belcastro, ``Two step
  based qos scheduler for dvb-s2 satellite system,'' Jun. 2011.

\bibitem{4114275}
M.~A. {Vazquez Castro} and G.~S. {Granados}, ``Cross-layer packet scheduler
  design of a multibeam broadband satellite system with adaptive coding and
  modulation,'' \emph{IEEE Transactions on Wireless Communications}, vol.~6,
  no.~1, pp. 248--258, Jan 2007.

\bibitem{RendonMorales2011}
E.~Rendon-Morales, J.~Mata-D{\'i}az, J.~Alins, J.~Mu{\~{n}}oz, and O.~Esparza,
  ``{Adaptive Packet Scheduling for the Support of QoS over DVB-S2 Satellite
  Systems},'' \emph{Int. Conf. Wired/Wireless Internet Communications (WWIC),
  Vilanova i la Geltr{\'u}, Spain}, pp. 15--26, Jun. 2011.

\bibitem{Neely2003}
M.~Neely, E.~Modiano, and C.~Rohrs, ``Power allocation and routing in multibeam
  satellites with time-varying channels,'' \emph{IEEE Wireless Communications},
  vol.~11, no.~1, pp. 138--152, Feb. 2003.

\bibitem{NeelyPhd2003}
M.~Neely, ``{Dynamic Power Allocation and Routing for Satellite and Wireless
  Networks with Time Varying Channels},'' \emph{PhD, Massachusetts Institute of
  Technology, Cambridge, MA, USA}, Nov. 2003.

\bibitem{Du2009}
H.~Du, L.~Fan, and B.~Evans, ``{Efficient Packet Scheduling for Heterogeneous
  Multimedia Provisioning over Broadband Satellite Networks: An Adaptive
  Multidimensional QoS-Based Design},'' \emph{International Journal of
  Satellite Communications and Networking}, vol.~27, no.~1, pp. 53--85, 2009.

\bibitem{Fairhurst2008}
G.~Fairhurst, G.~Giambene, G.~Giannetti, C.~Parraga, and A.~Sali, ``{Multimedia
  traffic scheduling in DVB-S2 networks with mobile users},'' \emph{IEEE Int.
  Workshop on Satellite and Space Communications, Toulouse, France}, pp.
  211--215, Oct. 2008.

\bibitem{joroughi}
V.~{Joroughi}, M.~. {Vázquez}, and A.~I. {Pérez-Neira}, ``Generalized
  multicast multibeam precoding for satellite communications,'' \emph{IEEE
  Transactions on Wireless Communications}, vol.~16, no.~2, pp. 952--966, Feb
  2017.

\bibitem{taricco}
G.~Taricco, ``{Linear Precoding Methods for Multibeam Broadband Satellite
  Systems},'' \emph{European Wireless Conf.}, May 2014.

\bibitem{guidotti}
A.~{Guidotti} and A.~{Vanelli-Coralli}, ``Geographical scheduling for multicast
  precoding in multi-beam satellite systems,'' in \emph{2018 9th Advanced
  Satellite Multimedia Systems Conference and the 15th Signal Processing for
  Space Communications Workshop (ASMS/SPSC)}, Sep. 2018, pp. 1--8.

\bibitem{lagunas3}
E.~{Lagunas}, S.~{Andrenacci}, S.~{Chatzinotas}, and B.~{Ottersten},
  ``Cross-layer forward packet scheduling for emerging precoded broadband
  multibeam satellite system,'' in \emph{2018 9th Advanced Satellite Multimedia
  Systems Conference and the 15th Signal Processing for Space Communications
  Workshop (ASMS/SPSC)}, Sep. 2018, pp. 1--8.

\bibitem{5286351}
R.~{De Gaudenzi} and O.~{del Rio Herrero}, ``Advances in random access
  protocols for satellite networks,'' in \emph{2009 International Workshop on
  Satellite and Space Communications}, Sep. 2009, pp. 331--336.

\bibitem{cadsat}
{ESA funded project}, ``{Carrier Aggregation in Satellite Communication
  Networks},'' \url{https://artes.esa.int/projects/cadsat}.

\bibitem{choi}
J.~P. {Choi} and V.~W.~S. {Chan}, ``Optimum power and beam allocation based on
  traffic demands and channel conditions over satellite downlinks,'' \emph{IEEE
  Transactions on Wireless Communications}, vol.~4, no.~6, pp. 2983--2993, Nov
  2005.

\bibitem{lei}
J.~{Lei} and M.~. {Vázquez-Castro}, ``Multibeam satellite frequency/time
  duality study and capacity optimization,'' \emph{Journal of Communications
  and Networks}, vol.~13, no.~5, pp. 472--480, Oct 2011.

\bibitem{aravanis}
A.~I. {Aravanis}, B.~{Shankar M. R.}, P.~{Arapoglou}, G.~{Danoy}, P.~G.
  {Cottis}, and B.~{Ottersten}, ``Power allocation in multibeam satellite
  systems: A two-stage multi-objective optimization,'' \emph{IEEE Transactions
  on Wireless Communications}, vol.~14, no.~6, pp. 3171--3182, June 2015.

\bibitem{cocco}
G.~{Cocco}, T.~{de Cola}, M.~{Angelone}, Z.~{Katona}, and S.~{Erl}, ``Radio
  resource management optimization of flexible satellite payloads for dvb-s2
  systems,'' \emph{IEEE Transactions on Broadcasting}, vol.~64, no.~2, pp.
  266--280, June 2018.

\bibitem{alberti}
X.~{Alberti}, J.~M. {Cebrian}, A.~{Del Bianco}, Z.~{Katona}, J.~{Lei}, M.~A.
  {Vazquez-Castro}, A.~{Zanus}, L.~{Gilbert}, and N.~{Alagha}, ``System
  capacity optimization in time and frequency for multibeam multi-media
  satellite systems,'' in \emph{2010 5th Advanced Satellite Multimedia Systems
  Conference and the 11th Signal Processing for Space Communications Workshop},
  Sep. 2010, pp. 226--233.

\bibitem{freedman}
A.~Freedman, D.~Rainish, and Y.~Gat, ``{Beam Hopping – How To Make it
  Possible},'' \emph{Ka and Broadband Communication Conference, Bologna,
  Italy}, Oct. 2015.

\bibitem{eva_patent}
E.~Lagunas, S.~Chatzinotas, V.~Joroughi, and S.~Andrenacci, ``{Method for
  controlling the transmission of signals of a multibeam broadband
  satellite},'' \emph{{Filed in Luxembourg Patent Office ref. LU 100757}}, Apr.
  2018.

\bibitem{ginesi_patent}
A.~Ginesi, P.~Arapoglou, and E.~Re, ``{Interference-resilient flexible
  techniques for payload resource allocation in broadband satellites},''
  \emph{{PCT/EP2016/063358}}, Dec. 2017.

\bibitem{giraud}
X.~{Giraud}, G.~{Lesthievent}, and H.~{Méric}, ``Receiver synchronisation
  based on a single dummy frame for dvb-s2/s2x beam hopping systems,'' in
  \emph{2018 25th International Conference on Telecommunications (ICT)}, June
  2018, pp. 634--638.

\bibitem{kibria}
M.~Kibria, E.~Lagunas, N.~Maturo, D.~Spano, and S.~Chatzinotas, ``{Cluster
  Hopping in Precoded Multi-Beam High Throughput Satellite System},''
  \emph{{IEEE GlobeCom, Hawaii, USA}}, Dec. 2019.

\bibitem{khan}
Z.~{Khan}, H.~{Ahmadi}, E.~{Hossain}, M.~{Coupechoux}, L.~A. {Dasilva}, and
  J.~J. {Lehtomäki}, ``Carrier aggregation/channel bonding in next generation
  cellular networks: methods and challenges,'' \emph{IEEE Network}, vol.~28,
  no.~6, pp. 34--40, Nov 2014.

\bibitem{kibria2}
M.~Kibria, E.~Lagunas, N.~Maturo, D.~Spano, H.~Al-Hraishawi, and
  S.~Chatzinotas, ``{Carrier Aggregation in Multi-Beam Satellite
  Communications},'' \emph{{IEEE GlobeCom, Hawaii, USA}}, Dec. 2019.

\bibitem{sat_iot_mac12}
O.~{Kodheli}, S.~{Andrenacci}, N.~{Maturo}, S.~{Chatzinotas}, and F.~{Zimmer},
  ``An uplink ue group-based scheduling technique for 5g mmtc systems over leo
  satellite,'' \emph{IEEE Access}, vol.~7, pp. 67\,413--67\,427, 2019.

\bibitem{sat_iot_mac9}
R.~{De Gaudenzi} and O.~{del Rio Herrero}, ``Advances in random access
  protocols for satellite networks,'' in \emph{2009 International Workshop on
  Satellite and Space Communications}, Sep. 2009, pp. 331--336.

\bibitem{sat_iot_mac10}
C.~{Kissling} and A.~M. {Dlr}, ``On the integration of random access and dama
  channels for the return link of satellite networks,'' in \emph{2013 IEEE
  International Conference on Communications (ICC)}, June 2013, pp. 4282--4287.

\bibitem{sat_iot_mac1}
\BIBentryALTinterwordspacing
T.~Ferrer, S.~Céspedes, and A.~Becerra, ``Review and evaluation of mac
  protocols for satellite iot systems using nanosatellites,'' \emph{Sensors},
  vol.~19, no.~8, 2019. [Online]. Available:
  \url{https://www.mdpi.com/1424-8220/19/8/1947}
\BIBentrySTDinterwordspacing

\bibitem{sat_iot_mac14}
S.~{Cluzel}, M.~{Dervin}, J.~{Radzik}, S.~{Cazalens}, C.~{Baudoin}, and
  D.~{Dragomirescu}, ``Physical layer abstraction for performance evaluation of
  leo satellite systems for iot using time-frequency aloha scheme,'' in
  \emph{2018 IEEE Global Conference on Signal and Information Processing
  (GlobalSIP)}, Nov 2018, pp. 1076--1080.

\bibitem{sat_iot_mac2}
\BIBentryALTinterwordspacing
M.~Krondorf, M.~Goblirsch, R.~de~Gaudenzi, G.~Cocco, N.~Toptsidis, and G.~Acar,
  ``Towards the implementation of advanced random access schemes for satellite
  iot,'' \emph{International Journal of Satellite Communications and
  Networking}, vol. n/a, no. n/a. [Online]. Available:
  \url{https://onlinelibrary.wiley.com/doi/abs/10.1002/sat.1331}
\BIBentrySTDinterwordspacing

\bibitem{sat_iot_mac3}
\BIBentryALTinterwordspacing
R.~De~Gaudenzi, O.~Del Rio~Herrero, G.~Gallinaro, S.~Cioni, and P.-D.
  Arapoglou, ``Random access schemes for satellite networks, from vsat to m2m:
  a survey,'' \emph{International Journal of Satellite Communications and
  Networking}, vol.~36, no.~1, pp. 66--107, 2018. [Online]. Available:
  \url{https://onlinelibrary.wiley.com/doi/abs/10.1002/sat.1204}
\BIBentrySTDinterwordspacing

\bibitem{sat_iot_mac4}
A.~{Mengali}, R.~{De Gaudenzi}, and .~{Stefanović}, ``On the modeling and
  performance assessment of random access with sic,'' \emph{IEEE Journal on
  Selected Areas in Communications}, vol.~36, no.~2, pp. 292--303, Feb 2018.

\bibitem{sat_iot_mac6}
O.~{Del Rio Herrero} and R.~{De Gaudenzi}, ``High efficiency satellite multiple
  access scheme for machine-to-machine communications,'' \emph{IEEE
  Transactions on Aerospace and Electronic Systems}, vol.~48, no.~4, pp.
  2961--2989, October 2012.

\bibitem{sat_iot_mac5}
E.~{Casini}, R.~{De Gaudenzi}, and O.~{Del Rio Herrero}, ``Contention
  resolution diversity slotted aloha (crdsa): An enhanced random access
  schemefor satellite access packet networks,'' \emph{IEEE Transactions on
  Wireless Communications}, vol.~6, no.~4, pp. 1408--1419, April 2007.

\bibitem{sat_iot_mac7}
R.~{De Gaudenzi}, O.~{del Río Herrero}, G.~{Acar}, and E.~{Garrido Barrabés},
  ``Asynchronous contention resolution diversity aloha: Making crdsa truly
  asynchronous,'' \emph{IEEE Transactions on Wireless Communications}, vol.~13,
  no.~11, pp. 6193--6206, Nov 2014.

\bibitem{sat_iot_mac8}
A.~{Mengali}, R.~{De Gaudenzi}, and P.~{Arapoglou}, ``Enhancing the physical
  layer of contention resolution diversity slotted aloha,'' \emph{IEEE
  Transactions on Communications}, vol.~65, no.~10, pp. 4295--4308, Oct 2017.

\bibitem{sat_iot_mac11}
\BIBentryALTinterwordspacing
K.~Zidane, R.~De~Gaudenzi, N.~Alagha, and S.~Cioni, ``Phase noise impact on the
  performance of contention resolution slotted random access schemes,''
  \emph{International Journal of Satellite Communications and Networking}, vol.
  n/a, no. n/a. [Online]. Available:
  \url{https://onlinelibrary.wiley.com/doi/abs/10.1002/sat.1325}
\BIBentrySTDinterwordspacing

\bibitem{Sharma2014ASMS}
S.~K. {Sharma}, S.~{Maleki}, S.~{Chatzinotas}, J.~{Grotz}, and B.~{Ottersten},
  ``Implementation issues of cognitive radio techniques for ka-band
  (17.7–19.7 ghz) satcoms,'' in \emph{2014 7th Advanced Satellite Multimedia
  Systems Conference and the 13th Signal Processing for Space Communications
  Workshop (ASMS/SPSC)}, Sep. 2014, pp. 241--248.

\bibitem{Sharma2013IA}
S.~K. Sharma, S.~Chatzinotas, and B.~Ottersten, ``Interference alignment for
  spectral coexistence of heterogeneous networks,'' \emph{EURASIP Journal on
  Wireless Communications and Networking}, vol. 2013, no.~1, p.~46, Feb 2013.

\bibitem{Freqpacking2013VTC}
S.~{Chatzinotas}, S.~K. {Sharma}, and B.~{Ottersten}, ``Frequency packing for
  interference alignment-based cognitive dual satellite systems,'' in
  \emph{2013 IEEE 78th Vehicular Technology Conference (VTC Fall)}, Sep. 2013,
  pp. 1--7.

\bibitem{Hassan2017exclusive}
M.~R. {Hassan}, G.~C. {Karmakar}, J.~{Kamruzzaman}, and B.~{Srinivasan},
  ``Exclusive use spectrum access trading models in cognitive radio networks: A
  survey,'' \emph{IEEE Commun. Surveys Tuts.}, vol.~19, no.~4, pp. 2192--2231,
  Fourthquarter 2017.

\bibitem{Sharma2015CR}
S.~K. {Sharma}, T.~E. {Bogale}, S.~{Chatzinotas}, B.~{Ottersten}, L.~B. {Le},
  and X.~{Wang}, ``Cognitive radio techniques under practical imperfections: A
  survey,'' \emph{IEEE Commun. Surveys Tuts.}, vol.~17, no.~4, pp. 1858--1884,
  Fourthquarter 2015.

\bibitem{SharmaFD2018}
S.~K. {Sharma}, T.~E. {Bogale}, L.~B. {Le}, S.~{Chatzinotas}, X.~{Wang}, and
  B.~{Ottersten}, ``Dynamic spectrum sharing in {5G} wireless networks with
  full-duplex technology: Recent advances and research challenges,'' \emph{IEEE
  Commun. Surveys Tuts.}, vol.~20, no.~1, pp. 674--707, Feb. 2018.

\bibitem{Yang2016advanced}
C.~{Yang}, J.~{Li}, M.~{Guizani}, A.~{Anpalagan}, and M.~{Elkashlan},
  ``Advanced spectrum sharing in 5g cognitive heterogeneous networks,''
  \emph{IEEE Wireless Commun.}, vol.~23, no.~2, pp. 94--101, April 2016.

\bibitem{Christdualsat}
D.~Christopoulos, S.~K. Sharma, S.~Chatzinotas, J.~Krause, and B.~Ottersten,
  ``Coordinated multibeam satellite co-location: The dual satellite paradigm,''
  \emph{arXiv preprint arXiv: 1503.06981v1}, 2015.

\bibitem{Sharmacogbeamhop}
S.~K. Sharma, S.~Chatzinotas, and B.~Ottersten, ``Cognitive beamhop-ping for
  spectral coexistence of multibeam satellites,'' \emph{Int. J. Satellite
  Commun. and Networking}, vol.~33, no.~1, pp. 69--91, Mar. 2014.

\bibitem{SharmaInline2014}
------, ``Inline interference mitigation techniques for spectral coexistence of
  {GEO} and {NGEO} satellites,'' \emph{Int. J. Satellite Commun. and
  Networking}, vol.~34, no.~1, pp. 11--39, Sept. 2014.

\bibitem{onftr1}
ONFTR-521, ``Sdn architecture,'' \emph{Issue 1.1}, Feb. 2016.

\bibitem{haleplidis1}
E.~{Haleplidis} and K.~P. (Editors), ``Software-defined networking (sdn):
  Layers and architecture terminology,'' \emph{IRTF RFC 7426}, Jan. 2015.

\bibitem{feamster1}
N.~{Feamster}, J.~{Rexford}, and E.~{Zegura}, ``The road to sdn: An
  intellectual history of programmable networks,'' \emph{ACM SIGCOMM Computer
  Communication Review}, vol.~44, no.~2, pp. 87--98, Apr. 2014.

\bibitem{tennenhouse1}
D.~{Tennenhouse} and D.~{Wetherall}, ``Towards an active network
  architecture,'' \emph{ACM SIGCOMM Computer Communication Review}, vol.~26,
  no.~2, pp. 5--18, Apr. 2016.

\bibitem{calvert1}
K.~{Calvert}, S.~{Bhattacharjee}, E.~{Zegura}, and J.~{Sterbenz}, ``Directions
  in active networks,'' in \emph{IEEE Communications Magazine}, Oct. 1998, pp.
  72--78.

\bibitem{van1}
J.~{van der Merwe}, S.~{Rooney}, L.~{Leslie}, and S.~{Crosby}, ``The tempest :
  A practical framework for network programmability,'' \emph{IEEE Network},
  vol.~12, no.~3, pp. 20--28, 1998.

\bibitem{smith1}
J.~{Smith}, K.~{Calvert}, S.~{Murphy}, H.~{Orman}, and L.~{Peterson},
  ``Activating networks: a progress report,'' \emph{Computer}, vol.~32, no.~4,
  pp. 32--41, Apr. 1999.

\bibitem{feamster2}
N.~{Feamster}, H.~{Balakrishman}, J.~{Rexford}, A.~{Shaikh}, and K.~{van der
  Merwe}, ``The case for separating routing from routers,'' \emph{ACM SIGCOMM
  Workshop on Future Directions in Network Architecture}, Sep. 2004.

\bibitem{caesar1}
M.~{Caesar}, N.~{Feamster}, J.~{Rexford}, A.~{Shaikh}, and J.~{van der Merwe},
  ``Design and implementation of a routing control platform,'' \emph{Proc. 2nd
  USENIX NSDI}, May. 2005.

\bibitem{salim1}
J.~{Salim}, H.~{Khosravi}, A.~{Klen}, and A.~{Kuznetsov}, ``Linux netlink as an
  ip services protocol,'' \emph{Internet Engineering Task Force, RFC 3549.},
  Jul. 2003.

\bibitem{khosravi1}
H.~{Khosravi} and T.~{Anderson}, ``Requirements for separation of ip control
  and forwarding,'' \emph{Internet Engineering Task Force, RFC 3654.}, Nov.
  2003.

\bibitem{yang1}
L.~{Yang}, R.~{Dantu}, T.~{Anderson}, and R.~{Gopal}, ``Forwarding and control
  element separation (forces) framework,'' \emph{Internet Engineering Task
  Force, RFC 3746.}, Apr. 2004.

\bibitem{casado1}
M.~{Casado}, M.~{Freedman}, J.~{Pettit}, J.~{Luo}, N.~{McKeown}, and
  S.~{Shenker}, ``Ethane: Taking control of the enterprise,'' \emph{ACM SIGCOMM
  Computer Communication Review}, vol.~37, no.~4, pp. 1--12, Oct. 2007.

\bibitem{mcKeown1}
N.~{McKeown}, T.~{Anderson}, H.~{Alakrishnan}, G.~{Parulkar}, L.~{Peterson},
  J.~{Rexford}, S.~{Shenker}, and T.~{Turner}, ``Openflow: Enabling innovation
  in campus networks,'' \emph{ACM SIGCOMM Computer Communication Review},
  vol.~38, no.~2, pp. 69--74, Apr. 2008.

\bibitem{sama1}
M.~{Sama}, L.~{Contreras}, J.~{Kaippallimalil}, I.~{Akiyoshi}, H.~{Qian}, and
  H.~{Ni}, ``Software-defined control of the virtualized mobile packet core,''
  \emph{IEEE Communications}, pp. 107--115, Feb. 2015.

\bibitem{bojic1}
D.~{Bojic}, E.~{Sasaki}, and N.~C. et~al., ``Advanced wireless and optical
  technologies for small-cell mobile backhaul with dynamic software-defined
  management,'' \emph{IEEE Communications Magazine}, pp. 86--93, Sep. 2013.

\bibitem{3gpp1}
``Architecture enhancements for control and user plane separation of epc
  nodes,'' \emph{3GPP TS 23.214, Release 14}, Jun. 2016.

\bibitem{3gpp2}
``System architecture for the 5g system,'' \emph{3GPP TS 23.501, Release 15},
  Dec. 2016.

\bibitem{onf1}
``Openflow-enabled transport sdn,'' \emph{Open Networking Foundation (ONF), ONF
  Solution Brief}, May. 2014.

\bibitem{onf2}
\BIBentryALTinterwordspacing
Open networking foundation (onf). wireless \& mobile working group. [Online].
  Available:
  \url{www.opennetworking.org/images/stories/downloads/working-groups/charter-wireless-mobile.pdf}
\BIBentrySTDinterwordspacing

\bibitem{netWorld1}
\BIBentryALTinterwordspacing
Networld2020 – satcom working group. the role of satellites in 5g. Version 5.
  [Online]. Available:
  \url{https://www.networld2020.eu/wp-content/uploads/2014/02/SatCom-in-5G\_v5.pdf}
\BIBentrySTDinterwordspacing

\bibitem{giambene1}
G.~{Giambene}, S.~{Kota}, and P.~{Pillai}, ``Satellite-5g integration: a
  network perspective,'' \emph{IEEE Netw}, vol.~32, pp. 25--31, 2018.

\bibitem{vital1}
\BIBentryALTinterwordspacing
H.~V.~R. Project. (2015) Satellite communication services: An integral part of
  the 5g ecosystem. [Online]. Available: \url{http://www.ict-vital.eu/}
\BIBentrySTDinterwordspacing

\bibitem{rossi1}
T.~{Rossi}, M.~{De Sanctis}, E.~{Cianca}, C.~{Fragale}, M.~{Ruggieri}, and
  H.~{Fenech}, ``Future space-based communications infrastructures based on
  high throughput satellites and software defined networking,'' \emph{IEEE
  International Symposium on Systems Engineering (ISSE)}, pp. 332--337, Sep.
  2015.

\bibitem{mendoza1}
F.~{Mendoza}, R.~{Ferrús}, and O.~{Sallent}, ``Experimental proof of concept
  of an sdn-based traffic engineering solution for hybrid satellite-terrestrial
  mobile backhauling,'' \emph{Int J Satell Commun Network}, pp. 1--16, 2019.

\bibitem{bertaux1}
L.~{Bertaux}, S.~{Medjiah}, P.~{Berthou}, and et~al., ``Software defined
  networking and virtualization for broadband satellite networks,'' \emph{IEEE
  Communications Magazine}, vol.~53, pp. 54--60, 2015.

\bibitem{bao1}
J.~{Bao}, B.~{Zhao}, W.~{Yu}, Z.~{Feng}, C.~{Wu}, and Z.~{Gong}, ``Opensan: A
  software-defined satellite network architecture,'' \emph{ACM SIGCOMM
  Computer Communication Review}, vol.~44, no.~4, pp. 347--348, Aug. 2014.

\bibitem{Kapovits1}
A.~{Kapovits}, S.~{Covaci}, C.~{Ververidis}, V.~{Siris}, and M.~{Guta},
  ``Advanced topics in service delivery over integrated satellite terrestrial
  networks,'' \emph{7th Advanced Satellite Multimedia Systems Conference and
  the 13th Signal Processing for Space Communications Workshop (ASMS/SPSC)},
  pp. 92--98, Sep. 2014.

\bibitem{xu1}
S.~{Xu}, X.~{Wang}, and M.~{Huang}, ``Software-defined next-generation
  satellite networks: architecture, challenges, and solutions,'' \emph{IEEE
  Access}, vol.~6, pp. 4027--4041, 2018.

\bibitem{akyildiz1}
I.~{Akyildiz}, P.~{Wang}, and S.-C. {Lin}, ``Softair: A software defined
  networking architecture for 5g wireless systems,'' \emph{Comput. Netw.},
  vol.~85, pp. 1--18, Jul. 2015.

\bibitem{ahmed2}
T.~{Ahmed}, E.~{Dubois}, J.-B. {Dupé}, R.~{Ferrús}, P.~{Gélard}, and
  N.~{Kuhn}, ``Software-defined satellite cloud ran,'' \emph{International
  Journal of Satellite Communications and Networking}, vol.~36, no.~1, pp.
  108--133, Jan. 2018.

\bibitem{ferrus4}
R.~{Ferrús}, O.~{Sallent}, T.~{Ahmed}, and R.~{Fedrizzi}, ``Towards
  sdn/nfv-enabled satellite ground segment systems: end-to-end traffic
  engineering use case,'' \emph{IEEE International Conference on Communications
  Workshops (ICC Workshops)}, 2017.

\bibitem{artes1}
\BIBentryALTinterwordspacing
Artes telecommunications \& integrated applications. satellite for 5g.
  [Online]. Available:
  \url{http://www.esa.int/Our\_Activities/Telecommunications\_Integrated\_Applications/Satellite\_for\_5G}
\BIBentrySTDinterwordspacing

\bibitem{mendoza2}
F.~{Mendoza}, R.~{Ferrús}, and O.~{Sallent}, ``Sdn-enabled satcom networks for
  satellite-terrestrial integration,'' \emph{Satellite Communications in the 5G
  Era, Chapter 3, IET Digital Library; 2018:61-99. ISBN: 978-1-78561-427-9.}

\bibitem{mendoza3}
------, ``Sdn-based traffic engineering for improved resilience in integrated
  satellite-terrestrial backhaul networks,'' \emph{4th International Conference
  on Information and Communication Technologies for Disaster Management
  (ICT-DM)}, Dec. 2017.

\bibitem{multi_cache_2001}
S.~{Ramesh}, I.~{Rhee}, and K.~{Guo}, ``Multicast with cache (mcache): an
  adaptive zero-delay video-on-demand service,'' \emph{IEEE Transactions on
  Circuits and Systems for Video Technology}, vol.~11, no.~3, pp. 440--456,
  March 2001.

\bibitem{Brinton13}
C.~G. Brinton, E.~Aryafar, S.~Corda, S.~Russo, R.~Reinoso, and M.~Chiang, ``An
  intelligent satellite multicast and caching overlay for cdns to improve
  performance in video applications,'' in \emph{Proc. 31st AIAA Int. Commun.
  Satellite Systems Conf.}, October 2013, pp. 2013--5664.

\bibitem{Kalan17}
A.~Kalantari, M.~Fittipaldi, S.~Chatzinotas, T.~X. Vu, and B.~Ottersten,
  ``Cache-assisted hybrid satellite-terrestrial backhauling for 5g cellular
  networks,'' in \emph{Proc. IEEE Global Commun. Conf.}, 2017, pp. 1--6.

\bibitem{VuKa}
{T. X. Vu et al}, ``Efficient 5g edge caching over satellite,'' in \emph{Proc.
  Int. Commun.Satellite Syst. Conf.}, 2018, pp. 1--5.

\bibitem{VuSAT19}
T.~X. Vu, Y.~Poirier, S.~Chatzinotas, N.~Maturo, J.~Grotz, and F.~Roelens,
  ``Modeling and implementation of 5g edge caching over satellite,''
  \emph{International Journal of Satellite Communications and Networking}, Jan.
  2020.

\bibitem{sat_wire_2005}
B.~{Evans}, M.~{Werner}, E.~{Lutz}, M.~{Bousquet}, G.~E. {Corazza}, G.~{Maral},
  and R.~{Rumeau}, ``Integration of satellite and terrestrial systems in future
  multimedia communications,'' \emph{IEEE Wireless Communications}, vol.~12,
  no.~5, pp. 72--80, Oct 2005.

\bibitem{Satellite_sate_2000}
H.~{Linder}, H.~D. {Clausen}, and B.~{Collini-Nocker}, ``Satellite internet
  services using dvb/mpeg-2 and multicast web caching,'' \emph{IEEE
  Communications Magazine}, vol.~38, no.~6, pp. 156--161, June 2000.

\bibitem{CS1}
E.~{Wang}, H.~{Li}, and S.~{Zhang}, ``Load balancing based on cache resource
  allocation in satellite networks,'' \emph{IEEE Access}, vol.~7, pp. 56
  864--56\,879, 2019.

\bibitem{CS2}
H.~L. H.~{Wu}, J.~{Li} and P.~{Hong}, ``A two-layer caching model for content
  delivery services in satellite-terrestrial networks,'' in \emph{IEEE Global
  Communications Conference}, Dec. 2016, pp. 1--6.

\bibitem{CS3}
C.~R. M.~{Luglio}, S.~P.~{Romano} and F.~{Zampognaro}, ``Service delivery
  models for converged satellite-terrestrial 5g network deployment: A
  satellite-assisted cdn use-case,'' \emph{IEEE Network}, vol.~33, no.~1, pp.
  142--150, January 2019.

\bibitem{CS4}
P.~Y. Y.~{Li}, Q.~{Zhang} and Z.~{Yang}, ``A back-tracing partition based
  on-path caching distribution strategy over integrated leo satellite and
  terrestrial networks,'' in \emph{2018 10th International Conference on
  Wireless Communications and Signal Processing (WCSP)}, Oct. 2018, pp. 1--6.

\bibitem{CS5}
M.~S. A.~A.~{Dowhuszko} and A.~I. {Pérez-Neira}, ``Integration of optical and
  satellite communication technologies to improve the cache filling time in
  future 5g edge networks,'' in \emph{2019 21st International Conference on
  Transparent Optical Networks (ICTON)}, July 2019, pp. 1--5.

\bibitem{CS6}
Y.~W.~G.~C. S.~{Liu}, X.~{Hu} and W.~{Wang}, ``Distributed caching based on
  matching game in leo satellite constellation networks,'' \emph{IEEE
  Communications Letters}, vol.~22, no.~2, pp. 300--303, Feb. 2018.

\bibitem{CS7}
X.~Y. K.~{An}, Y.~{Li} and T.~{Liang}, ``On the performance of cache-enabled
  hybrid satellite-terrestrial relay networks,'' \emph{IEEE Wireless
  Communications Letters}, p. 1—1, 2019.

\bibitem{mec_sat1}
Z.~{Zhang}, W.~{Zhang}, and F.~{Tseng}, ``Satellite mobile edge computing:
  Improving qos of high-speed satellite-terrestrial networks using edge
  computing techniques,'' \emph{IEEE Network}, vol.~33, no.~1, pp. 70--76,
  January 2019.

\bibitem{mec_sat2}
L.~Yan, S.~Cao, Y.~Gong, H.~Han, J.~Wei, Y.~Zhao, and S.~Yang, ``Satec: A 5g
  satellite edge computing framework based on microservice architecture,''
  \emph{Sensors (Basel, Switzerland)}, vol.~19, no. 4 831, Feb. 2019.

\bibitem{juan_Maheshwarappa2017}
M.~R. Maheshwarappa, M.~D. Bowyer, and C.~P. Bridges, ``{Improvements in CPU
  {\&} FPGA Performance for Small Satellite SDR Applications},'' \emph{IEEE
  Transactions on Aerospace and Electronic Systems}, vol.~53, no.~1, pp.
  310--322, 2017.

\bibitem{juan_Kozlowski2018}
S.~Kozlowski, K.~Kurek, J.~Skarzynski, K.~Szczygielska, and M.~Darmetko,
  ``{Investigation on adaptive satellite communication system performance using
  SDR technique},'' \emph{MIKON 2018 - 22nd International Microwave and Radar
  Conference}, pp. 363--366, 2018.

\bibitem{juan_8566051}
M.~R. {Maheshwarappa}, M.~D.~J. {Bowyer}, and C.~P. {Bridges}, ``A
  reconfigurable sdr architecture for parallel satellite reception,''
  \emph{IEEE Aerospace and Electronic Systems Magazine}, vol.~33, no.~11, pp.
  40--53, November 2018.

\bibitem{juan_GOMspace}
\BIBentryALTinterwordspacing
GOMspace, ``{GOMspace | Home}.'' [Online]. Available:
  \url{https://gomspace.com/home.aspx}
\BIBentrySTDinterwordspacing

\bibitem{juan_nanoavionics}
\BIBentryALTinterwordspacing
NanoAvionics, ``{Nanosatellites {\&} CubeSats | NanoAvionics}.'' [Online].
  Available: \url{https://nanoavionics.com/}
\BIBentrySTDinterwordspacing

\bibitem{juan_alen}
\BIBentryALTinterwordspacing
``{Al{\'{e}}n Space | Nanosatellites - CubeSats - Small Satellites}.''
  [Online]. Available: \url{https://alen.space/}
\BIBentrySTDinterwordspacing

\bibitem{juan_isis}
\BIBentryALTinterwordspacing
``{Turn-key CubeSat and nanosat solutions | Innovative Solutions In Space}.''
  [Online]. Available: \url{https://www.isispace.nl/satellite-solutions/}
\BIBentrySTDinterwordspacing

\bibitem{juan_tyvak}
\BIBentryALTinterwordspacing
I.~Tyvak Nano-Satellite~Systems, ``{Tyvak Nano-Satellite Systems, Inc |
  Home}.'' [Online]. Available: \url{https://www.tyvak.com/}
\BIBentrySTDinterwordspacing

\bibitem{juan_7811224}
M.~R. {Maheshwarappa}, M.~D.~J. {Bowyer}, and C.~P. {Bridges}, ``Improvements
  in cpu fpga performance for small satellite sdr applications,'' \emph{IEEE
  Transactions on Aerospace and Electronic Systems}, vol.~53, no.~1, pp.
  310--322, Feb 2017.

\bibitem{juan_sdr_fec1}
D.~{Digdarsini}, D.~{Mishra}, S.~{Mehta}, and T.~V.~S. {Ram}, ``Fpga
  implementation of fec encoder with bch ldpc codes for dvb s2 system,'' in
  \emph{2019 6th International Conference on Signal Processing and Integrated
  Networks (SPIN)}, March 2019, pp. 78--81.

\bibitem{juan_sdr_fec2}
R.~{Purnamasari}, H.~{Wijanto}, and I.~{Hidayat}, ``Design and implementation
  of ldpc(low density parity check) coding technique on fpga (field
  programmable gate array) for dvb-s2 (digital video broadcasting-satellite),''
  in \emph{2014 IEEE International Conference on Aerospace Electronics and
  Remote Sensing Technology}, Nov 2014, pp. 83--88.

\bibitem{juan_sdr_fec3}
Y.~{Zhu} and C.~{Chakrabarti}, ``Memory efficient ldpc code design for high
  throughput software defined radio (sdr) systems,'' in \emph{2007 IEEE
  International Conference on Acoustics, Speech and Signal Processing - ICASSP
  '07}, vol.~2, April 2007, pp. II--9--II--12.

\bibitem{juan_sdr_mcs}
P.~N.~T. {H.} and A.~{Gelgor}, ``Means to enhance the bandwidth gain from
  applying multicomponent signals in dvb-s2,'' in \emph{2019 IEEE International
  Conference on Electrical Engineering and Photonics (EExPolytech)}, Oct 2019,
  pp. 173--176.

\bibitem{juan_sdr_dvb}
R.~{Palisetty}, V.~K. {Sinha}, S.~{Mallick}, and K.~C. {Ray}, ``Fpga
  prototyping of energy dispersal and improved error efficiency techniques for
  dvb-satellite standard,'' in \emph{2015 International Conference on VLSI
  Systems, Architecture, Technology and Applications (VLSI-SATA)}, Jan 2015,
  pp. 1--5.

\bibitem{juan_sdr_dvbs2}
E.~R. {de Lima}, A.~F.~R. {Queiroz}, D.~C. {Alves}, G.~S. {da Silva}, C.~G.
  {Chaves}, J.~G. {Mertes}, and T.~M. {Marson}, ``A detailed dvb-s2 receiver
  implementation: Fpga prototyping and preliminary asic resource estimation,''
  in \emph{2014 IEEE Latin-America Conference on Communications (LATINCOM)},
  Nov 2014, pp. 1--6.

\bibitem{juan_6865984}
H.~C. {Bui} and L.~{Franck}, ``Cost effective emulation of geostationary
  satellite channels by means of software-defined radio,'' in \emph{2014 IEEE
  Metrology for Aerospace (MetroAeroSpace)}, May 2014, pp. 538--542.

\bibitem{juan_ni}
\BIBentryALTinterwordspacing
N.~Instruments, ``{USRP Software Defined Radio Device}.'' [Online]. Available:
  \url{http://ni.com/usrp}
\BIBentrySTDinterwordspacing

\bibitem{juan_lassena}
\BIBentryALTinterwordspacing
``{{\'{E}}TS : Laboratoire Sp{\'{e}}cialis{\'{e}} en Syst{\`{e}}mes
  Embarqu{\'{e}}s, Navigation et Avionique}.'' [Online]. Available:
  \url{https://lassena.etsmtl.ca/}
\BIBentrySTDinterwordspacing

\bibitem{juan_kratos}
\BIBentryALTinterwordspacing
``{RF Channel Simulator | Kratos}.'' [Online]. Available:
  \url{https://www.kratosdefense.com/products/space/signals/test-and-simulation/rf-channel-simulator?r=krtl}
\BIBentrySTDinterwordspacing

\bibitem{juan_nutaq}
\BIBentryALTinterwordspacing
``{PicoSDR Series for Wireless Multi-Standard Prototyping.}'' [Online].
  Available: \url{https://www.nutaq.com/products/picosdr/}
\BIBentrySTDinterwordspacing

\bibitem{juan_sdr_Getu2017}
T.~M. Getu, W.~Ajib, and O.~A. Yeste-Ojeda, ``{Tensor-Based Efficient
  Multi-Interferer RFI Excision Algorithms for SIMO Systems},'' \emph{IEEE
  Transactions on Communications}, vol.~65, no.~7, pp. 3037--3052, 2017.

\bibitem{juan_sdr_Getu2018}
T.~M. Getu, W.~Ajib, O.~A. Yeste-Ojeda, and R.~Landry, ``{Tensor-Based
  Efficient Multi-Interferer RFI Excision: Results Using Real-World Data},'' in
  \emph{2018 International Conference on Computing, Networking and
  Communications, ICNC 2018}.\hskip 1em plus 0.5em minus 0.4em\relax Institute
  of Electrical and Electronics Engineers Inc., jun 2018, pp. 917--921.

\bibitem{juan_sdr_Zhang2017}
E.~Zhang, J.~Zambrano, A.~Amrhar, R.~Landry, and W.~Ajib, ``{Design and
  implementation of a Wideband Radio using SDR for avionic applications},''
  \emph{ICNS 2017 - ICNS: CNS/ATM Challenges for UAS Integration}, pp. 1--9,
  2017.

\bibitem{juan_serenade7}
J.~Duncan, J.~Querol, N.~Maturo, J.~Krivochiza, D.~Spano, N.~Saba, L.~Marrero,
  and S.~Chatzinotas, ``{Hardware Precoding Demonstration in Multi-Beam UHTS
  Communications under Realistic Payload Characteristics},'' in \emph{37th
  International Communications Satellite Systems Conference (ICSSC 2019)}, nov
  2019.

\bibitem{juan_serenade6}
J.~{Duncan}, J.~{Krivochiza}, S.~{Andrenacci}, S.~{Chatzinotas}, and
  B.~{Ottersten}, ``Hardware demonstration of precoded communications in
  multi-beam uhts systems,'' in \emph{36th International Communications
  Satellite Systems Conference (ICSSC 2018)}, Oct 2018, pp. 1--5.

\bibitem{juan_serenade5}
N.~{Maturo}, J.~C.~M. {Duncan}, J.~{Krivochiza}, J.~{Querol}, D.~{Spano},
  S.~{Chatzinotas}, and B.~{Ottersten}, ``Demonstrator of precoding technique
  for a multi-beams satellite system,'' in \emph{2019 8th International
  Workshop on Tracking, Telemetry and Command Systems for Space Applications
  (TTC)}, Sep. 2019, pp. 1--8.

\bibitem{juan_8340048}
C.~{Politis}, S.~{Maleki}, J.~M. {Duncan}, J.~{Krivochiza}, S.~{Chatzinotas},
  and B.~{Ottesten}, ``Sdr implementation of a testbed for real-time
  interference detection with signal cancellation,'' \emph{IEEE Access},
  vol.~6, pp. 20\,807--20\,821, 2018.

\bibitem{juan_serenade8}
\BIBentryALTinterwordspacing
L.~{Martinez Marrero}, J.~C. {Merlano Duncan}, J.~Querol, S.~Chatzinotas, A.~J.
  {Camps Carmona}, and B.~Ottersten, ``{Effects of multiple oscillator phase
  noise in precoding performance},'' in \emph{ICSSC2019}, 2019. [Online].
  Available: \url{http://orbilu.uni.lu/handle/10993/40625}
\BIBentrySTDinterwordspacing

\bibitem{satis1}
\BIBentryALTinterwordspacing
SATis5. [Online]. Available: \url{https://satis5.eurescom.eu/}
\BIBentrySTDinterwordspacing

\bibitem{juan_serenade4}
J.~{Krivochiza}, J.~C. {Merlano-Duncan}, S.~{Chatzinotas}, and B.~{Ottersten},
  ``M-qam modulation symbol-level precoding for power minimization: Closed-form
  solution,'' in \emph{2019 16th International Symposium on Wireless
  Communication Systems (ISWCS)}, Aug 2019, pp. 395--399.

\bibitem{juan_serenade3}
J.~{Krivochiza}, J.~{Merlano Duncan}, S.~{Andrenacci}, S.~{Chatzinotas}, and
  B.~{Ottersten}, ``Fpga acceleration for computationally efficient
  symbol-level precoding in multi-user multi-antenna communication systems,''
  \emph{IEEE Access}, vol.~7, pp. 15\,509--15\,520, 2019.

\bibitem{juan_serenade2}
J.~C. {Merlano-Duncan}, J.~{Krivochiza}, S.~{Andrenacci}, S.~{Chatzinotas}, and
  B.~{Ottersten}, ``Computationally efficient symbol-level precoding
  communications demonstrator,'' in \emph{2017 IEEE 28th Annual International
  Symposium on Personal, Indoor, and Mobile Radio Communications (PIMRC)}, Oct
  2017, pp. 1--5.

\bibitem{juan_serenade1}
J.~{Krivochiza}, J.~C. {Merlano-Duncan}, S.~{Andrenacci}, S.~{Chatzinotas}, and
  B.~{Ottersten}, ``Closed-form solution for computationally efficient
  symbol-level precoding,'' in \emph{2018 IEEE Global Communications Conference
  (GLOBECOM)}, Dec 2018, pp. 1--6.

\bibitem{itrinegy1}
\BIBentryALTinterwordspacing
Itrinegy. Satellite for 5g. [Online]. Available:
  \url{https://itrinegy.com/networkemulators/}
\BIBentrySTDinterwordspacing

\bibitem{datasoft1}
\BIBentryALTinterwordspacing
Datasoft. [Online]. Available:
  \url{https://www.datasoft.com/products/embedded\_software/satellitesim/index.html}
\BIBentrySTDinterwordspacing

\bibitem{opensand1}
\BIBentryALTinterwordspacing
OpenSAND. [Online]. Available: \url{https://opensand.org/content/home.php}
\BIBentrySTDinterwordspacing

\bibitem{sns1}
\BIBentryALTinterwordspacing
SNS3. [Online]. Available: \url{https://www.sns3.org/content/home.php}
\BIBentrySTDinterwordspacing

\bibitem{artes2}
\BIBentryALTinterwordspacing
Artes real-time satellite network emulator. [Online]. Available:
  \url{https://artes.esa.int/funding/real-time-satellite-network-emulator-artes-3a082-0}
\BIBentrySTDinterwordspacing

\bibitem{allstar1}
\BIBentryALTinterwordspacing
5g-allstar project. [Online]. Available: \url{https://5g-allstar.eu/project/}
\BIBentrySTDinterwordspacing

\bibitem{sat5g1}
\BIBentryALTinterwordspacing
Sat-5g project. [Online]. Available: \url{https://www.sat5g-project.eu/}
\BIBentrySTDinterwordspacing

\bibitem{Digitaltwin1}
{IBM}, ``{Introduction to Digital Twin: Simple, but detailed},''
  Online:https://www.youtube.com/watch?v=RaOejcczPas, online; accessed 10th Jan
  2020.

\bibitem{Digitaltwin2}
E.~H. Glaessgen and D.~Stargel, ``The digital twin paradigm for future nasa and
  u.s. air force vehicles,'' in \emph{53rd Structures, Structural Dynamics, and
  Materials Conference: Special Session on the Digital Twin}, June 2012, pp.
  1--14.

\bibitem{Digitaltwin3}
S.~Carsten, ``Space-based ``digital twin'' of earth brings affordable insights,
  and web connectivity, to the other seven billion of us,''
  https://medium.com/@cstoecker/space-based-digital-twin-of-earth-brings-affordable-insights-and-web-connectivity-to-the-other-83428572b92a,
  online; accessed 10th Jan 2020.

\bibitem{juan_swarm_EGlennon2018a}
\BIBentryALTinterwordspacing
E.~A. {\&}. A.~D. {E Glennon}, E.~P. Glennon, A.~G. Dempster, and
  E.~Aboutanios, ``{Distributed Beamforming Architectures for Space {\&}
  Airborne Applications},'' 2018. [Online]. Available:
  \url{http://www.ignss2018.unsw.edu.au/sites/ignss2018/files/u80/Papers/IGNSS2018{\_}paper{\_}29.pdf
  http://www.ignss2018.unsw.edu.au/sites/ignss2018/files/u80/Slides/D1-S2-ThB-Glennon{\_}DistributedBeamforming.pdf}
\BIBentrySTDinterwordspacing

\bibitem{juan_8412572}
X.~{Cai}, M.~{Zhou}, T.~{Xia}, W.~H. {Fong}, W.~{Lee}, and X.~{Huang},
  ``Low-power sdr design on an fpga for intersatellite communications,''
  \emph{IEEE Transactions on Very Large Scale Integration (VLSI) Systems},
  vol.~26, no.~11, pp. 2419--2430, Nov 2018.

\bibitem{juan_Engelen2014}
S.~Engelen, E.~Gill, and C.~Verhoeven, ``{On the reliability, availability, and
  throughput of satellite swarms},'' \emph{IEEE Transactions on Aerospace and
  Electronic Systems}, vol.~50, no.~2, pp. 1027--1037, 2014.

\bibitem{juan_ISL2_6496947}
A.~{Budianu}, T.~J.~W. {Castro}, A.~{Meijerink}, and M.~J. {Bentum},
  ``Inter-satellite links for cubesats,'' in \emph{2013 IEEE Aerospace
  Conference}, March 2013, pp. 1--10.

\bibitem{juan_TANDEMX_Fiedler2008}
\BIBentryALTinterwordspacing
H.~Fiedler, G.~Krieger, M.~Zink, M.~Younis, M.~Bachmann, S.~Huber, I.~Hajnsek,
  and A.~Moreira, ``{The TanDEM-X mission: an overview},'' in \emph{2008
  International Conference on Radar}.\hskip 1em plus 0.5em minus 0.4em\relax
  IEEE, sep 2008, pp. 60--64. [Online]. Available:
  \url{http://ieeexplore.ieee.org/document/4653892/}
\BIBentrySTDinterwordspacing

\bibitem{juan_olfar_Bentum2009}
M.~J. Bentum, C.~J. Verhoeven, A.~J. Boonstra, A.~J. {Van Der Veen}, and E.~K.
  Gill, ``{A novel astronomical application for formation flying small
  satellites},'' in \emph{60th International Astronautical Congress 2009, IAC
  2009}, vol.~2.\hskip 1em plus 0.5em minus 0.4em\relax Press IAC, oct 2009,
  pp. 1254--1261.

\bibitem{juan_olfar_Rajan2011}
R.~T. Rajan, S.~Engelen, M.~Bentum, and C.~Verhoeven, ``{Orbiting low frequency
  array for radio astronomy},'' in \emph{IEEE Aerospace Conference
  Proceedings}.\hskip 1em plus 0.5em minus 0.4em\relax IEEE Aerospace and
  Electronic Systems Society, apr 2011, pp. 1--11.

\bibitem{juan_olfar_Willink-Castro2012}
T.~Willink-Castro, A.~Budianu, A.~Meijerink, and M.~J. Bentum, ``{Antenna
  system design for olfar's inter-satellite link},'' p. B2.3.5, oct 2012.

\bibitem{juan_olfar_82db56c8ed40472eb74af00b0512cb42}
A.~Budianu, T.~Willink-Castro, A.~Meijerink, and M.~Bentum,
  ``\BIBforeignlanguage{Undefined}{Communication schemes for olfar's
  inter-satellite links},'' in \emph{\BIBforeignlanguage{Undefined}{63rd
  International Astronautical Congress (IAC 2012)}}.\hskip 1em plus 0.5em minus
  0.4em\relax International Astronautical Federation (IAF), 10 2012, p. B2.5.2.

\bibitem{juan_QB50}
\BIBentryALTinterwordspacing
``{QB50 Project},'' 2016. [Online]. Available: \url{https://www.qb50.eu/}
\BIBentrySTDinterwordspacing

\bibitem{juan_qb50_e592ba6fe65043a2a4bd954d3a2b7b1e}
E.~Gill, P.~Sundaramoorthy, J.~Bouwmeester, and B.~Sanders,
  ``\BIBforeignlanguage{English}{Formation flying to enhance the qb50 space
  network},'' in \emph{\BIBforeignlanguage{English}{Proceedings of the 4S
  symposium}}, s.n., Ed.\hskip 1em plus 0.5em minus 0.4em\relax CNES/ESA, 2010,
  pp. 1--15.

\bibitem{juan_iSL1}
M.~{Werner}, A.~{Jahn}, E.~{Lutz}, and A.~{Bottcher}, ``Analysis of system
  parameters for leo/ico-satellite communication networks,'' \emph{IEEE Journal
  on Selected Areas in Communications}, vol.~13, no.~2, pp. 371--381, Feb 1995.

\bibitem{juan_iridium_5340513}
S.~R. {Pratt}, R.~A. {Raines}, C.~E. {Fossa}, and M.~A. {Temple}, ``An
  operational and performance overview of the iridium low earth orbit satellite
  system,'' \emph{IEEE Communications Surveys}, vol.~2, no.~2, pp. 2--10,
  Second 1999.

\bibitem{Ahamadi17}
H.~{Ahmadi}, K.~{Katzis}, and M.~Z. {Shakir}, ``A novel airborne
  self-organising architecture for 5g+ networks,'' in \emph{2017 IEEE 86th
  Vehicular Technology Conference (VTC-Fall)}, Sep. 2017, pp. 1--5.

\bibitem{NASAinternet}
J.~{Jackson}, ``The interplanetary internet [networked space communications],''
  \emph{IEEE Spectrum}, vol.~42, no.~8, pp. 30--35, Aug 2005.

\bibitem{David_FBS}
Z.~Becvar, M.~Vondra, P.~Mach, J.~Plachy, and D.~Gesbert, ``Performance of
  mobile networks with {UAV}s: Can flying base stations substitute ultra-dense
  small cells?'' in \emph{Proc. 23th European Wireless Conference}, 2017.

\bibitem{Vatalaroanalysis}
F.~{Vatalaro}, G.~E. {Corazza}, C.~{Caini}, and C.~{Ferrarelli}, ``Analysis of
  {LEO}, {MEO}, and {GEO} global mobile satellite systems in the presence of
  interference and fading,'' \emph{IEEE Journal on Selected Areas in
  Communications}, vol.~13, no.~2, pp. 291--300, Feb 1995.

\bibitem{Pourmoghadas2017}
A.~Pourmoghadas, S.~K. Sharma, S.~Chatzinotas, and B.~Ottersten, ``On the
  spectral coexistence of {GSO} and {NGSO} {FSS} systems: Power control
  mechanisms and a methodology for inter-site distance determination,''
  \emph{Int. J. Satellite Commun. and Networking}, vol.~35, no.~5, pp.
  443--459, Sept. 2017.

\bibitem{Park2010interference}
C.~{Park}, C.~{Kang}, Y.~{Choi}, and C.~{Oh}, ``Interference analysis of
  geostationary satellite networks in the presence of moving non-geostationary
  satellites,'' in \emph{2010 2nd International Conference on Information
  Technology Convergence and Services}, Aug 2010, pp. 1--5.

\bibitem{Terminalside2016}
S.~K. Sharma, S.~Chatzinotas, and B.~Ottersten, ``Terminal-side interference
  mitigation for spectral coexistence of satellite and terrestrial systems in
  non-exclusive ka-band,'' in \emph{Proc. 22nd Ka and 34th ICSSC conference},
  Oct. 2016, pp. 1--5.

\bibitem{QKD1}
C.~H. Bennett, ``Quantum cryptography: public key distribution and coin
  tossing,'' in \emph{Proc. IEEE Int. Conf. Comput. Syst. Signal Process},
  1984.

\bibitem{QKD3}
\BIBentryALTinterwordspacing
Quartz :quantum cryptography telecommunication system. [Online]. Available:
  \url{https://www.esa.int/Applications/Telecommunications_Integrated_Applications/Space_photons_bring_a_new_dimension_to_cryptography}
\BIBentrySTDinterwordspacing

\bibitem{SharmamassiveMTC}
S.~K. {Sharma} and X.~{Wang}, ``Towards massive machine type communications in
  ultra-dense cellular {IoT} networks: Current issues and machine
  learning-assisted solutions,'' \emph{IEEE Communications Surveys Tutorials},
  pp. 1--1, 2019.

\bibitem{Raj2019unsupervised}
S.~{Rajendran}, W.~{Meert}, V.~{Lenders}, and S.~{Pollin}, ``Unsupervised
  wireless spectrum anomaly detection with interpretable features,'' \emph{IEEE
  Transactions on Cognitive Communications and Networking}, vol.~5, no.~3, pp.
  637--647, 2019.

\end{thebibliography}

%








\end{document}